## Phonon dispersions and electronic structures of two-dimensional IV-V compounds

Wanxing Lin, [1,#] Shi-Dong Liang, [1] Jiesen Li, [2,*] Dao-Xin Yao[1,†]

[1] State Key Laboratory of Optoelectronic Materials and Technologies, School of Physics, Sun Yat-Sen University, Guangzhou, P. R. China

[2] School of Environment and Chemical Engineering, Foshan University, Foshan, P. R. China

Present address: # W. L.: Institute of Applied Physics and Materials Engineering, University of Macau, N23 Avenida da Universidade, Taipa, Macau, China

**Abstract**

One novel family of two-dimensional IV-V compounds have been proposed, whose dynamical stabilities and electronic properties have been systematically investigated using the density functional theory. Extending from our previous work, two phases of carbon phosphorus bilayers α- and β-$C_2P_2$ have been proposed. Both of them are dynamically stable and thermally stable at 300K. They possess intrinsic HSE gaps of 2.70 eV and 2.67 eV, respectively. Similar α- and β-$C_2Y_2$ (Y= As, Sb, and Bi) can be obtained if the phosphorus atoms in the α- and β-$C_2P_2$ replaced by other pnictogens, respectively. If the C atoms in the α- and β-$C_2Y_2$ (Y= P, As, Sb, and Bi) are further replaced by other IV elements X (X=Si, Ge, Sn, and Pb), respectively, more derivatives of α- and β-$X_2Y_2$ (Y=N, P, As, Sb, and Bi) also can be obtained. It was found that the majority of them are dynamically stable. The proposed compounds range from metal to insulators depending on their constitutions. All insulated compounds can undergo a transition from insulator to metal induced by biaxial strain. Some of them can undergo a transition from indirect band gap to direct band gap. These new compounds can become candidates as photovoltaic device, thermoelectric material field as well as lamellated superconductors.

**Keywords**: Two-dimensional, Carbon phosphorus, IV-V compounds, Phonon dispersions, Electronic structures, Biaxial strain

---

*Corresponding author. E-mail: ljs@fosu.edu.cn (Jiesen Li)

†Corresponding author. E-mail: yaodaox@mail.sysu.edu.cn (Dao-Xin Yao)

Wanxing Lin: 0000-0001-9763-6299.

Shi-Dong Liang: 0000-0001-7753-0024.

Jiesen Li: 0000-0002-5230-2874.

Dao-Xin Yao: 0000-0003-1097-3802.

## 1. Introduction

Since the discovery of graphene by the mechanical exfoliation in experiment [1], the research on this fantastic two-dimensional (2D) materials has become a focus due to its superior electronic [2] and transport properties [3-5]. Similarly, some other 2D materials comprised of group IV atoms, such as silicene, 2D germanium [6], stanene [7], and plumbene [8, 9], have been proposed theoretically. All of them can exhibit topological properties under certain conditions [7-12], and some phases have been realized in experiment [13-16]. On the other hand, various 2D monolayers comprised of V atoms also have been widely investigated both theoretically [17-20] and experimentally [21-23]. Remarkably, the few-layer phosphorene has an impressive potential application due to its high mobility [24, 25] and novel transport properties [26]. As large tunable bandgap insulators, the nitrogen atomic monolayers [27-31] may be applied in the straintronics field [32]. Following the astonishing superconductivity and correlated property that have been detected in the bilayer twisted graphene [33-39], the investigation of bilayer twisted materials enriched the potential application of 2D materials [40-43].

Besides the exploration of the 2D materials comprised of only one element in group IV or V, various binary compounds also received significant attention in recent years. For example, three phases of carbon nitride bilayers have been introduced in our previous work [44], which exhibit novel electronic and mechanical properties [45, 46]. As the nearest neighbor of nitrogen in group V, the phosphorus can also form various binaries with the carbon. The study on organophosphorus compounds contain both carbon and phosphorus is an active field in material science. The phosphorus-carbon binary compounds, which are referred to as phosphorus carbides (PCs), with various stoichiometries have been studied both experimentally and theoretically [47]. Recently, various 2D carbon-phosphorus (CPs) monolayers have been predicted by the first principle investigations [48-53]. The α-CP, β-CP, and γ-CP are promising 2D materials in optoelectronics as well as electronics [48], and the black phosphorus carbide (α-CP) monolayers is a potential material of gas sensors [54]. Furthermore, the few-layer black carbide phosphorus (α-CP) has a bright application of the field-effect transistor in experiment [55]. The stable phosphorus carbide ($\beta_0$-PC) monolayer [49] can exhibit robust superconducting behavior [56]. It has been found that the P-C bond could enhance the performance of battery anodes [57]. The carbon can also form monolayers with the other elements in group V, the 2D material $M_2C_3$ (M = As, Sb, and Bi) with high carrier mobility have potential applications in the photovoltaic field [58].

This work begins with two phases of carbon phosphorus monolayers, named α-$C_2P_2$ and β-$C_2P_2$, respectively. Based on the structures of two monolayers, a family monolayers α-$C_2Y_2$ and β-$C_2Y_2$ (Y = As, Sb, and Bi) have been proposed by replacing the phosphorus atoms with the others in group V. All the α-$C_2Y_2$ and β-$C_2Y_2$ (Y = P, As, Sb, and Bi) have been systematically investigated. These 2D materials are dynamically stable, and range from insulators to metals, in which the insulators can occur insulator-metal transition under strains. Furthermore, as the C atoms in α- and β-$C_2Y_2$ (Y = P, As, Sb, and Bi) replaced by the other elements X (X = Si, Ge, Sn, and Pb) in group IV, more monolayers have been proposed, the dynamical stabilities, the thermal stabilities, as well as the electronic properties, have been systematically investigated.

## 2. Calculation methods

The projector augmented wave and Perdew-Burke-Ernzerhof (PBE) functional exchange-correlation [59] have been used to optimize the crystal structures and calculate the electronic structures, as implemented in the Vienna Ab initio Simulation Package (VASP) code [60]. A screened Coulomb potential has been used in the Heyd-Scuseria-Ernzerhof (HSE06) hybrid functional calculation [61, 62]. A vacuum no less than 20 Å has been inserted to the bilayers to eliminate the coupling between periodic images. The structures have been relaxed until all the components of the net forces on the ions are no more than $10^{-4}$ eV/Å, and the convergence criterion in the electronic iteration is $10^{-8}$ eV. The Brillouin zone was sampled by the 20×20×1 k Γ-centered grid. The phonon dispersions have been calculated by using VASP + Phonopy based on the finite displacement method [63]. The *ab initio* molecular dynamics (AIMD) simulations were based on the Nosé thermostat for 5 ps with a time step of 1 femtosecond [64]. Similar to the $C_2N_2$ system [45], the canonical ensemble (NVT) system of (6×6) hexagonal supercell includes 144 atoms sampled on the (1×1×1) grid mesh that were simulated. The cut-off energy and the size of the supercells for calculations of the force constants are available in Table S9 of the Supplementary Material (SM).

## 3. Results and discussion

Enlightened by the previously predicated carbon nitride bilayers [44], we proposed two types of carbon phosphorus ($C_2P_2$) bilayers in this work. As the N atoms in the α- and β-$C_2N_2$ replaced by the P atoms, two similar bilayers can be obtained, named α-$C_2P_2$ and β-$C_2P_2$, respectively. The lattice structures of the 2D materials in the equilibrium state are shown in Fig. 1(a and b), which are consisted of two carbon phosphorus monolayers through the C-C bonds. Both phases are hexagonal structures with the same crystal lattice parameters, the lattice constants $a$, the length of interlayer C-P band $l_{C\text{-}P}$, and the thickness $\delta$ of them are 2.90 Å, 1.89 Å, and 3.30 Å, respectively, which are larger than the α- and β-$C_2N_2$ [44], as shown in Table 1. While the length of C-C bonds $l_{C\text{-}C}$ between layers is 1.55 Å, which is shorter than the $C_2N_2$ bilayers, but 0.13 Å longer than C-C bonds in the graphene. Both C-C bonding and C-P bonding are $sp^3$ hybridized. Furthermore, more similar bilayers $C_2Y_2$ (Y= As, Sb, and Bi) can be obtained as the P atoms in the α- and β-$C_2P_2$ replaced by the Y (Y=As, Sb, and Bi) atoms in group

V, which have the similar lattice structures and bonding characteristic of $C_2P_2$ bilayers. The fully optimized crystal lattice parameters of $C_2Y_2$ (Y=N, P, As, Sb, and Bi) in Table 1 indicate that as the atomic number of Y (Y=N, P, As, Sb, and Bi) atom increases, the lattice constant $a$, the length of interlayer C-Y bond $l_{C-Y,}$ and the thickness of the bilayers δ are monotonously increasing, while the distance between the nearest C atoms $l_{C-C}$ is monotonously decreasing. This trend is consistent with the periodic law of the elements. Similar 2D materials can be obtained if the C atoms in $C_2Y_2$ (Y=N, P, As, Sb, and Bi) bilayers are replaced by the group V elements (Si, Ge, Sn, and Pb), which denoted as $X_2Y_2$ (X=Si, Ge, Sn, and Pb; Y=N, P, As, Sb, and Bi). These derivatives have similar lattice structures and symmetry as α- or β-$C_2P_{2,}$ as shown in Fig. 1. The fully optimized crystal lattice parameters of the derivatives are available in Table S1-S4 of the SM. The crystal lattice parameters indicate that, for each particular X (X=Si, Ge, Sn, and Pb) element, as the atomic number of Y atom increases, the lattice constant $a$, the X-Y bond length $l_{X-Y,}$ and the thickness of the bilayers δ are monotonously increasing, while the distances between the two nearest atoms in group IV $l_{X-X}$ are monotonously decreasing. All the $X_2Y_2$ (X=C, Si, Ge, Sn, and Pb; Y=N, P, As, Sb, and Bi) bilayers are hexagonal structures, and the point group of both α and β types are $D_{3h}$ and $D_{3d}$, respectively. In each layer, the X (X=C, Si, Ge, Sn, and Pb) atoms and Y (Y= P, As, Sb, and Bi) atoms are $sp^3$ hybridized, and thus not coplanar. The Brillouin zone (BZ) of them are regular hexagons, as shown in Fig. 1(c).

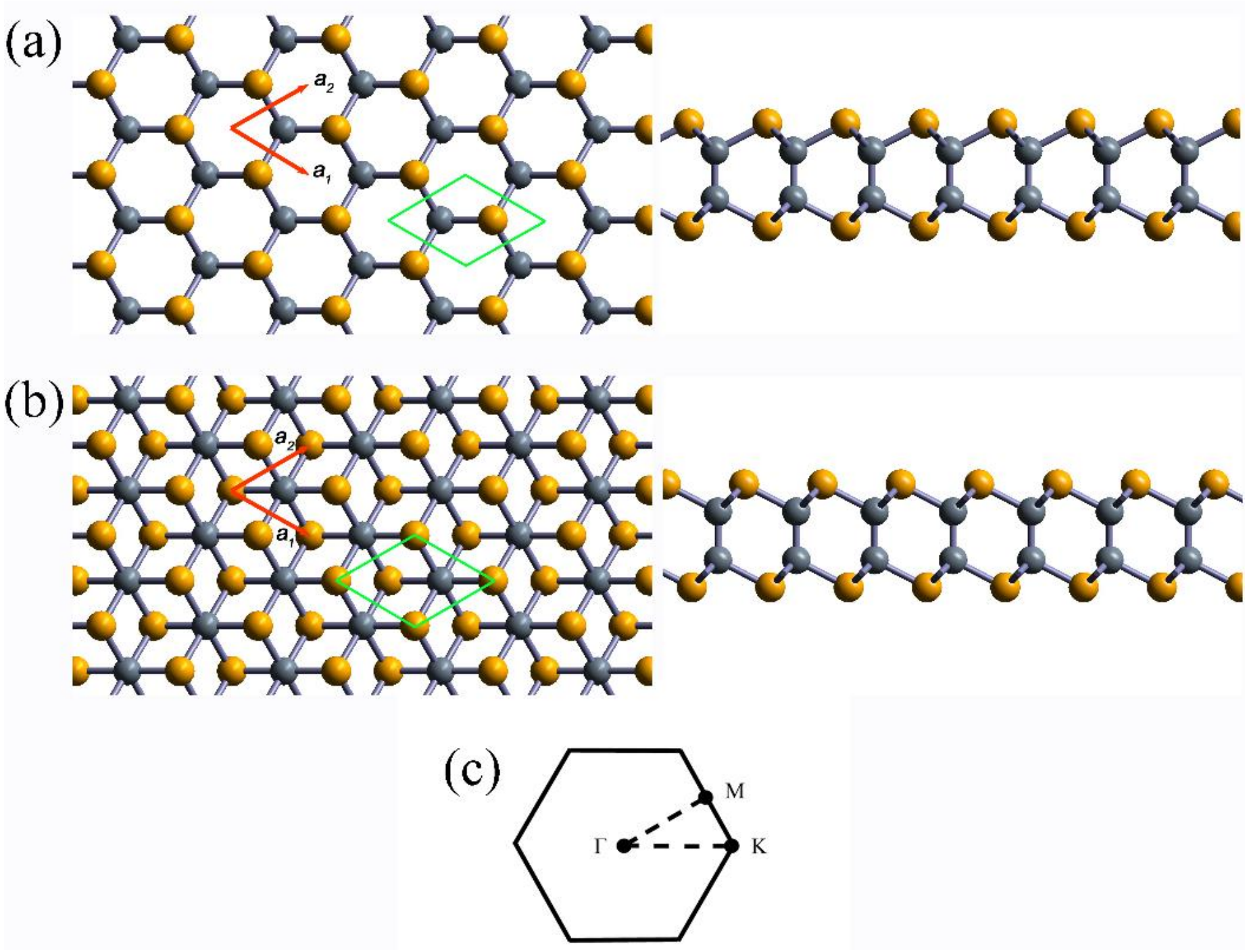

**Figure 1.** Top view and side view of (a) α-$C_2P_2$ and (b) β-$C_2P_2$ bilayer, respectively. The grey and light-yellow spheres denote carbon and phosphorus atoms, respectively. (c) The first Brillouin zone of the two phases.

**Table 1.** The crystal lattice parameters of $C_2Y_2$ (Y=N, P, As, Sb, and Bi) (Unit in Å)

| Phase | $a$ | $l_{C\text{-}C}$ | $l_{C\text{-}Y}$ | $\delta$ |
|---|---|---|---|---|
| $\alpha$-$C_2N_2$ [44] | 2.35 | 1.62 | 1.44 | 2.60 |
| $\beta$-$C_2N_2$ [44] | 2.35 | 1.62 | 1.44 | 2.60 |
| $\alpha$-$C_2P_2$ | 2.90 | 1.55 | 1.89 | 3.30 |
| $\beta$-$C_2P_2$ | 2.90 | 1.55 | 1.89 | 3.30 |
| $\alpha$-$C_2As_2$ | 3.11 | 1.53 | 2.04 | 3.46 |
| $\beta$-$C_2As_2$ | 3.11 | 1.53 | 2.04 | 3.46 |
| $\alpha$-$C_2Sb_2$ | 3.41 | 1.53 | 2.25 | 3.71 |
| $\beta$-$C_2Sb_2$ | 3.41 | 1.53 | 2.25 | 3.71 |
| $\alpha$-$C_2Bi_2$ | 3.60 | 1.49 | 2.38 | 3.79 |
| $\beta$-$C_2Bi_2$ | 3.60 | 1.49 | 2.38 | 3.79 |

To study the dynamical stabilities of all the 46 phases proposed bilayers, their phonon dispersions have been calculated by the first principle theory. The phonon dispersions of the α- and β-$C_2Y_2$ (Y=P, As, Sb, and Bi) bilayers, as shown in Fig. 2(a-h), exhibit no imaginary frequencies in the BZ. Modes with imaginary frequencies are not oscillatory during their time evolution. This leads to structural instability; in other words, the absence of imaginary frequencies suggests that these materials are dynamically stable [65]. The greater atomic mass of the Y (Y= P, As, Sb, Bi) atoms results in smaller acoustic velocities, higher optical frequencies, and broader frequency gaps. In order to test the evolution of the structures at finite temperature, the AIMD simulation of the α-and β-$C_2Y_2$ (Y=P, As, Sb, Bi) has been performed for 5 picoseconds at 300 K. The variation of the total energy with time in AIMD simulations at 300 K, and the final geometry structures are available in Figures S1-S8 of the SM. From the final geometry structures, all α-and β-$C_2Y_2$ (Y=P, As, Sb) as well as β-$C_2Bi_2$ have not reconstructed or structure damage after 5 picoseconds simulation, which manifests that these structures are stable at 300K. While the α-$C_2Bi_2$ structure has reconstructed and light structure damage after 5 picoseconds simulation, which indicates this structure may not stand free for a long time at 300K. Furthermore, for α-$Si_2P_2$, β-$Si_2P_2$, α-$Si_2As_2$, β-$Si_2As_2$, β-$Ge_2P_2$, α-$Ge_2As_2$, β-$Ge_2As_2$, β-$Sn_2N_2$, β-$Sn_2P_2$, β-$Sn_2As_2$, and α-$Sn_2Bi_2$, small imaginary frequency modes are present at the vicinity of Γ. All other derivatives of $X_2Y_2$ (X=Si, Ge, Sn, and Pb; Y= N, P, As, Sb, and Bi) bilayers are dynamically stable. The phonon dispersions of these materials are available in Figures S9-S27 of the SM. The lowest acoustic modes of α-$Pb_2N_2$, β-$Pb_2N_2$, α-$Pb_2Bi_2$, and β-$Pb_2Bi_2$ exhibit visible imaginary frequencies around K points or along the Γ-K line in the BZ, as shown in Figure S23 and Figure S27 of the SM, which means these bilayers may not be dynamically stable in the free state. However, we can infer that the β-$X_2Y_2$ (X=Si, Ge, Sn, and Pb; Y= N, P, As, Sb, and Bi) bilayers are much more stable than the α-$X_2Y_2$ bilayers at 300K due to the different stacking patterns. Furthermore, we infer that in a particular structure, the smaller mass difference of the two atoms is, the more thermally stable it is.

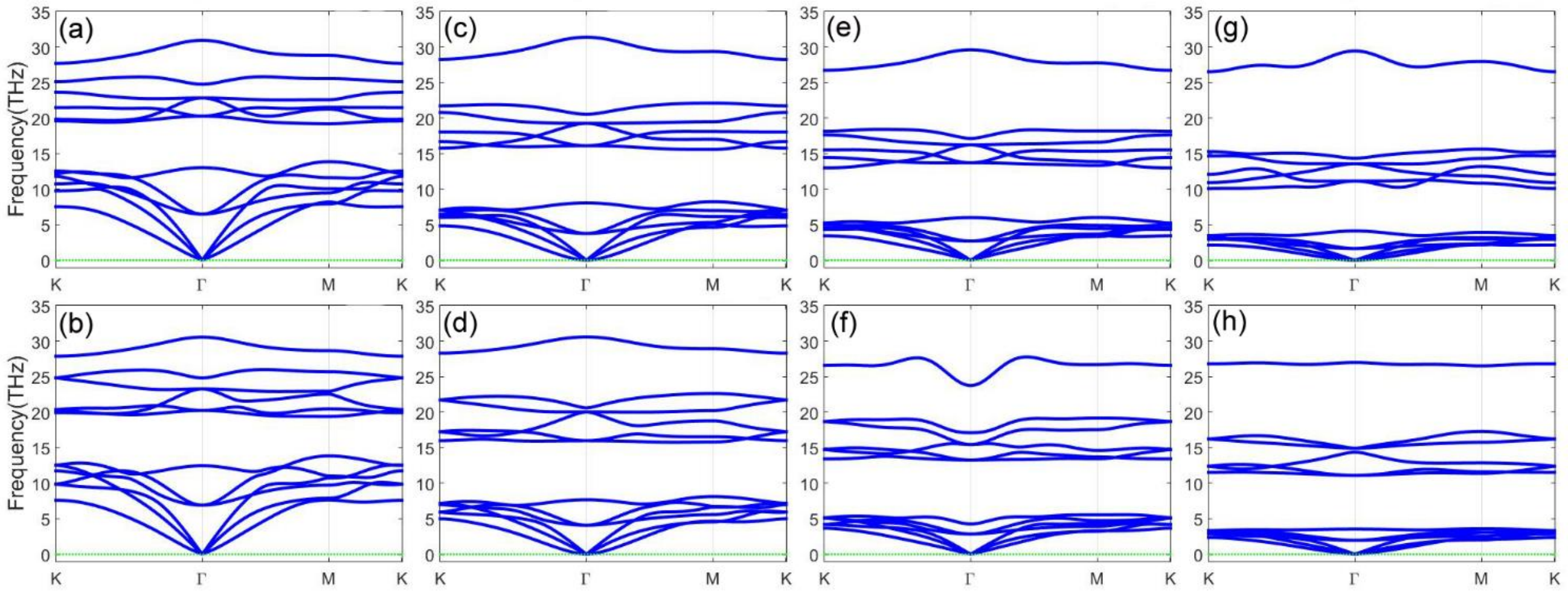

**Figure 2.** Phonon dispersions of the $C_2Y_2$ (Y= P, As, Sb, and Bi) bilayers. (a)~(b) are the phonon dispersions of α-$C_2P_2$, β-$C_2P_2$, α-$C_2As_2$, β-$C_2As_2$, α-$C_2Sb_2$, β-$C_2Sb_2$, α-$C_2Bi_2$, and β-$C_2Bi_2$, respectively.

The α- and β-$C_2P_2$ bilayers are indirect insulators with a gap of 1.79 eV and 1.77 eV on the PBE level, respectively, as shown in Fig. 3. On the HSE level, their band gaps become 2.70 eV and 2.67 eV, respectively, which are much narrower than the α- and β-$C_2N_2$ [44]. The conducting band minimums (CBM) of the α- and β-$C_2P_2$ at the M and K point in the BZ, respectively. And the valence band maximum (VBM) of both at the Γ points. The spin-orbital coupling (SOC) had not been included in the calculation of α- and β-$C_2P_2$, while for $C_2Y_2$ (Y=As, Sb, and Bi), the SOC have been included in the electronic structure calculations.

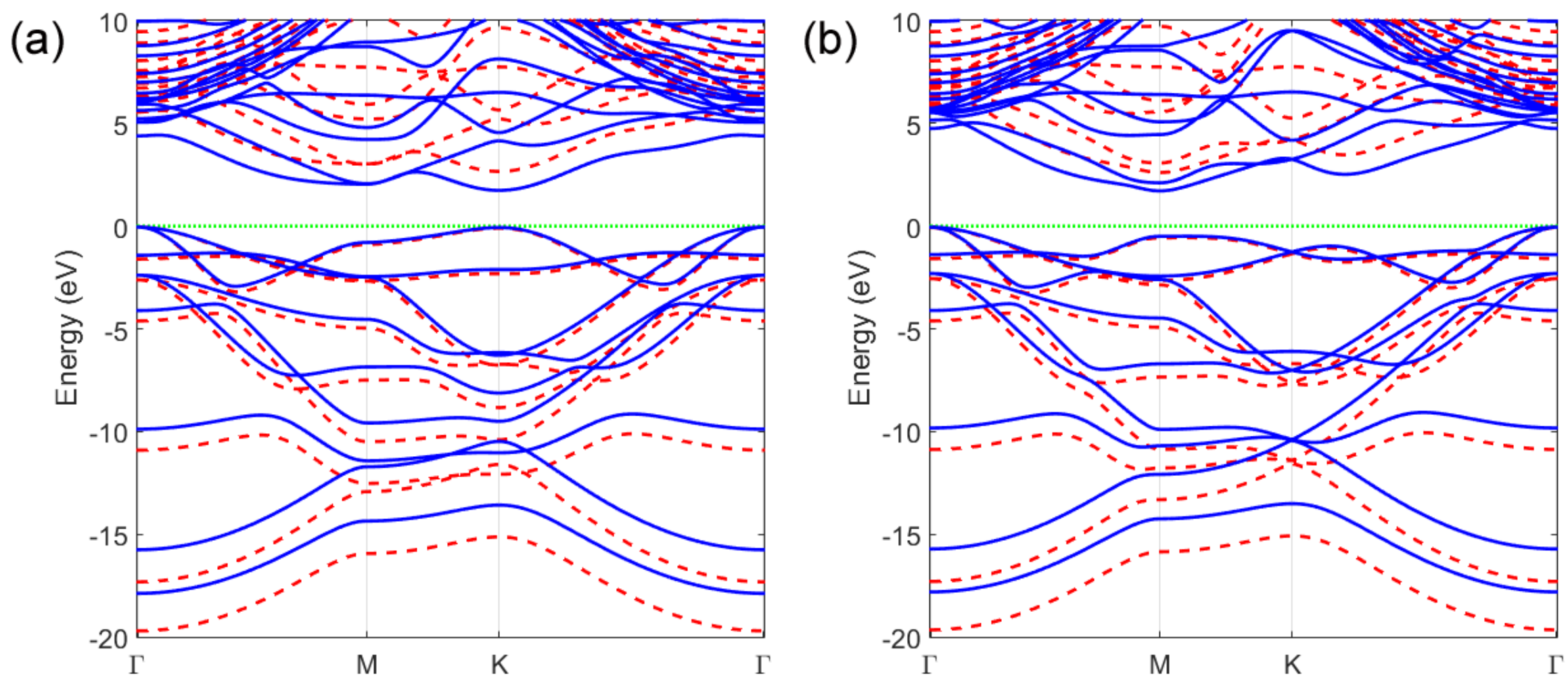

**Figure 3.** Band structure of α-$C_2P_2$ (a) and β-$C_2P_2$ (b), the solid blue lines denote results from PBE, and the red dash lines denote results from HSE06.

Even though the PBE method underestimates the band gap of the insulators, the physical properties and the general profile of the proposed materials should be similar. The electronic properties of the other structures are calculated using the PBE method. As shown in Fig. 4(a-c), all the α-$C_2As_2$, β-$C_2As_2$, and α-$C_2Sb_2$ are indirect gap insulators with or without SOC. The calculated gaps of the three phases from PBE are 1.18 eV, 1.02 eV, and 0.22 eV, respectively. And the ones from PBE+SOC calculations are 1.09 eV, 0.95 eV, and 0.07 eV, respectively. As the SOC breaks the degeneracy of the bands in the BZ, the gaps have been decreased by 0.09 eV, 0.07 eV, and 0.15 eV, respectively. The other three phases, β-$C_2Sb_2$, α-$C_2Bi_2$, and β-$C_2Bi_2$, are metals with or without SOC. Interestingly, for both α- and β-$C_2Bi_2$, there are four crossbands near the Γ points without the SOC, which disappear if the SOC is included. These results imply the possible topological superconductivity under some specific conditions. The values of the bandgaps, the positions of CBM, and VBM are listed in Table 2. From Table 2, we can conclude that for a particular α or β type lattice structure, as the atomic number of group V increases, the band gaps monotonously decrease, while the effect of SOC in the bands increases. Furthermore, the electronic bands of the derivatives $X_2Y_2$ (X=Si, Ge, Sn, and Pb; Y= N, P, As, Sb, and Bi) also have been calculated, as shown in Fig. S28-S41 of the SM, whose electronic band parameters (the gap value, the CBM, and the VBM positions) listed in Table S5-S8 of the SM.

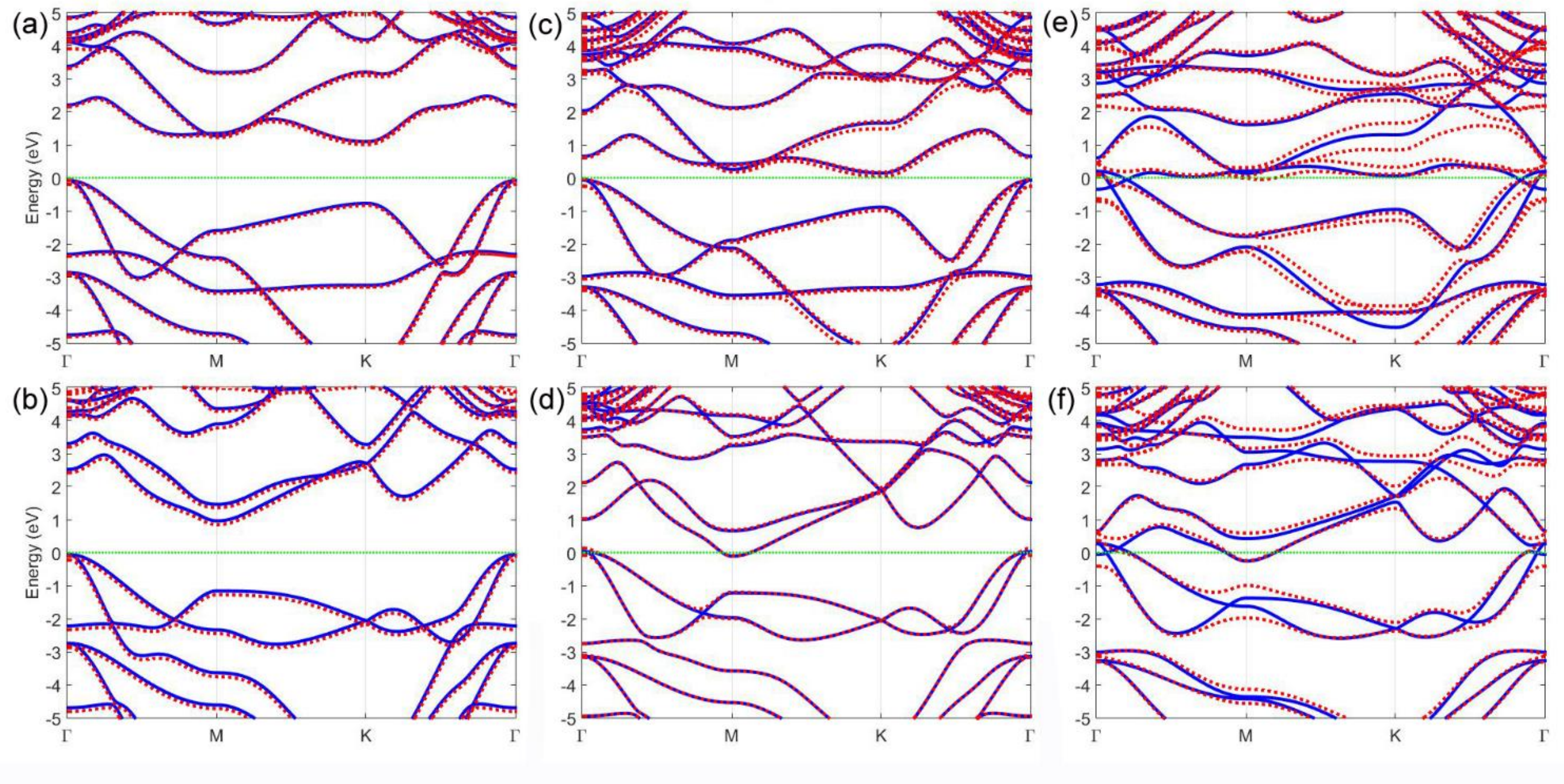

**Figure 4.** Band structure of (a) α-$C_2As_2$, (b) β-$C_2As_2$, (c) α-$C_2Sb_2$, (d) β-$C_2Sb_2$, (e) α-$C_2Bi_2$, (f) β-$C_2Bi_2$. The blue solid and broken red line curve denotes calculated by PBE and PBE+SOC, respectively.

**Table 2.** The electronic band parameters of $C_2Y_2$ (Y=N, P, As, Sb, and Bi) in the PBE background.

| Phase | PBE | | | PBE+SOC | | | △ | $\varepsilon_C$ (%) |
|---|---|---|---|---|---|---|---|---|
| | Gap1 (eV) | CBM position | VBM position | Gap2 (eV) | CBM position | VBM position | | |
| $\alpha$-$C_2N_2$ | 3.76 | M | K-Γ | / | / | / | / | 27[a] [44] |
| $\beta$-$C_2N_2$ | 4.23 | M | K-Γ | / | / | / | / | 22[a] [44] |
| $\alpha$-$C_2P_2$ | 1.79 | K | Γ | / | / | / | / | 19[a] |
| $\beta$-$C_2P_2$ | 1.77 | M | Γ | / | / | / | / | 24[a] |
| $\alpha$-$C_2As_2$ | 1.18 | K | Γ | 1.09 | K | Γ | -0.09 | 14[b] |
| $\beta$-$C_2As_2$ | 1.02 | M | Γ | 0.95 | M | Γ | -0.07 | ~16[b] |
| $\alpha$-$C_2Sb_2$ | 0.22 | K | Γ | 0.07 | K | Γ | -0.15 | 10[b] |
| $\beta$-$C_2Sb_2$ | 0 | M | Γ | 0 | M | Γ | 0 | 16[b] |
| $\alpha$-$C_2Bi_2$ | 0 | / | / | 0 | M-K | Γ | 0 | / |
| $\beta$-$C_2Bi_2$ | 0 | / | / | 0 | M | Γ | 0 | / |

Gap1 and Gap2 represent the band gaps calculated by PBE and PBE+SOC, respectively, △ = Gap2-Gap1. $\varepsilon_C$ represents the critical strain from insulativity to metallicity.

a Represents the critical stain in the PBE background.

b Represents the critical stain in the PBE+SOC background.

In the actual applications, the 2D materials are usually attached to the particular substrates, and they are subjected to inevitable lattice mismatch, in which the lattice is either strained or compressed. In other words, the bandgap can be tuned by strain, which is a useful method in the nano materials research field [66-69]. Because the crystal lattices are regular hexagons, the biaxial strain has been applied to tune the bandgap. The strain defined as

$$\varepsilon = \frac{a - a_0}{a_0} \times 100\%$$

where a is the lattice constant under the strain, and $a_0$ is the lattice constant for the free standing state. As shown in Fig. 6(a), as the biaxial strain increases, both bandgap of α- and β-$C_2P_2$ monotonically decrease and exhibit linear relations in the range 2-12% and 4-13%, respectively. The gaps of α- and β-$C_2P_2$ close as the strain reach 19% and 24%, respectively. Because of the realignment of bands in the strain, there are two maxima at 2% and 3-4%, respectively. From the PBE+SOC calculation, the CBM of α-$C_2As_2$ locates at the K point in the free state, as shown in Fig. 4(a). As the strain increases, the CBM shifts from K point to Γ point, which changes the indirect gap into a direct one, as shown in Fig. 5(a). The gap will close at strain 14%, if the strain increases further, the conducting band and valence band will cross the Fermi level, as shown in Fig. 5(b). Similar to α-$C_2As_2$, the CBM of β-$C_2As_2$ at M points in the free state, as shown in Fig. 4(b). As the strain increases, the CBM shifts from K point to Γ point, which also induces the transition from the indirect gap to a direct gap, as shown in Fig. 5(c). The β-$C_2As_2$ bilayer will become a gapless semimetal at 16% strain, whose CBM and VBM degenerate at Γ point. Amusingly, the band gap will reopen if the strain increases further, as shown in Fig. 5(d). There are also two maxima at 2% and 6% in the gap-strain relations of α-$C_2As_2$ and β-$C_2As_2$, as shown in Fig. 6(b), respectively. The band gaps of α-$C_2Sb_2$ and β-$C_2Sb_2$ also close as the strain reach 10% and 16%, respectively. There are two maxima at 2% and 5%, respectively, as shown in Fig. 6(c). Each total energies of the α- and β-$C_2Y_2$ (Y=P, As, and Sb) monotonically increase as the strain increases, as shown in Fig. 6(d), and the total energy of each β phase is a slightly lower than the α one for any strains. Furthermore, the other insulators, α-$Si_2P_2$, β-$Si_2P_2$, α-$Si_2As_2$, β-$Si_2As_2$, α-$Si_2Sb_2$, β-$Si_2Sb_2$, α-$Si_2Bi_2$, β-$Si_2Bi_2$, α-$Ge_2N_2$, β-$Ge_2N_2$, α-$Ge_2P_2$, β-$Ge_2P_2$,α-$Ge_2As_2$, β-$Ge_2As_2$, α-$Ge_2Sb_2$, β-$Ge_2Sb_2$, α-$Sn_2P_2$, β-$Sn_2P_2$, α-$Sn_2As_2$, β-$Sn_2As_2$, α-$Sn_2Sb_2$, β-$Sn_2Sb_2$, α-$Pb_2P_2$, β-$Pb_2P_2$, α-$Pb_2As_2$, and β-$Pb_2As_2$ also can undergo insulator-metal transition by the biaxial strain, as shown in Figure S47-S59 of the SM, and the critical strains of the bilayers listed in Table S5-S8 of the SM.

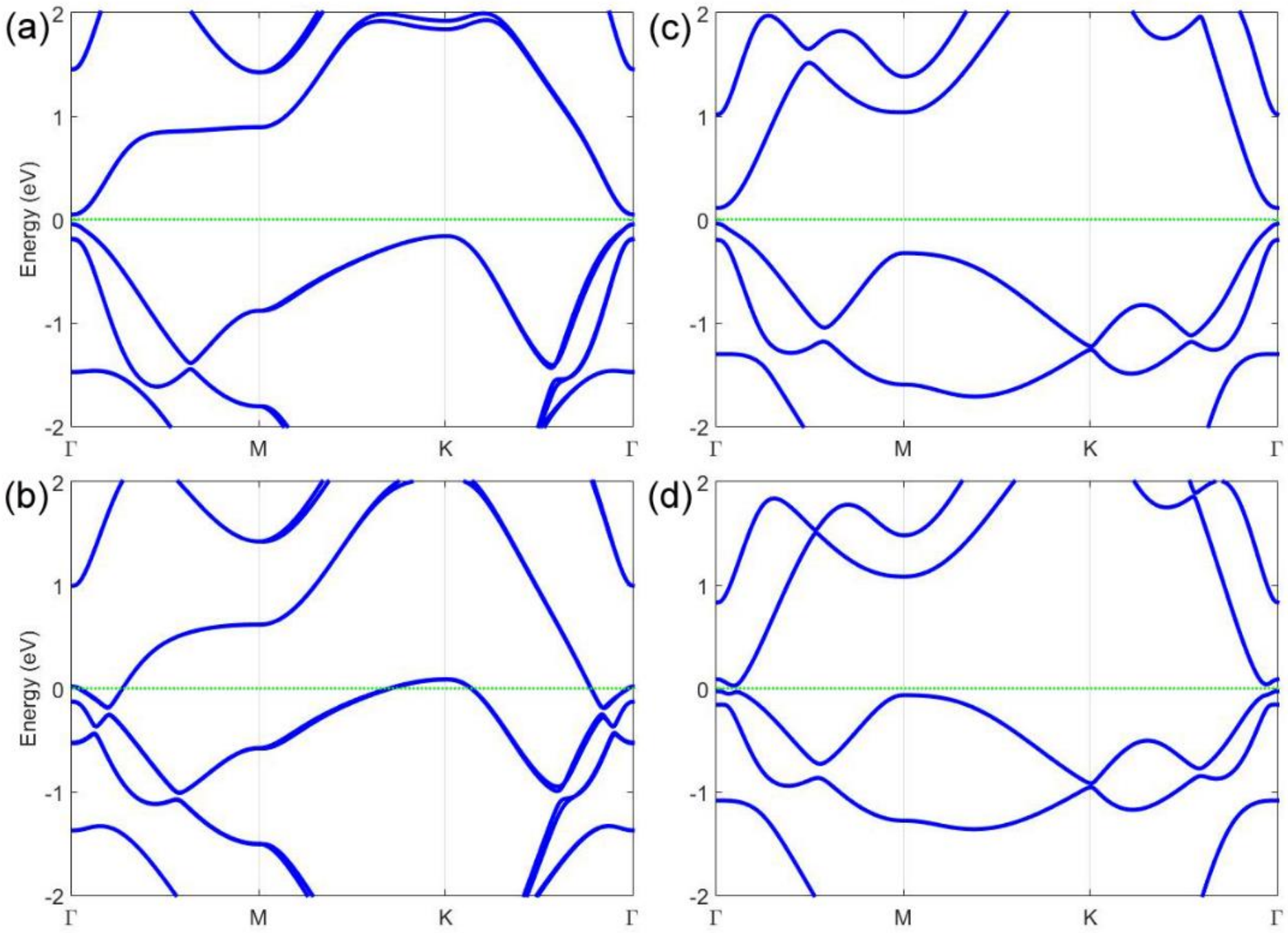

**Figure 5.** Band structure of α-$C_2As_2$ under the strain of (a) 12% and (b) 18%, β-$C_2As_2$ under the strain of (c) 14%, and (d) 18% calculated by PBE+SOC.

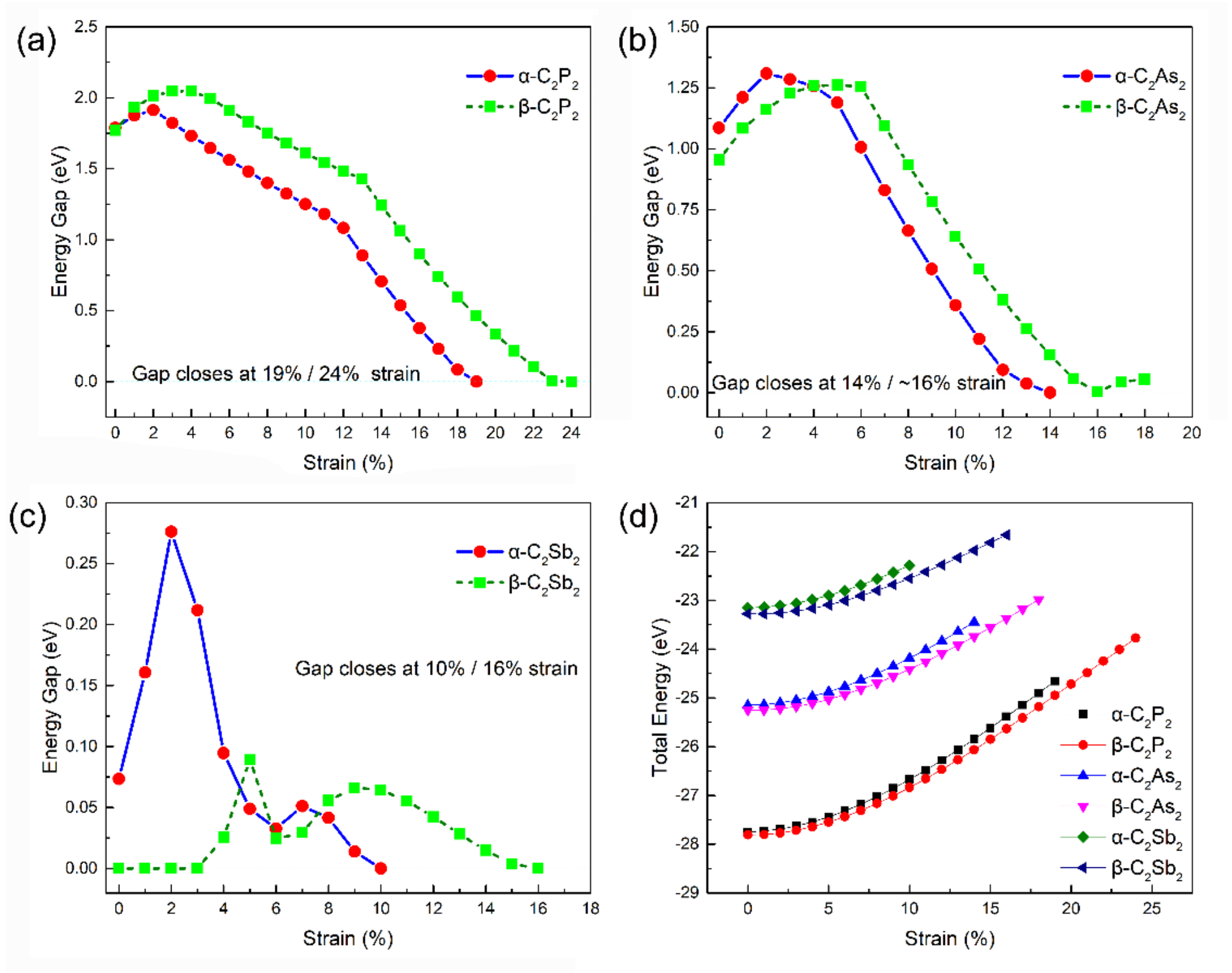

**Figure 6.** Dependence of energy gaps on the strain of (a) $C_2P_2$, (b) $C_2As_2$, (c) $C_2Sb_2$, and (d) the total energies of them.

## 4. Conclusions

Forty-six phases of 2D IV-V compounds have been proposed as an extension of our previous work. The dynamical stabilities and electronic properties have been investigated using the density functional theory. The fully optimized structural parameters of the bilayers exhibit trends that are consistent with the periodic law of elements. The phonon dispersions indicate that the majority of the bilayers are dynamically stable in the free state. From the phonon dispersions of the 2D IV-V compounds, one can find that, as the atomic number in the bilayers increases, the highest optical frequency of the phonon decreases. The greater atomic mass of the bilayers implies stronger stability. The AIMD simulations demonstrate that most of the proposed bilayers could be thermally stable even at room temperature. The calculated electronic structures indicate that the bilayers range from insulators to metals with various band gaps, depending on the constitution. Usually, the bilayers containing lighter elements in group IV and lighter elements in group V exhibit broader bandgaps. With a higher atomic number, the SOC effect on the electronic band becomes more significant. For a particular element in group IV, the bilayers own familiar band structures, and the band gap decreases as the atomic number increases. The band gap of the majority bilayers can decrease under strain, even closes, and exhibit transition from insulator to metal. Besides, the band gap of some special bilayers will increase in the beginning, and then decrease and eventually close under the biaxial stain. The wide-gap bilayers may have potential applications in the photovoltaic devices, and the narrow bandgap ones may become thermoelectric materials. Furthermore, the superconductivity can be investigated by analyzing the electroacoustic coupling effect of the gapless bilayers. In a word, the new 2D materials proposed in this work perhaps become competent candidates in the spintronics and straintronics field. The parallel works of α-$C_2P_2$ [70] and IV-V compounds [71] have also been reported.

**CRediT authorship contribution statement**

**Wanxing Lin**: Conceptualization, Methodology, Validation, Formal analysis, Investigation, Data curation, Writing - original draft, Visualization. **Shi-Dong Liang**: Formal analysis, Resources, Writing - review & editing, Funding acquisition. **Jiesen Li**: Conceptualization, Methodology, Software, Formal analysis, Investigation, Resources, Data curation, Writing - review & editing, Funding acquisition. **Dao-Xin Yao**: Conceptualization, Formal analysis, Resources, Data curation, Writing - review & editing, Supervision, Project administration, Funding acquisition.

**Declaration of competing interest**

The authors declare that they have no known competing financial interests or personal relationships that could have appeared to influence the work reported in this paper.

**Acknowledgments**

One of the authors Wanxing Lin, would like to thank Hai-Feng Li, Rui-Qin Zhang, Yu-Jun Zhao, Ji-Hai Liao, Peng-Fei Liu, Lufeng Ruan, Matthew J. Lake, and Vincent Meunier for helpful discussions. W.L. and D.X.Y. are supported by the National Key R&D Program of China (2017YFA0206203, 2018YFA0306001), NSFC-11974432, Guangdong Basic and Applied Basic Research Foundation (2019A1515011337), and the Leading Talent Program of Guangdong Special Projects. S.D.L. is supported by the Natural Science Foundation of Guangdong Province (2016A030313313). J. L. is supported by the NSFC-11747108, the Opening Project of Guangdong Province Key Laboratory of Computational Science at the Sun Yat-Sen University (2018015), Opening Project of Guangdong High Performance Computing Society (2017060103), and High-Level Talent Start-Up Research Project of Foshan University (Gg040934). Most calculations in this work were performed on the Tianhe-2 supercomputer with the help of engineers from the National Supercomputer Center in Guangzhou.

Supplementary Material for

# Phonon dispersions and electronic structures of two-dimensional IV-V compounds

Wanxing Lin, [1,#] Shi-Dong Liang, [1] Jiesen Li, [2,*] Dao-Xin Yao[1,†]

[1] State Key Laboratory of Optoelectronic Materials and Technologies, School of Physics, Sun Yat-Sen University, Guangzhou, P. R. China

[2] School of Environment and Chemical Engineering, Foshan University, Foshan, P. R. China

Present address: # W. L.: Institute of Applied Physics and Materials Engineering, University of Macau, N23 Avenida da Universidade, Taipa, Macau, China

*Corresponding author. E-mail: ljs@fosu.edu.cn (Jiesen Li)

†Corresponding author. E-mail: yaodaox@mail.sysu.edu.cn (Dao-Xin Yao)

CONTENTS

## I. Crystal lattice parameters

The lattice constants and the distance between the nearest carbon atoms $l_{Si\text{-}Y}$, the Si-Y bond, and the thick of the bilayers $\delta$ can be obtained by fully optimized.

Table S1. Crystal lattice parameters of $Si_2Y_2$ (Y=N, P, As, Sb, Bi) (Unit in Å)

| Phase | $a$ | $l_{X\text{-}X}$ | $l_{X\text{-}Y}$ | $\delta$ |
|---|---|---|---|---|
| $\alpha$-$Si_2N_2$ [1] | 2.90 | 2.43 | 1.76 | 3.54 |
| $\beta$-$Si_2N_2$ [1] | 2.90 | 2.43 | 1.76 | 3.54 |
| $\alpha$-$Si_2P_2$ | 3.53 | 2.37 | 2.28 | 4.41 |
| $\beta$-$Si_2P_2$ | 3.53 | 2.37 | 2.28 | 4.41 |
| $\alpha$-$Si_2As_2$ | 3.69 | 2.36 | 2.40 | 4.58 |
| $\beta$-$Si_2As_2$ | 3.69 | 2.36 | 2.40 | 4.58 |
| $\alpha$-$Si_2Sb_2$ | 4.02 | 2.36 | 2.62 | 4.82 |
| $\beta$-$Si_2Sb_2$ | 4.02 | 2.36 | 2.62 | 4.82 |
| $\alpha$-$Si_2Bi_2$ | 4.17 | 2.35 | 2.73 | 4.92 |
| $\beta$-$Si_2Bi_2$ | 4.17 | 2.35 | 2.73 | 4.92 |

Table S2. Crystal lattice parameters of $Ge_2Y_2$ (Y=N, P, As, Sb, Bi) (Unit in Å)

| Phase | $a$ | $l_{X\text{-}X}$ | $l_{X\text{-}Y}$ | $\delta$ |
|---|---|---|---|---|
| $\alpha$-$Ge_2N_2$ | 3.10 | 2.57 | 1.91 | 3.90 |
| $\beta$-$Ge_2N_2$ | 3.10 | 2.57 | 1.91 | 3.90 |
| $\alpha$-$Ge_2P_2$ | 3.66 | 2.51 | 2.37 | 4.65 |
| $\beta$-$Ge_2P_2$ | 3.66 | 2.51 | 2.37 | 4.65 |
| $\alpha$-$Ge_2As_2$ | 3.82 | 2.50 | 2.49 | 4.80 |
| $\beta$-$Ge_2As_2$ | 3.82 | 2.50 | 2.49 | 4.80 |
| $\alpha$-$Ge_2Sb_2$ | 4.12 | 2.50 | 2.69 | 5.01 |
| $\beta$-$Ge_2Sb_2$ | 4.12 | 2.50 | 2.69 | 5.01 |
| $\alpha$-$Ge_2Bi_2$ | 4.26 | 2.49 | 2.78 | 5.09 |
| $\beta$-$Ge_2Bi_2$ | 4.26 | 2.49 | 2.78 | 5.09 |

Table S3. Crystal lattice parameters of $Sn_2Y_2$ (Y=N, P, As, Sb, Bi) (Unit in Å)

| Phase | $a$ | $l_{X\text{-}X}$ | $l_{X\text{-}Y}$ | $\delta$ |
|---|---|---|---|---|
| $\alpha$-$Sn_2N_2$ | 3.42 | 2.98 | 2.11 | 4.44 |
| $\beta$-$Sn_2N_2$ | 3.42 | 2.98 | 2.11 | 4.44 |
| $\alpha$-$Sn_2P_2$ | 3.95 | 2.89 | 2.56 | 5.22 |
| $\beta$-$Sn_2P_2$ | 3.95 | 2.89 | 2.56 | 5.22 |
| $\alpha$-$Sn_2As_2$ | 4.09 | 2.88 | 2.67 | 5.36 |
| $\beta$-$Sn_2As_2$ | 4.09 | 2.88 | 2.67 | 5.36 |
| $\alpha$-$Sn_2Sb_2$ | 4.38 | 2.87 | 2.87 | 5.58 |
| $\beta$-$Sn_2Sb_2$ | 4.38 | 2.87 | 2.87 | 5.58 |
| $\alpha$-$Sn_2Bi_2$ | 4.51 | 2.86 | 2.96 | 5.66 |
| $\beta$-$Sn_2Bi_2$ | 4.51 | 2.86 | 2.96 | 5.66 |

Table S4. Crystal lattice parameters of $Pb_2Y_2$ (Y=N, P, As, Sb, Bi) (Unit in Å)

| Phase | $a$ | $l_{X\text{-}X}$ | $l_{X\text{-}Y}$ | $\delta$ |
|---|---|---|---|---|
| $\alpha$-$Pb_2N_2$ | 3.64 | 3.17 | 2.24 | 4.75 |
| $\beta$-$Pb_2N_2$ | 3.64 | 3.17 | 2.24 | 4.75 |
| $\alpha$-$Pb_2P_2$ | 4.12 | 3.06 | 2.67 | 5.49 |
| $\beta$-$Pb_2P_2$ | 3.64 | 3.17 | 2.24 | 4.75 |
| $\alpha$-$Pb_2As_2$ | 4.25 | 3.05 | 2.77 | 5.62 |
| $\beta$-$Pb_2As_2$ | 4.25 | 3.05 | 2.77 | 5.62 |
| $\alpha$-$Pb_2Sb_2$ | 4.53 | 3.04 | 2.96 | 5.81 |
| $\beta$-$Pb_2Sb_2$ | 4.53 | 3.04 | 2.96 | 5.81 |
| $\alpha$-$Pb_2Bi_2$ | 4.63 | 3.02 | 3.03 | 5.89 |
| $\beta$-$Pb_2Bi_2$ | 4.63 | 3.02 | 3.03 | 5.89 |

## II. AIMD simulations

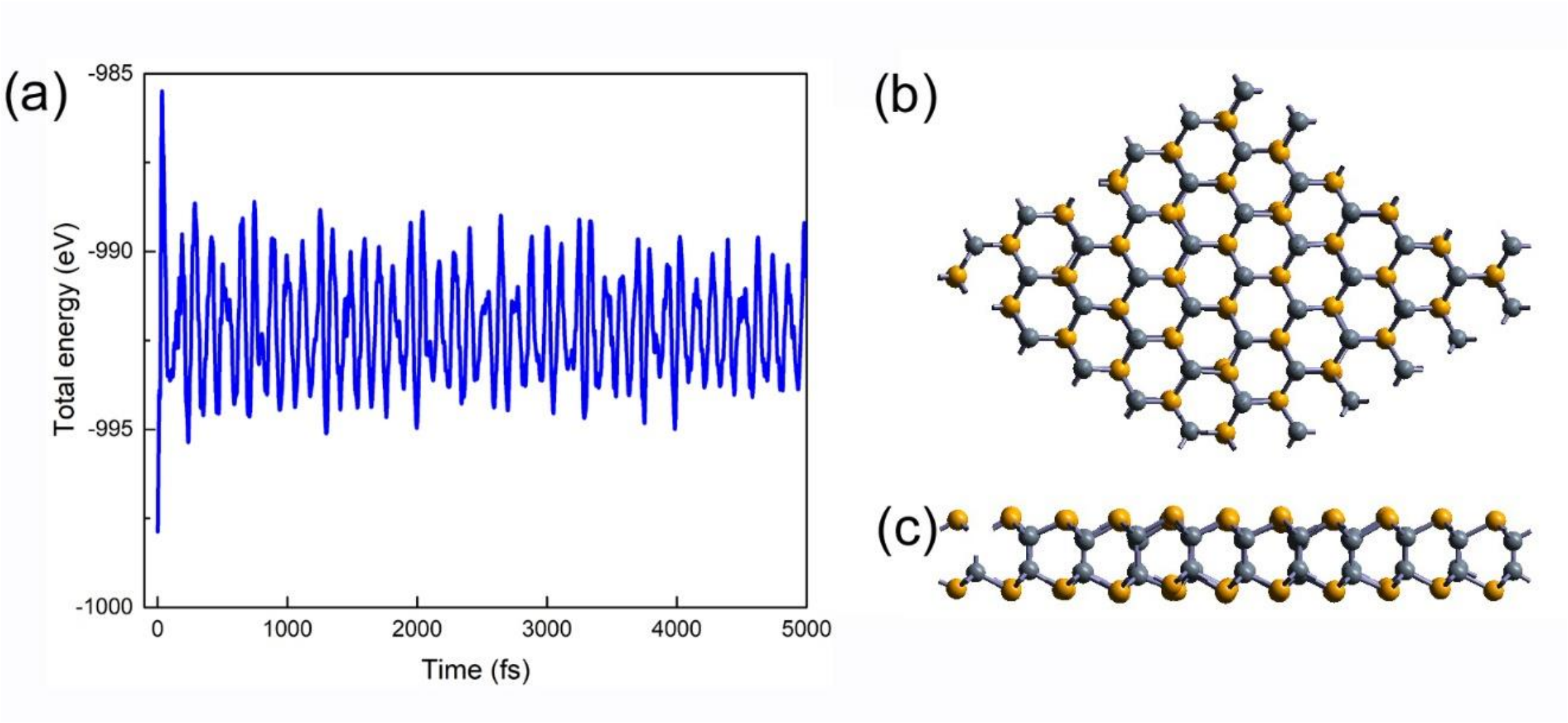

Figure S1. (a) The variation of the total energy of α-$C_2P_2$ with time in AIMD simulations at 300 K, (b) and (c) show the top view and side view of the final geometry structure, respectively.

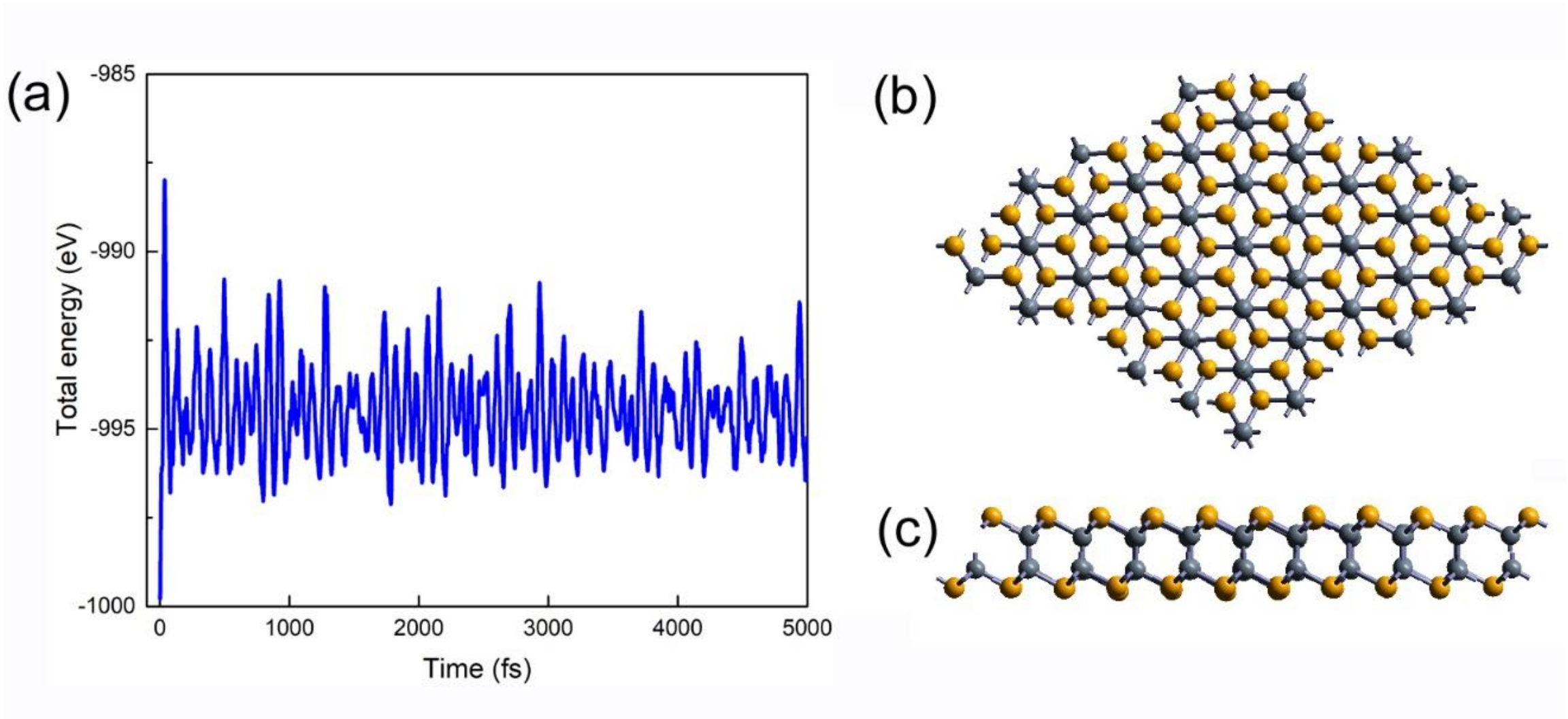

Figure S2. (a) The variation of the total energy of β-$C_2P_2$ with time in AIMD simulations at 300 K, (b) and (c) show the top view and side view of the final geometry structure, respectively.

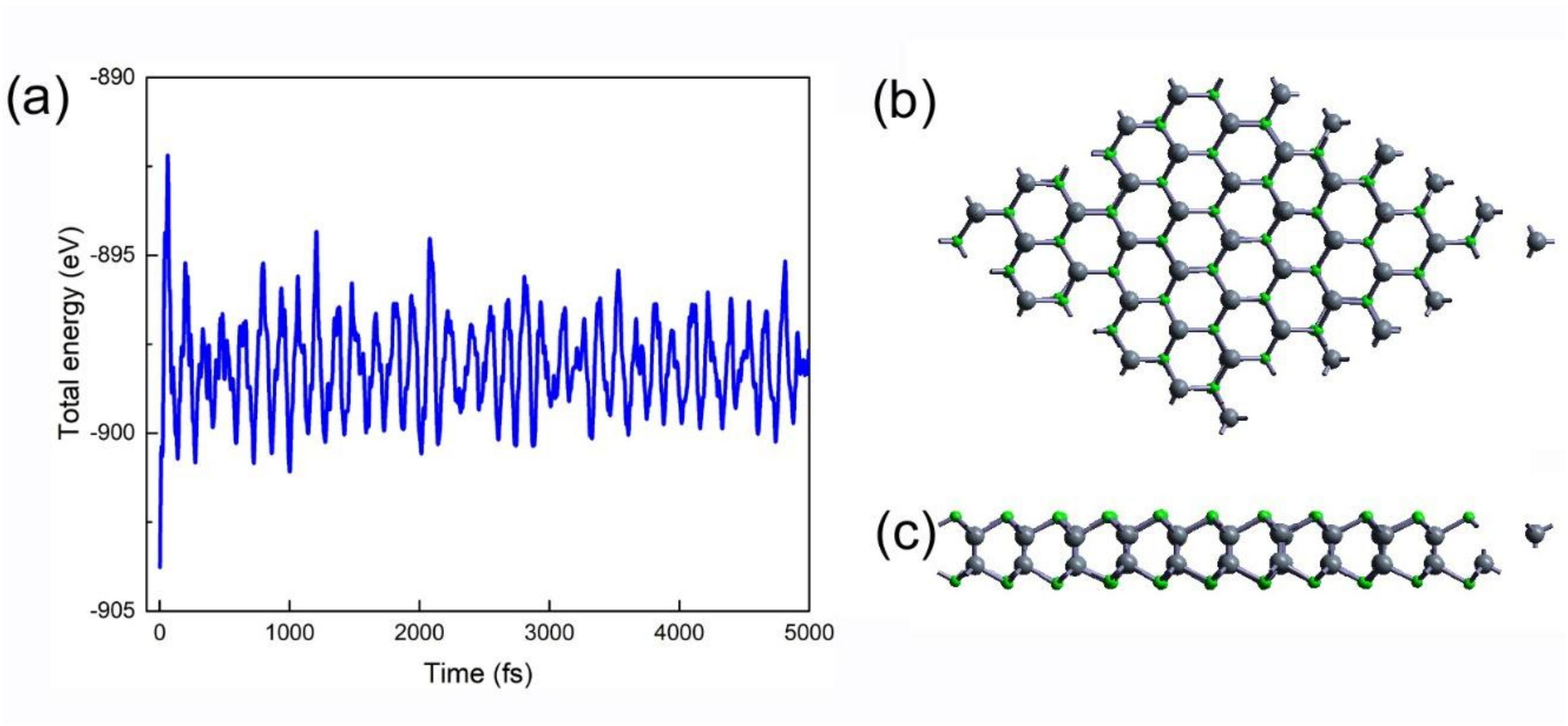

Figure S3. (a) The variation of the total energy of α-$C_2As_2$ with time in AIMD simulations at 300 K, (b) and (c) show the top view and side view of the final geometry structure, respectively.

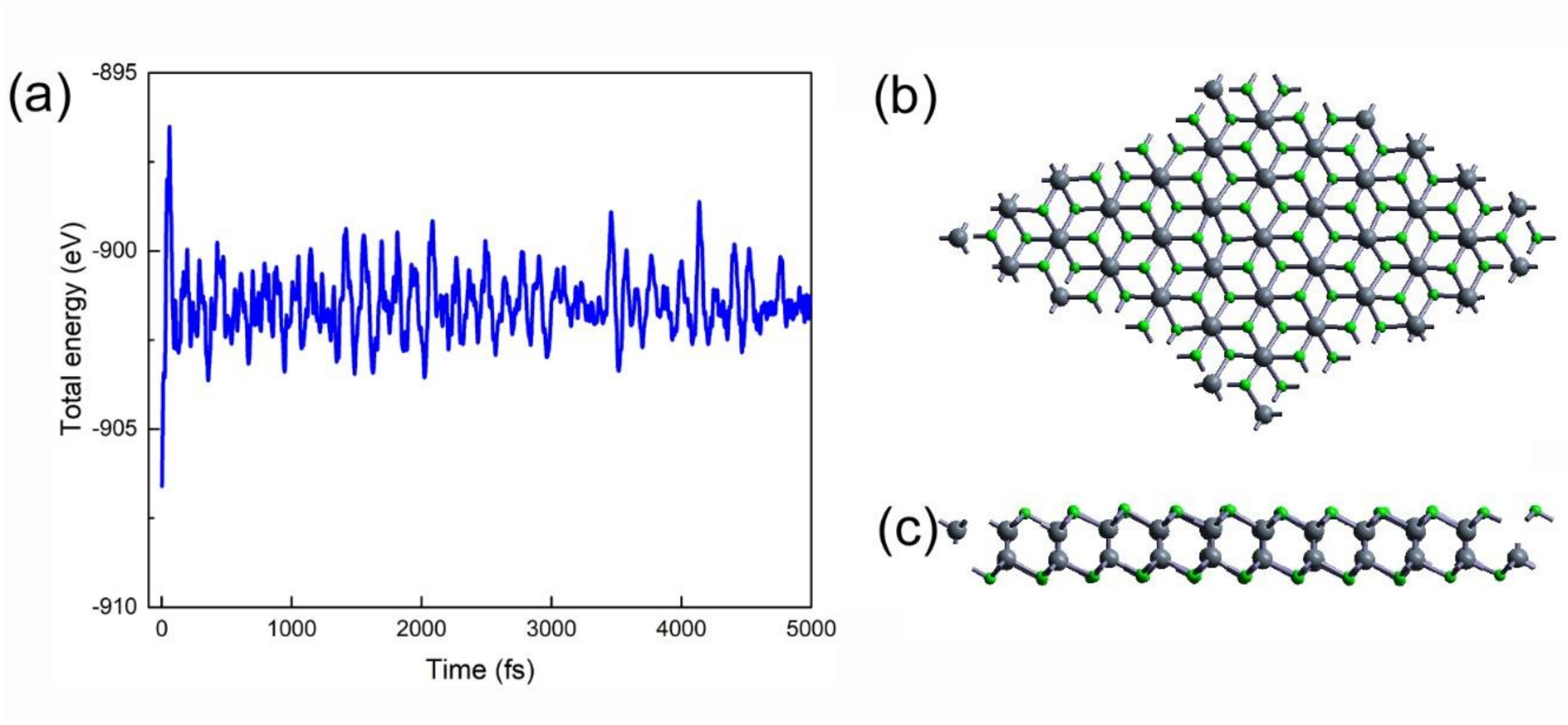

Figure S4. (a) The variation of the total energy of β-$C_2As_2$ with time in AIMD simulations at 300 K, (b) and (c) show the top view and side view of the final geometry structure, respectively.

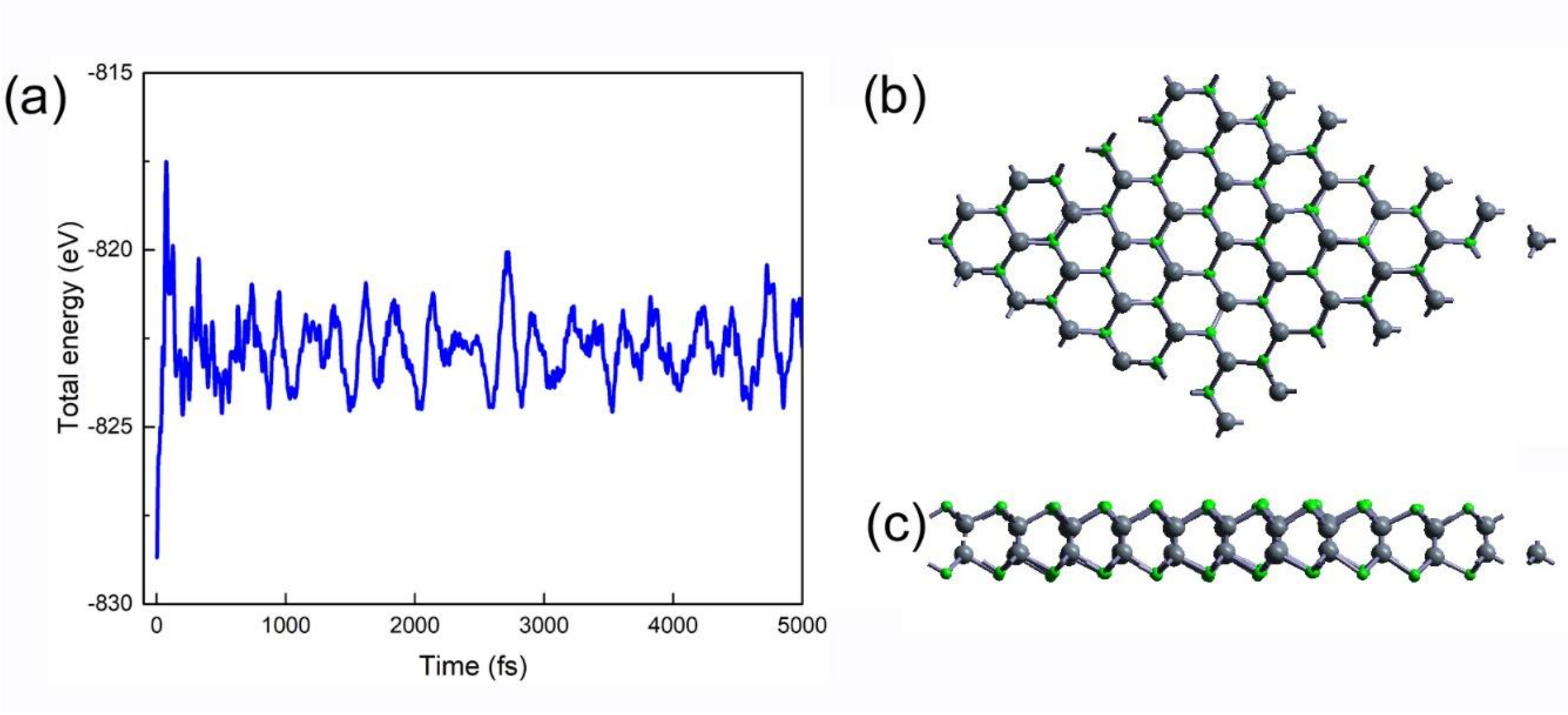

Figure S5. (a) The variation of the total energy of α-$C_2Sb_2$ with time in AIMD simulations at 300 K, (b) and (c) show the top view and side view of the final geometry structure, respectively.

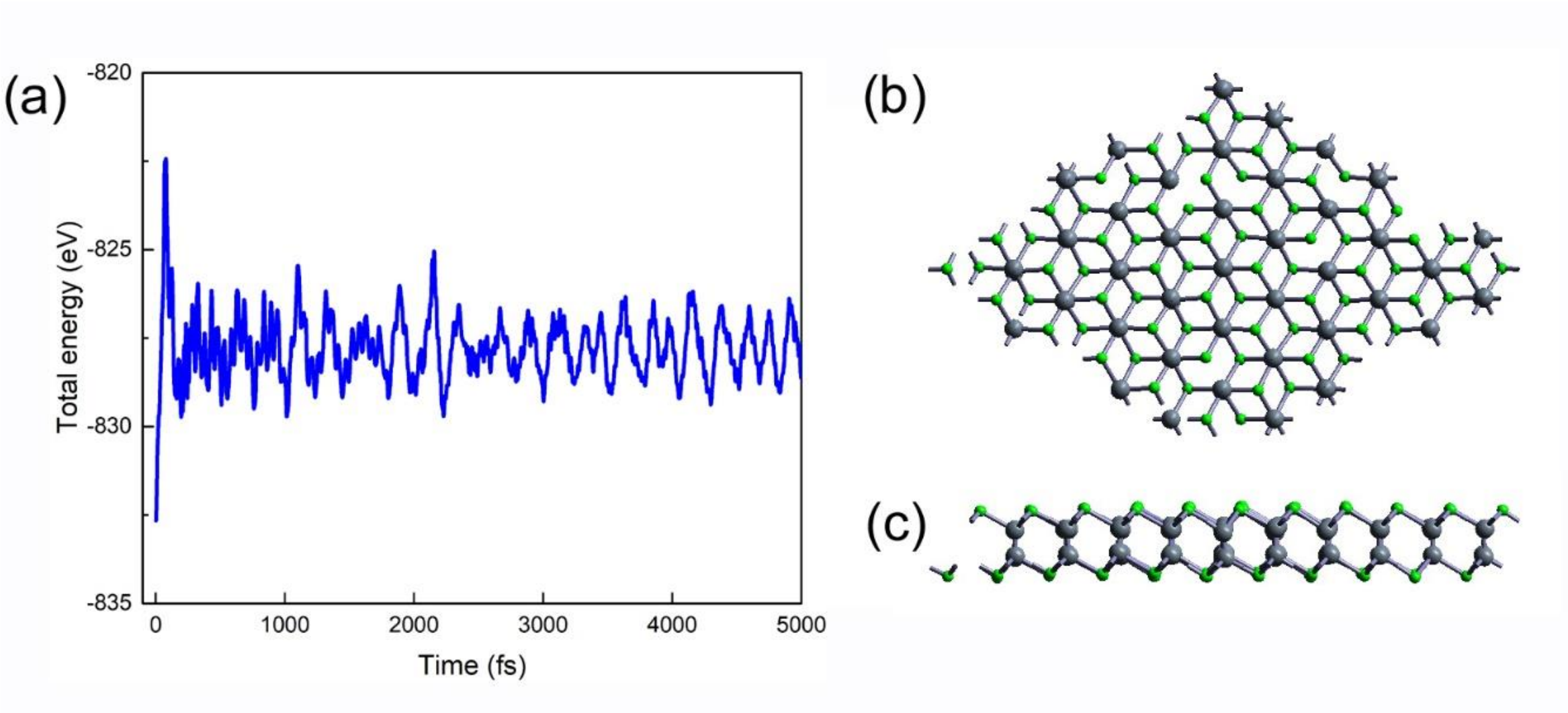

Figure S6. (a) The variation of the total energy of β-$C_2Sb_2$ with time in AIMD simulations at 300 K, (b) and (c) show the top view and side view of the final geometry structure, respectively.

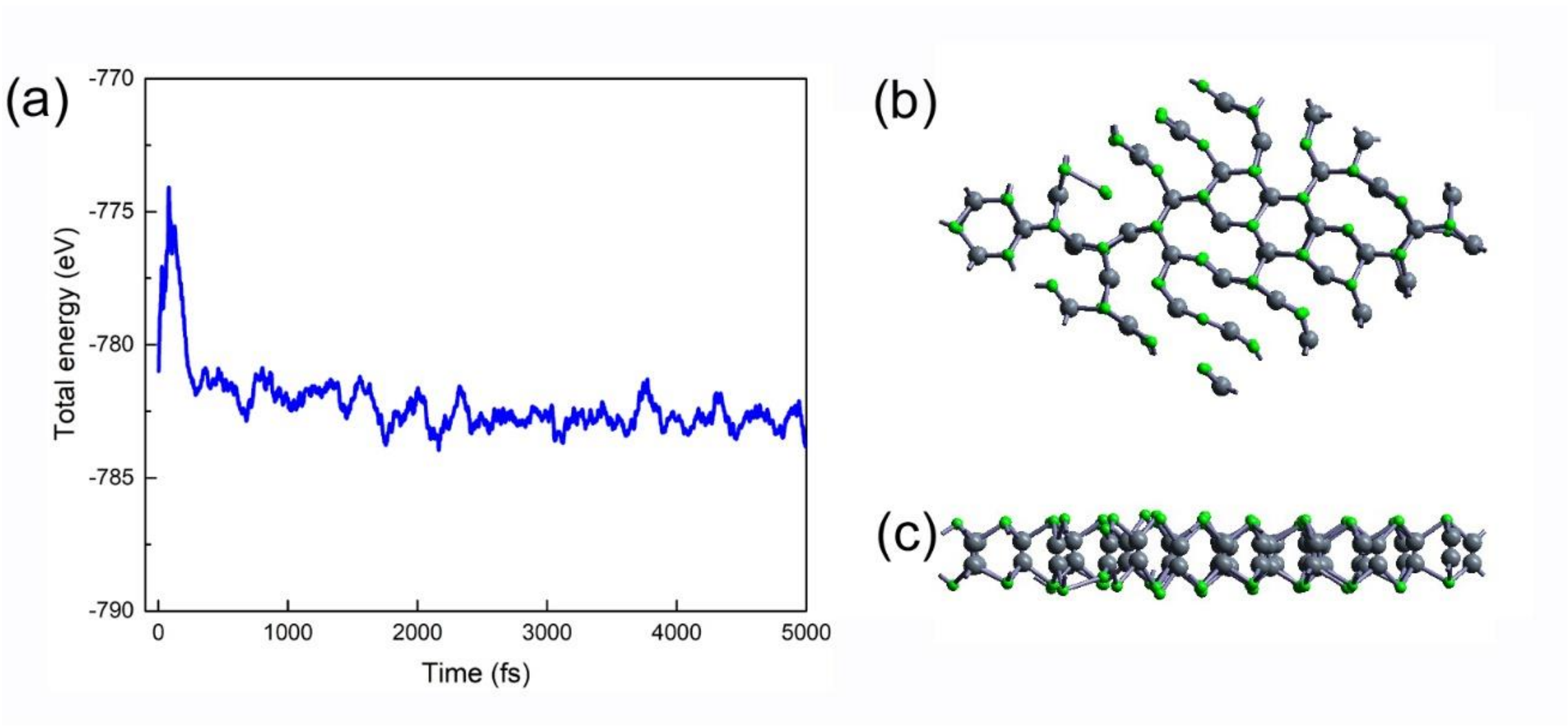

Figure S7. (a) The variation of the total energy of α-$C_2Bi_2$ with time in AIMD simulations at 300 K, (b) and (c) show the top view and side view of the final geometry structure, respectively.

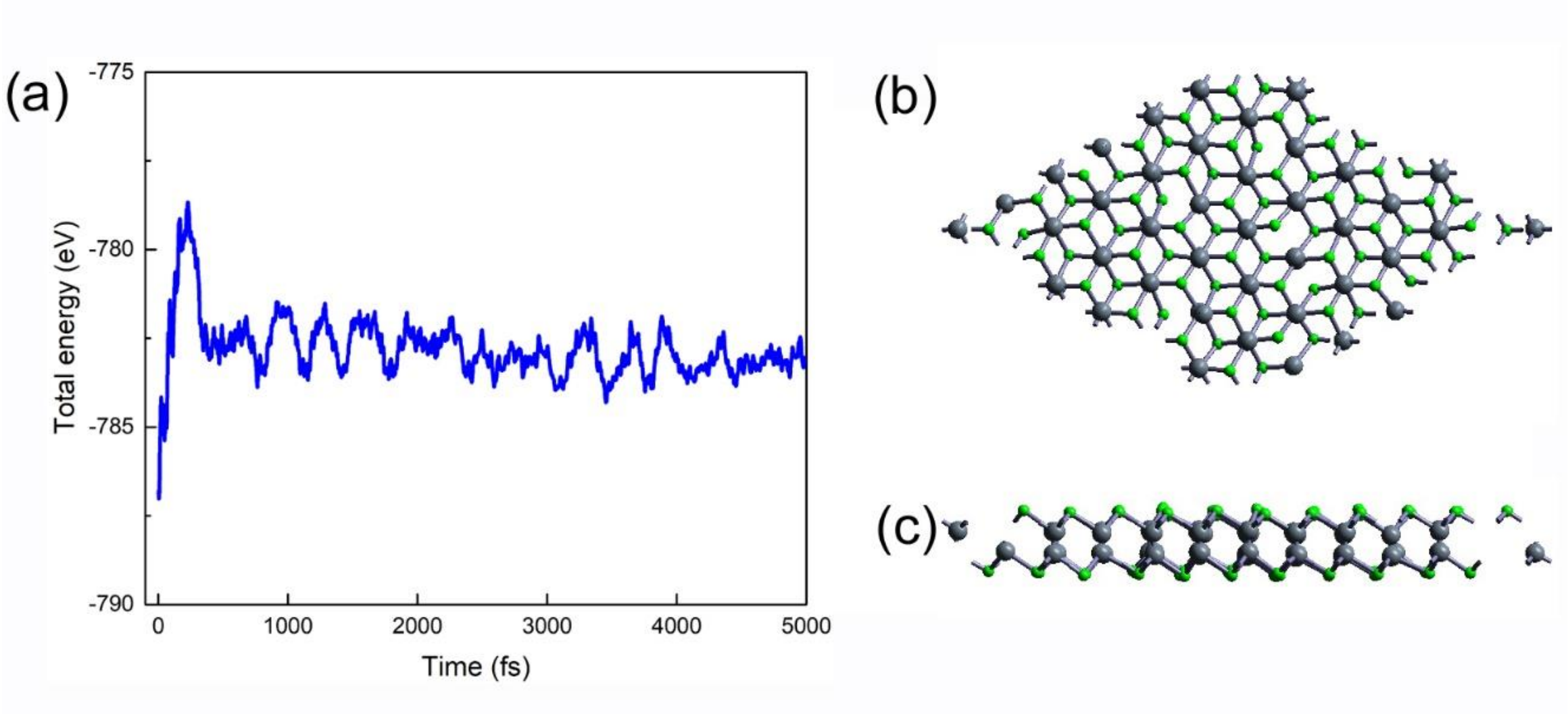

Figure S8. (a) The variation of the total energy of β-$C_2Bi_2$ with time in AIMD simulations at 300 K, (b) and (c) show the top view and side view of the final geometry structure, respectively.

## III. Phonon dispersions

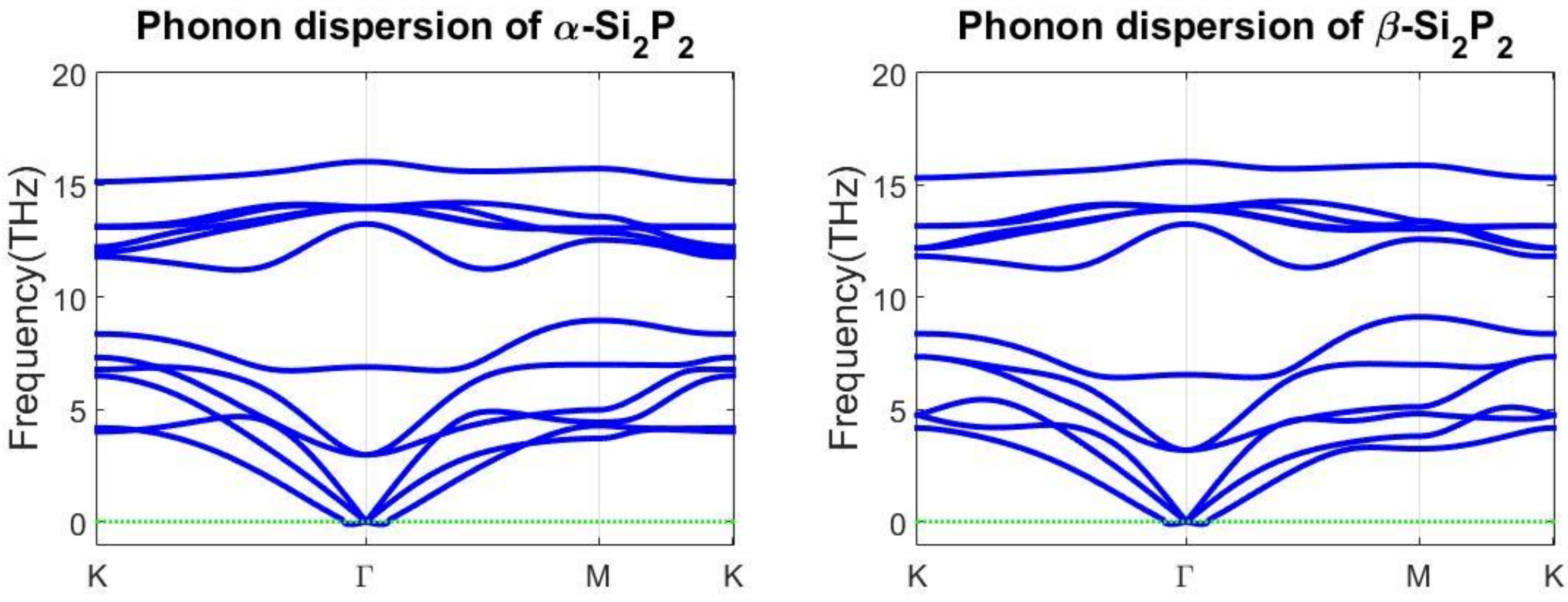

Figure S9

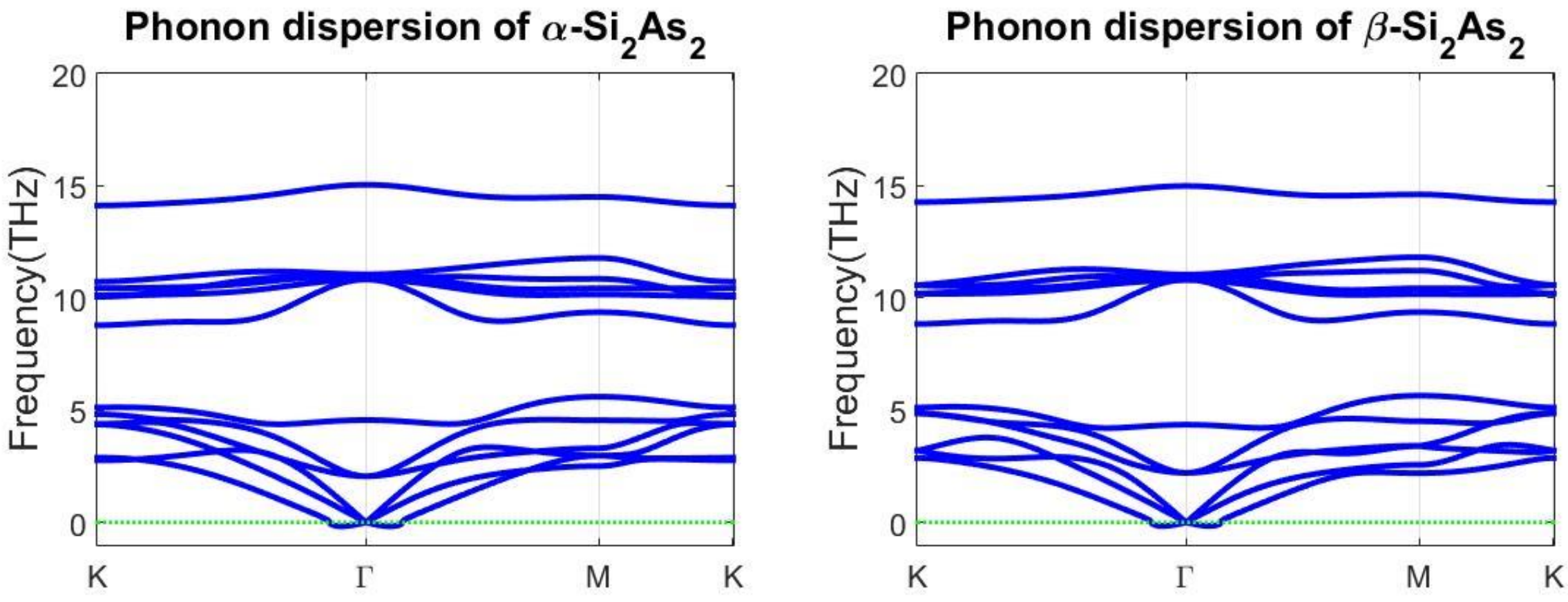

Figure S10

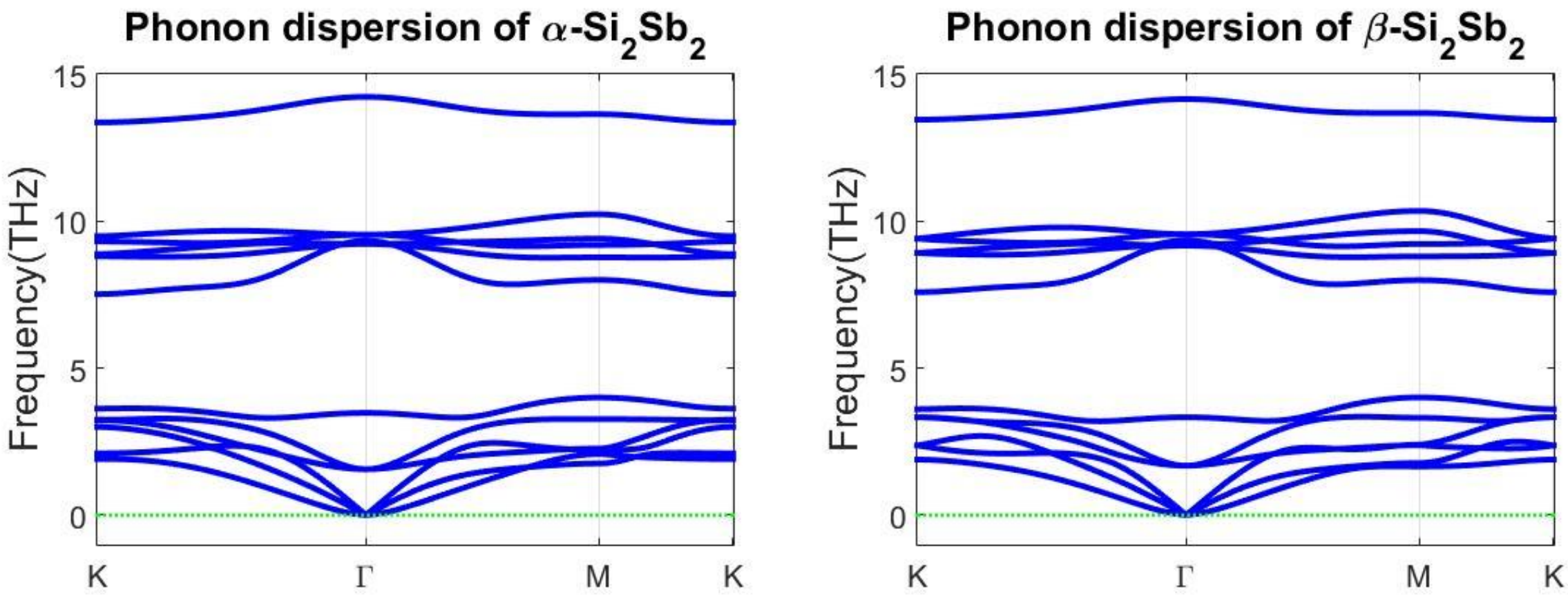

Figure S11

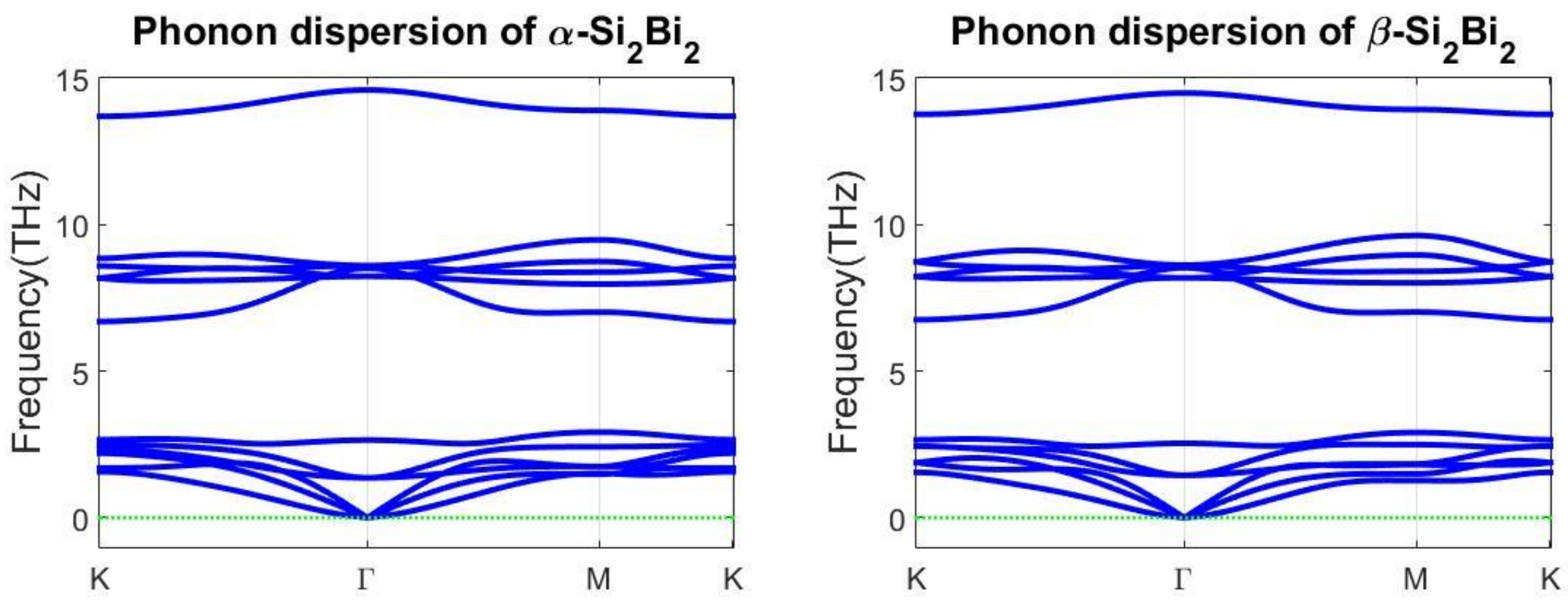

Figure S12

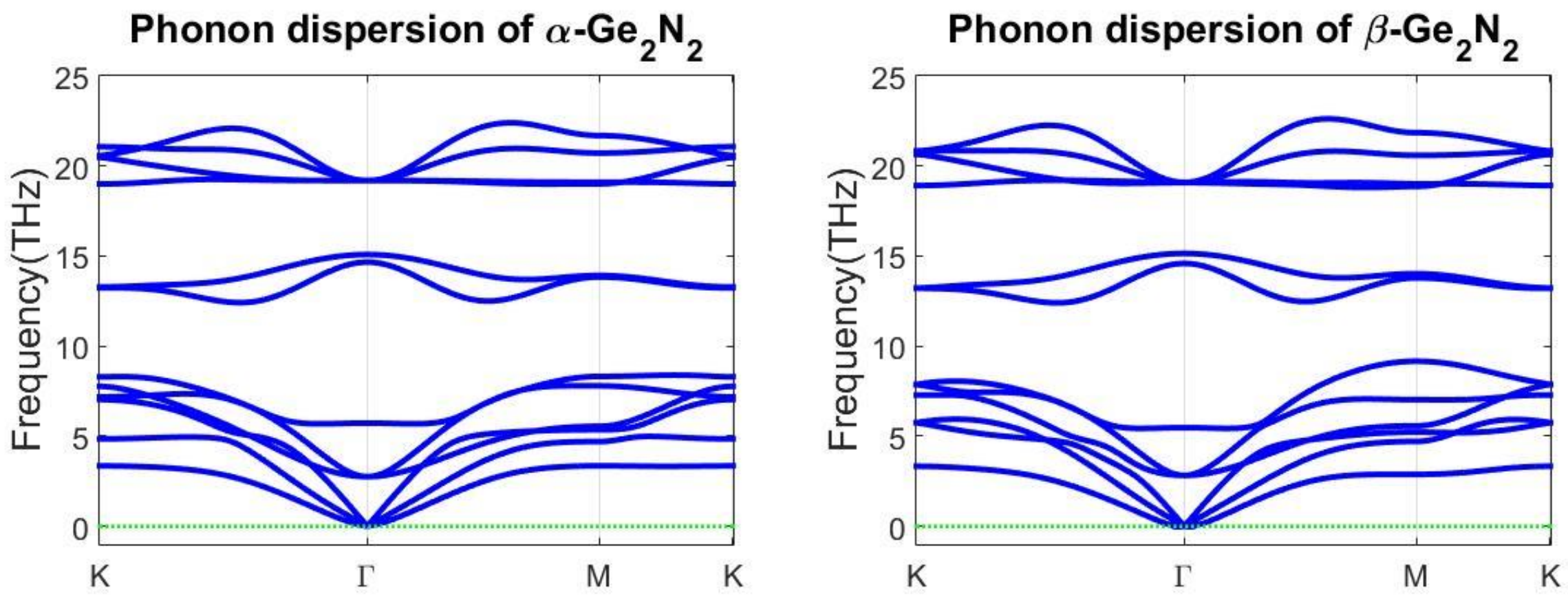

Figure S13

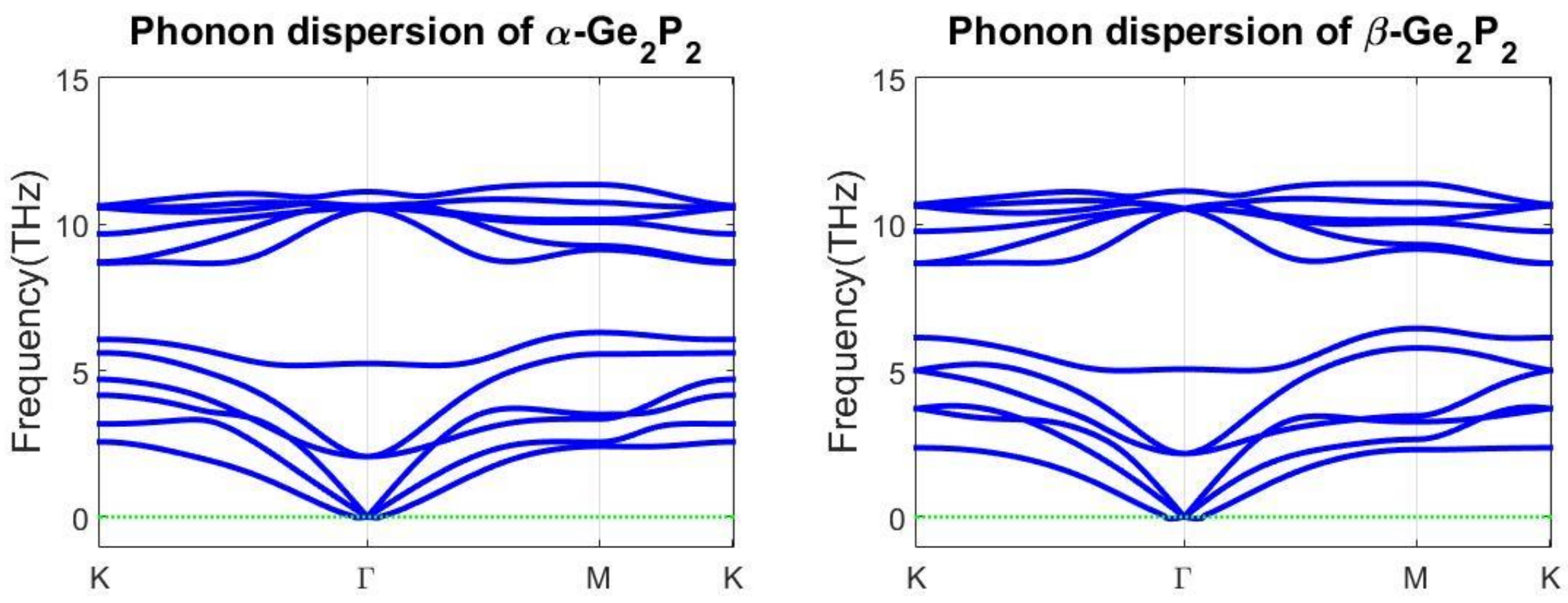

Figure S14

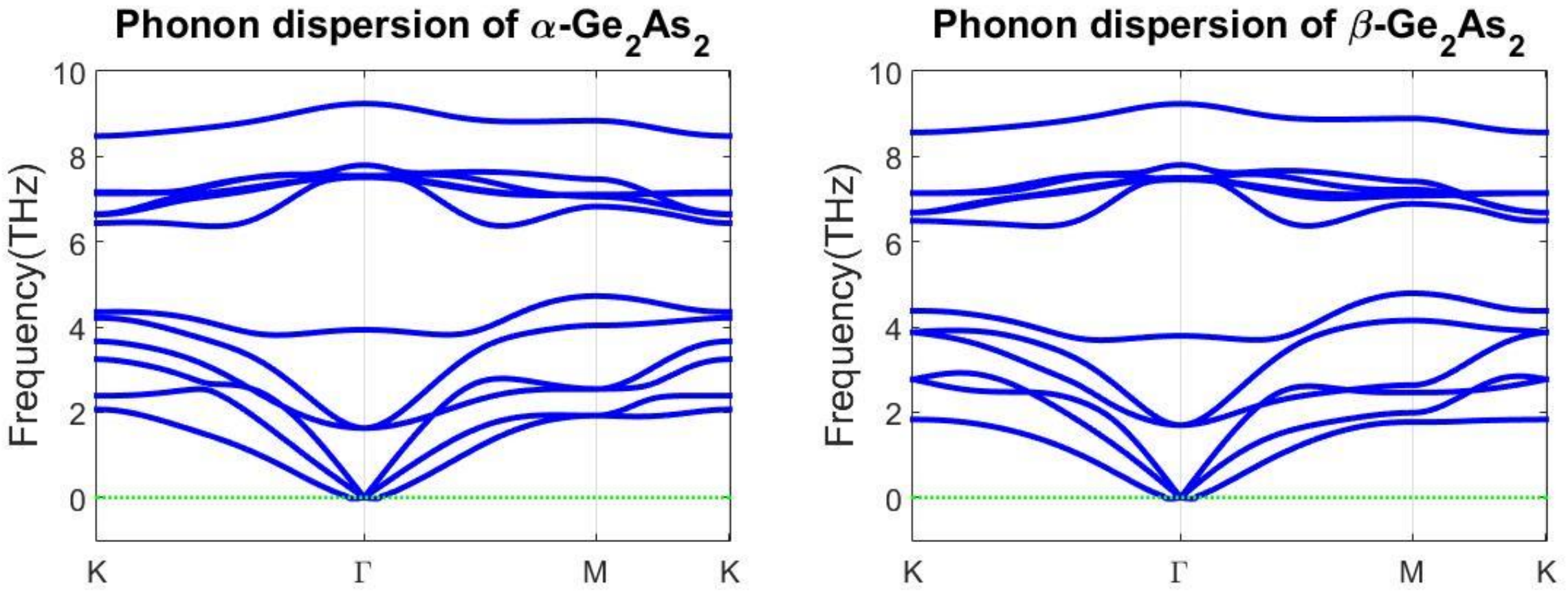

Figure S15

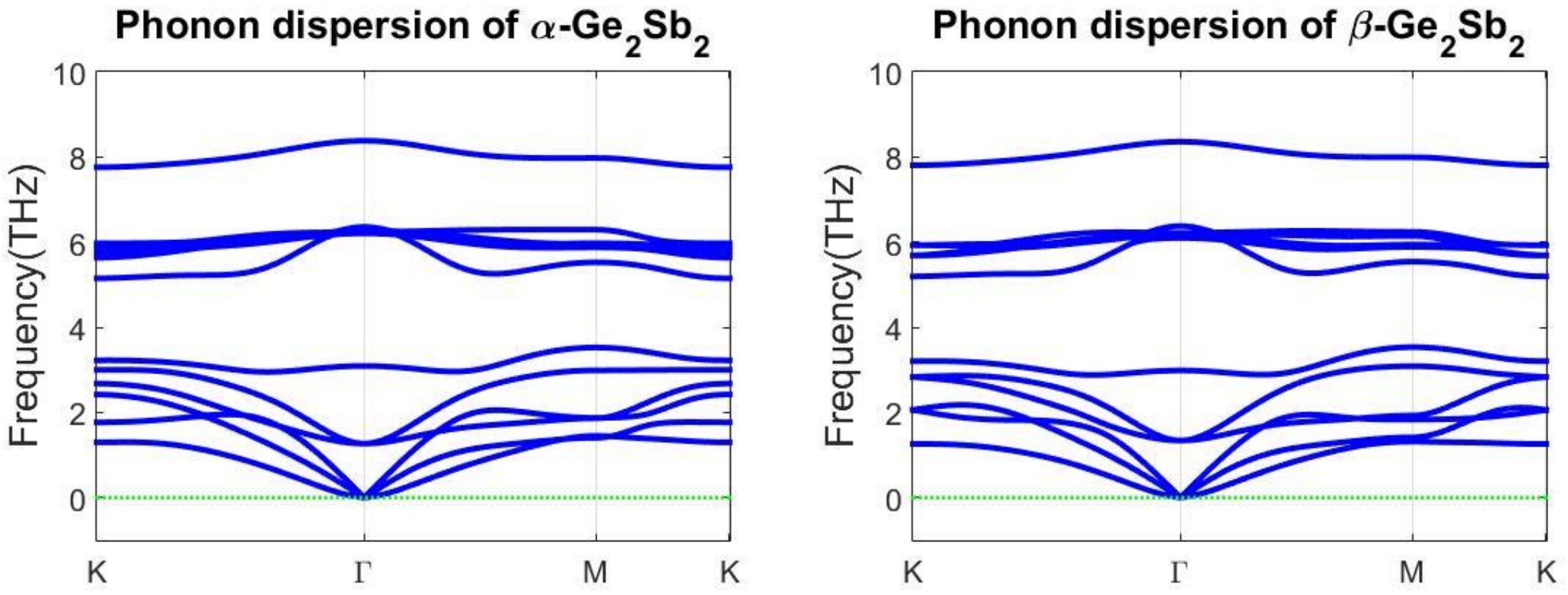

Figure S16

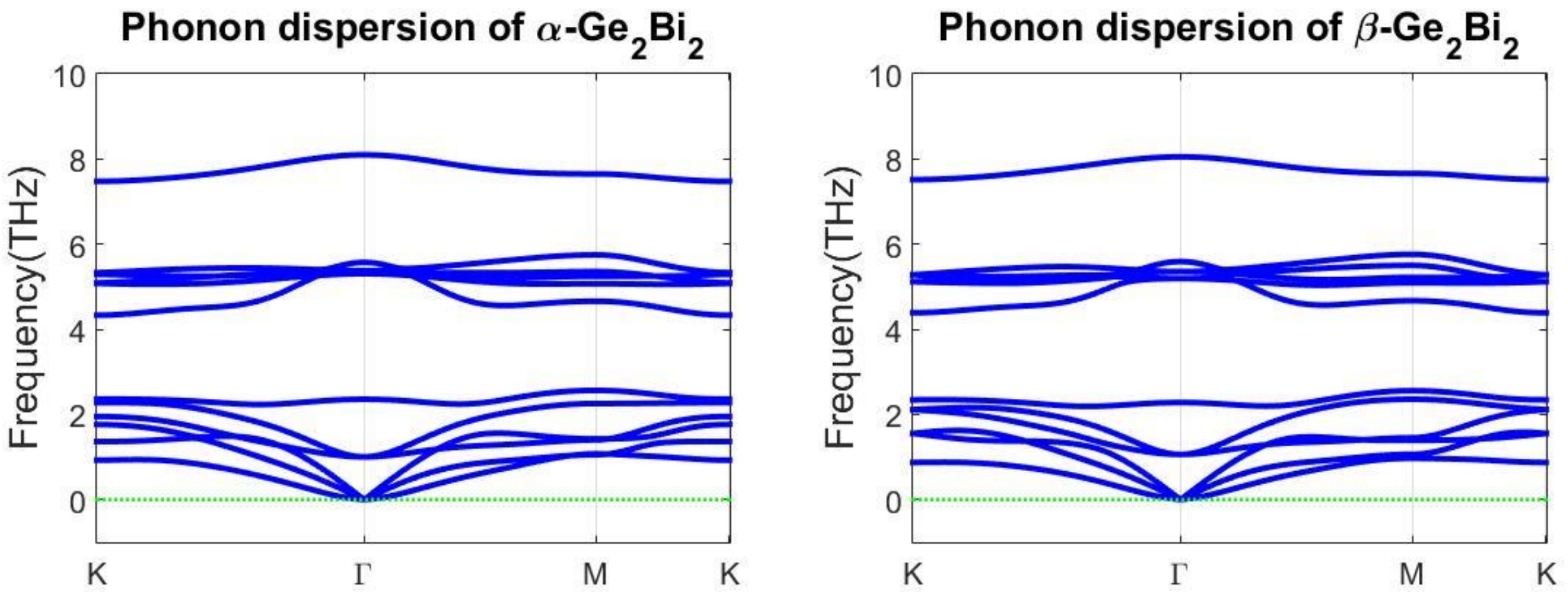

Figure S17

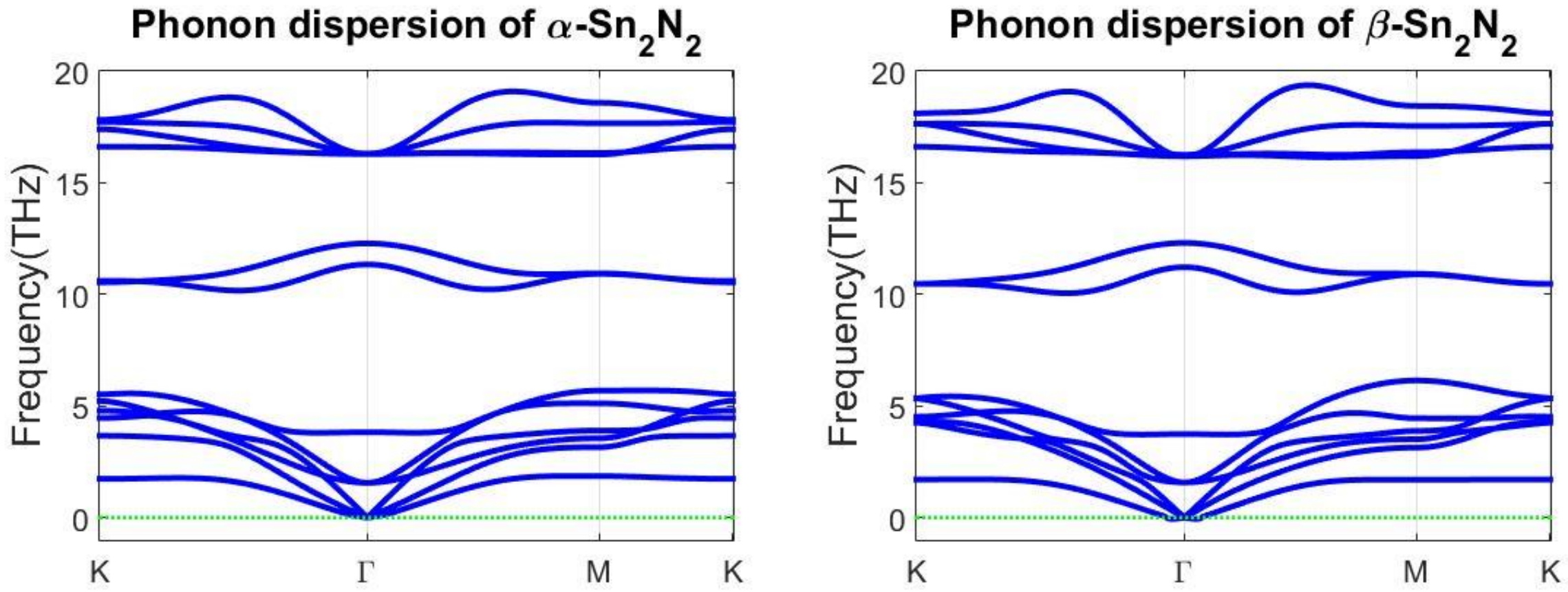

Figure S18

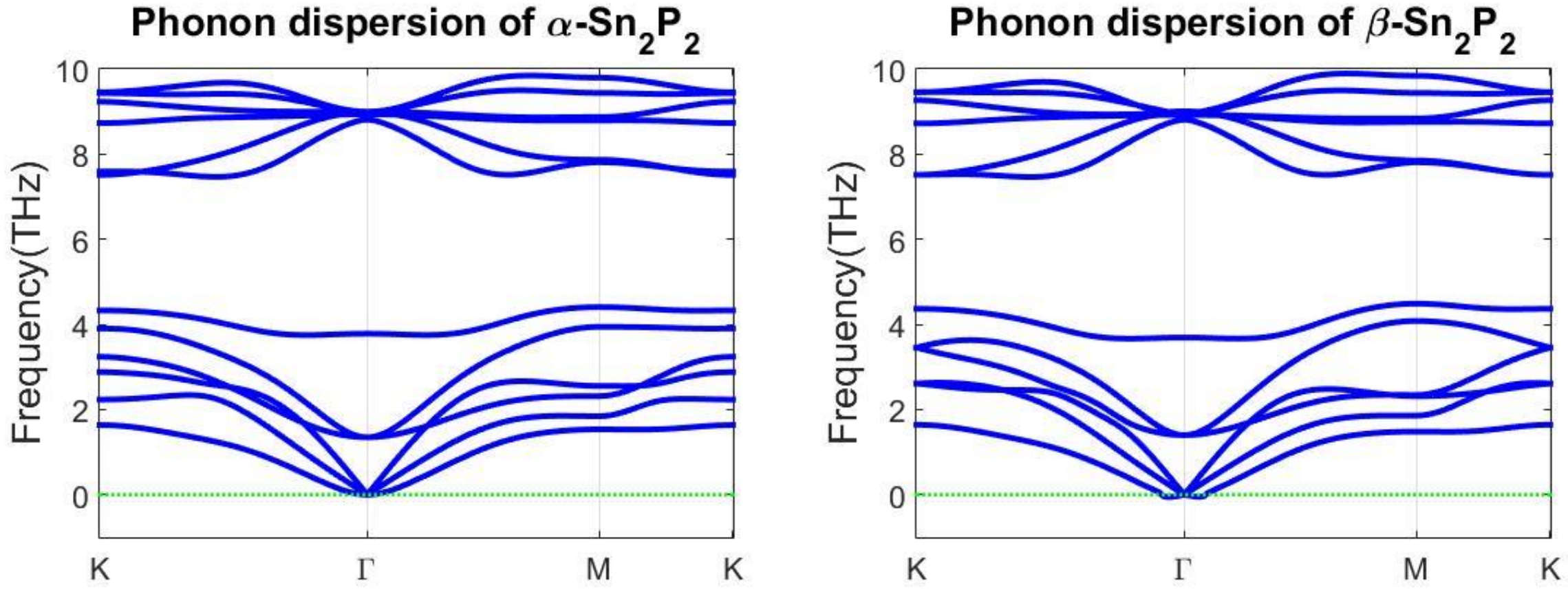

Figure S19

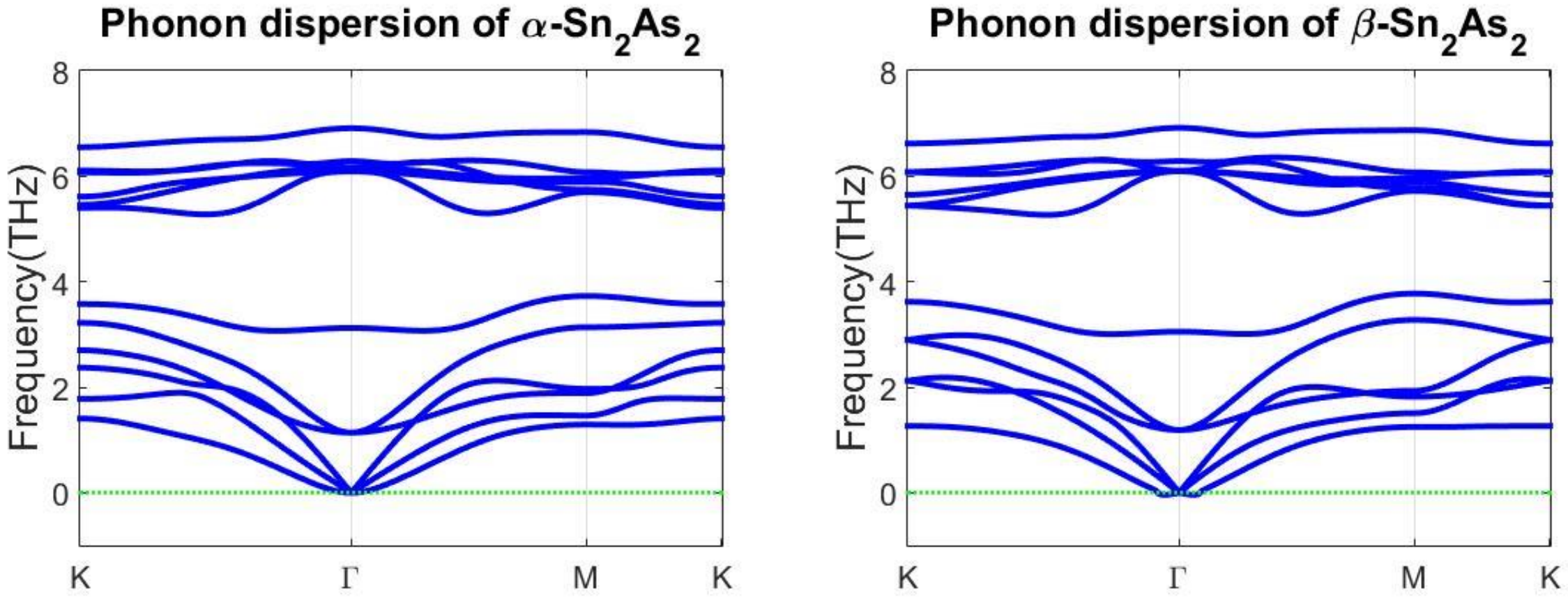

Figure S20

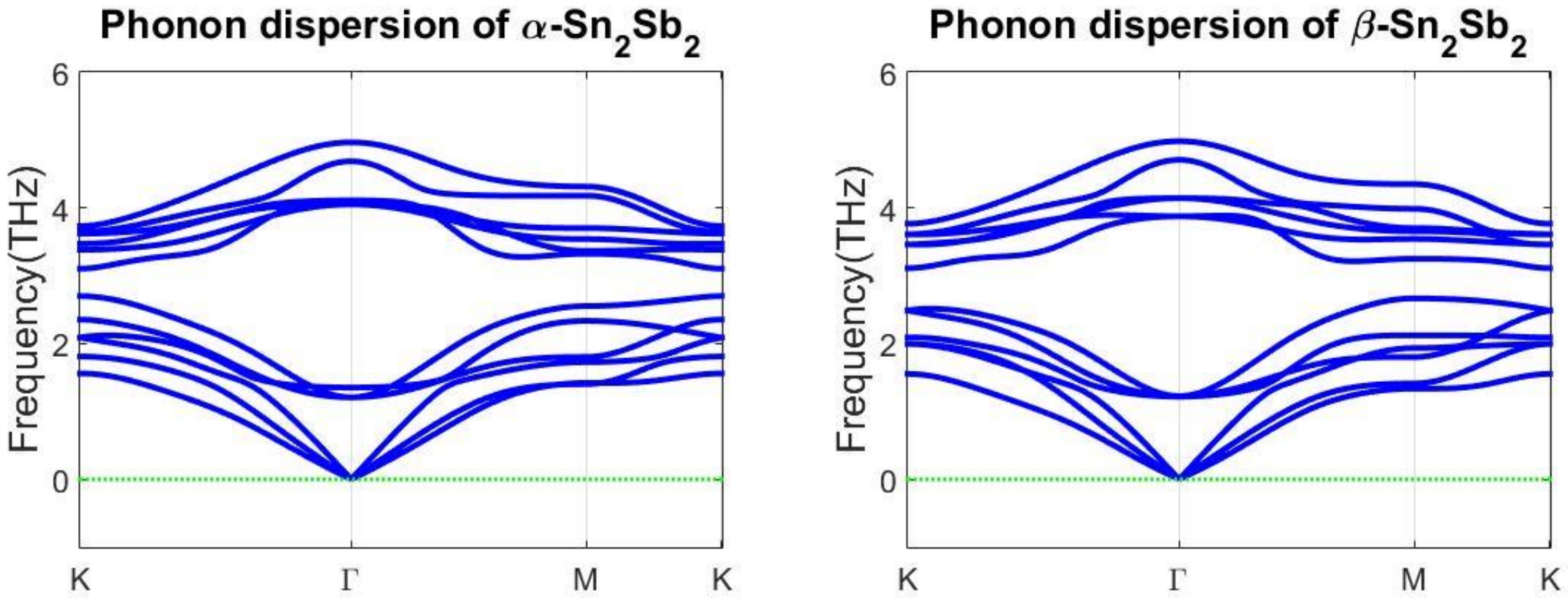

Figure S21

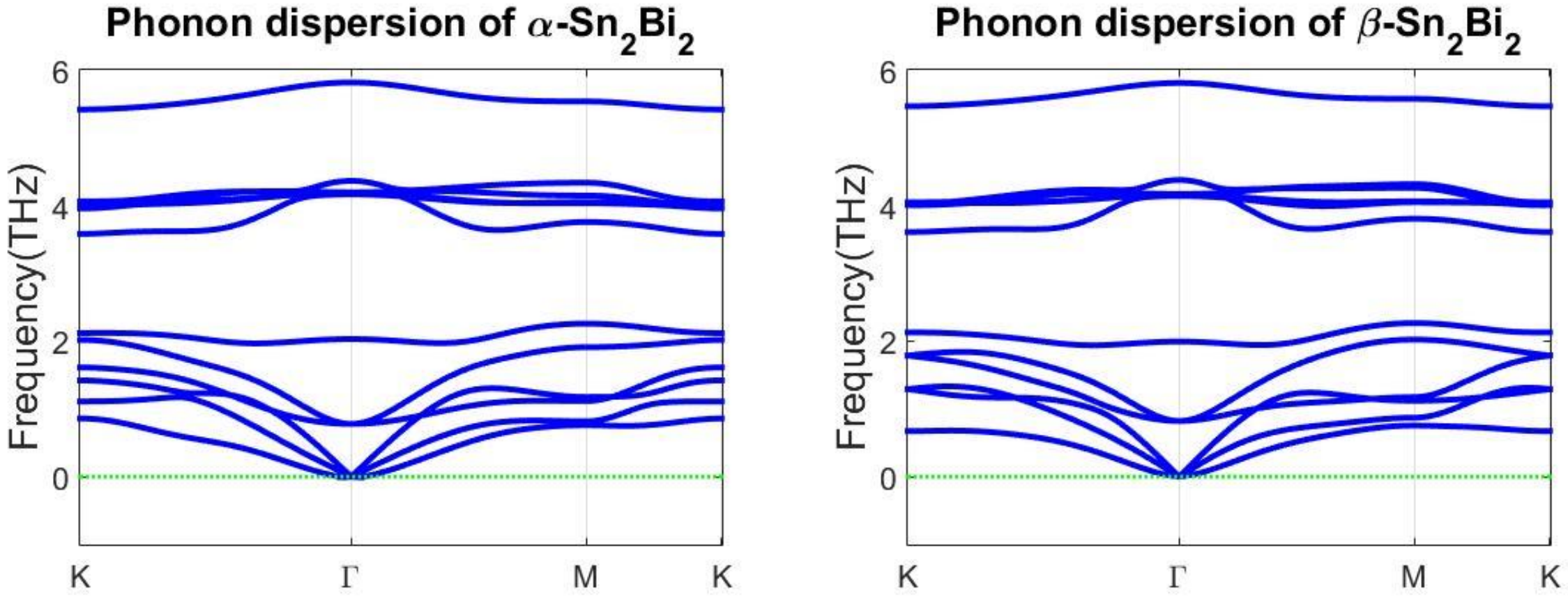

Figure S22

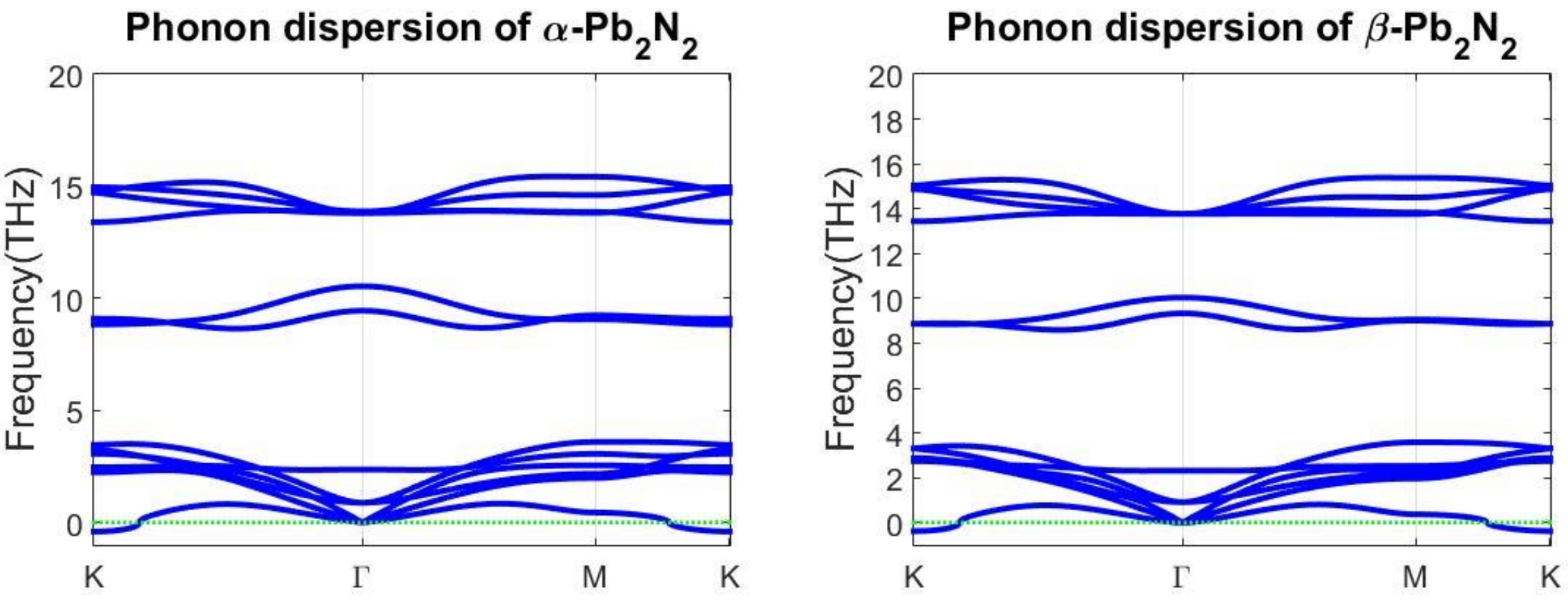

Figure S23

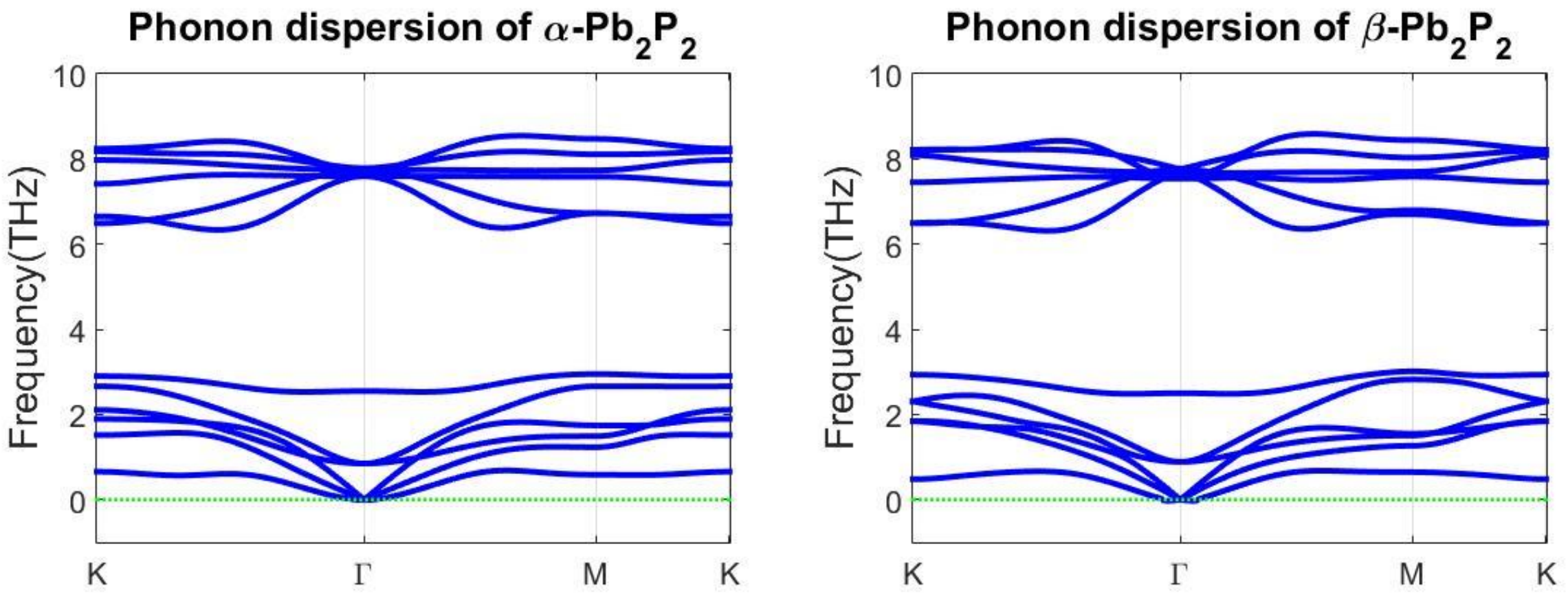

Figure S24

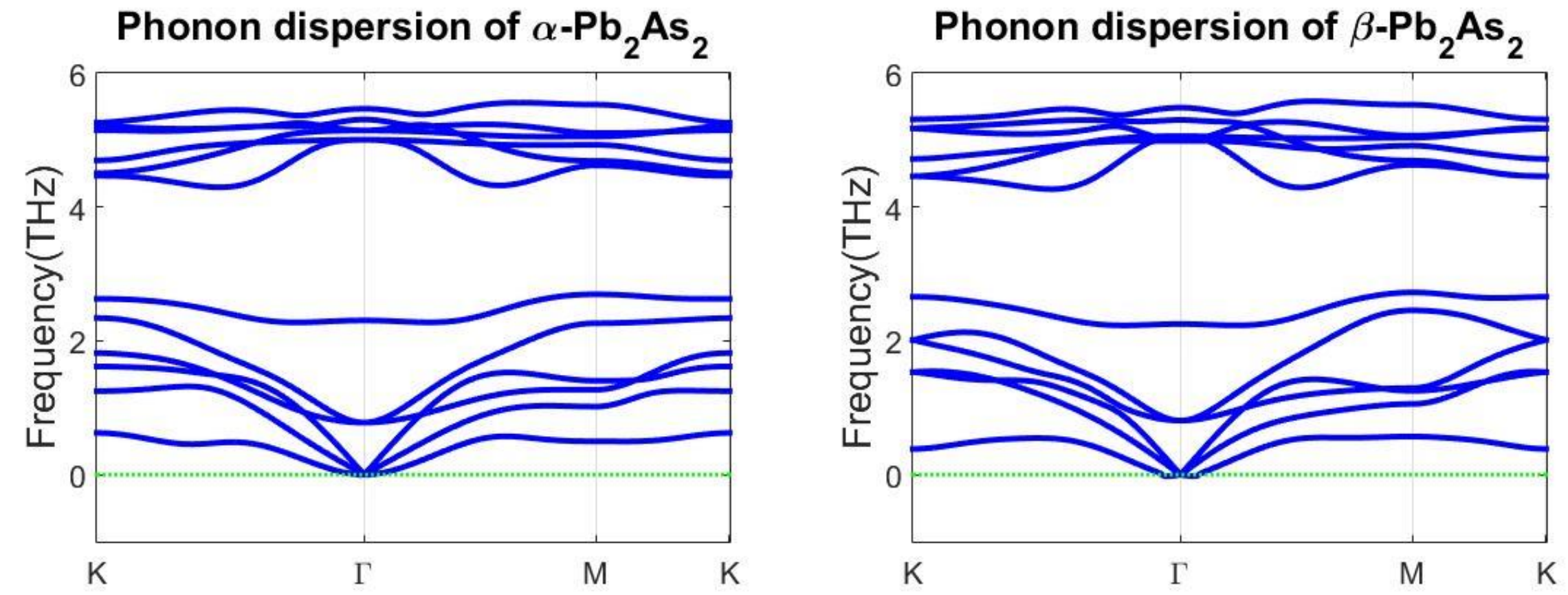

Figure S25

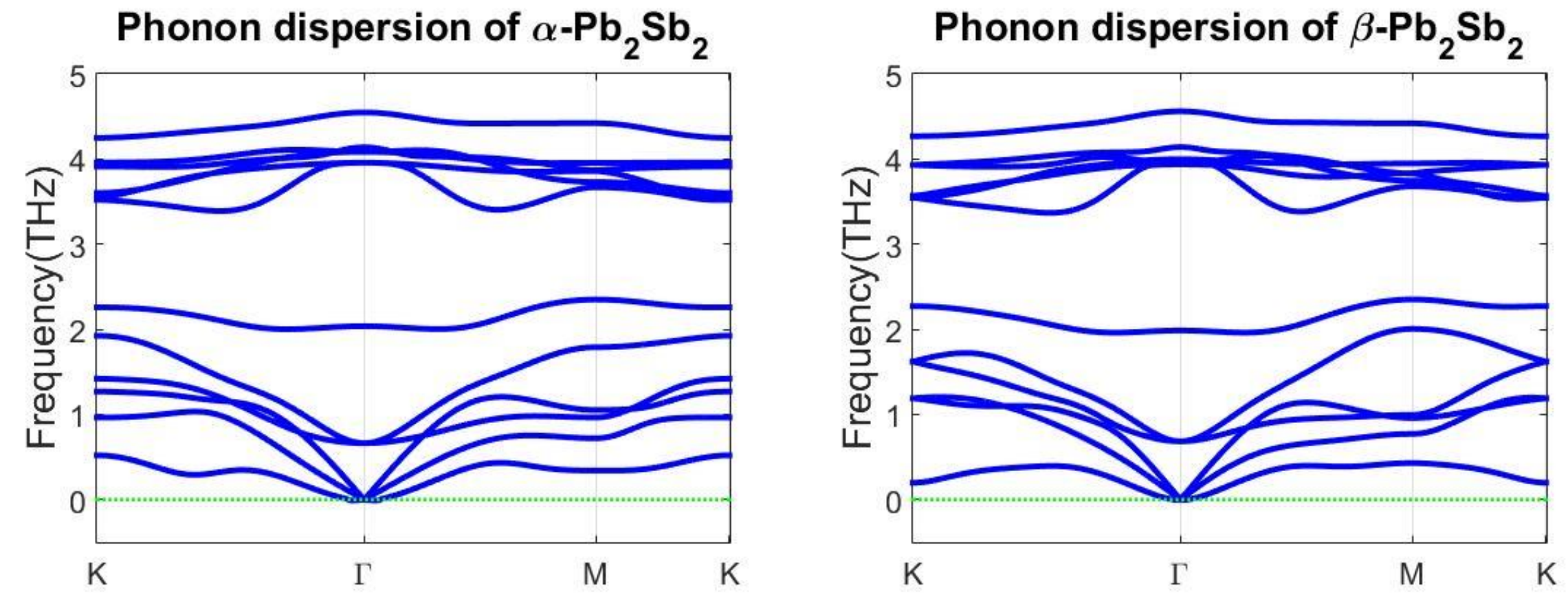

Figure S26

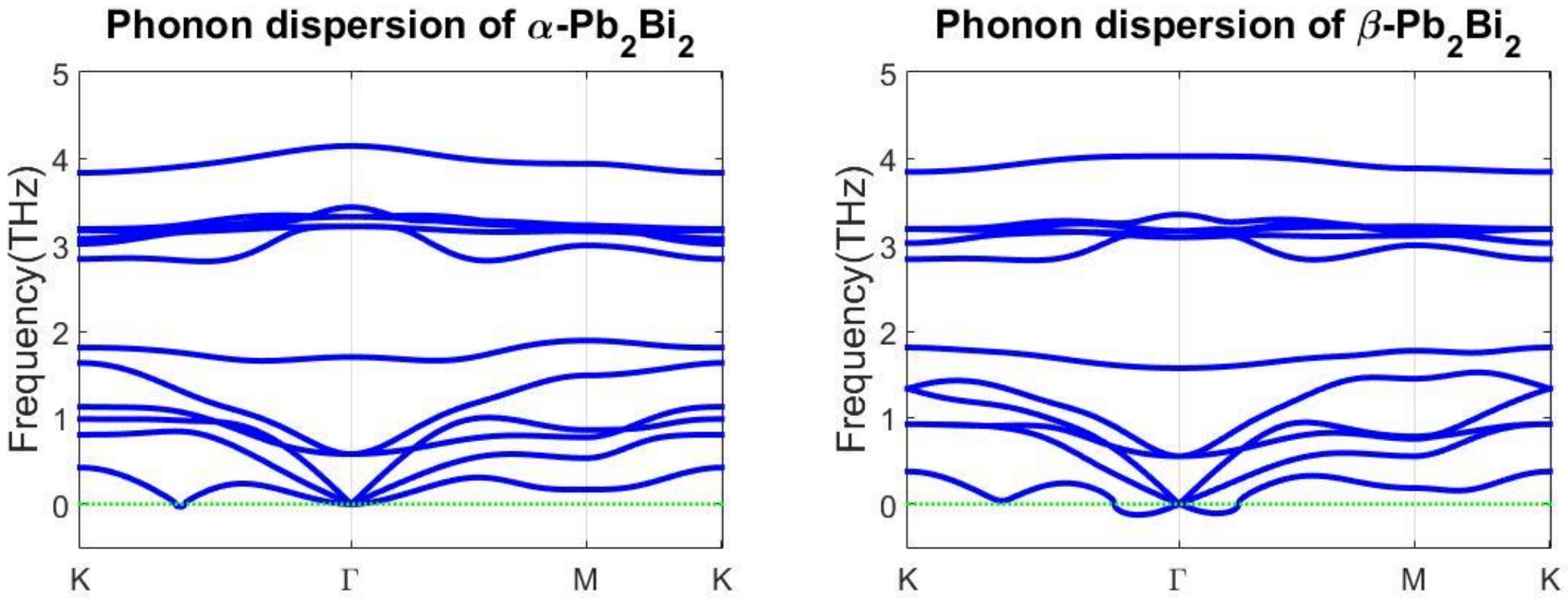

Figure S27

## IV. Electronic bands

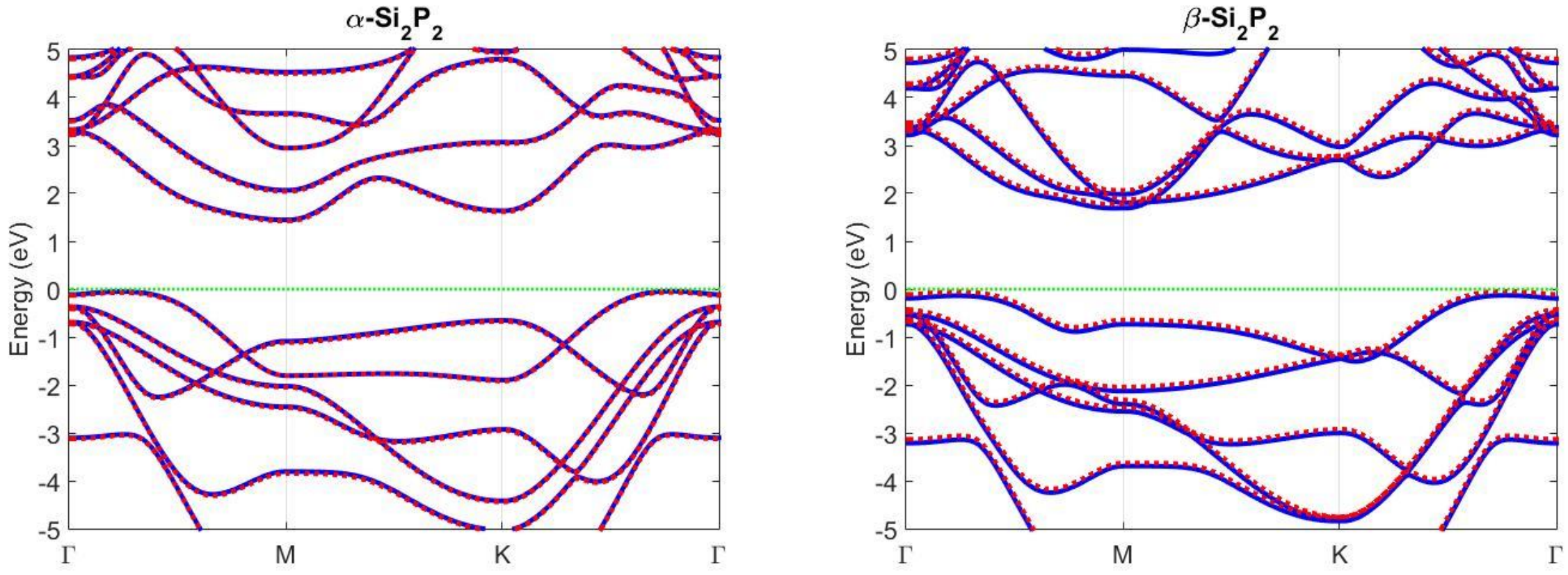

Figure S28. Electronic band of $\alpha$-$Si_2P_2$ (left) and $\beta$-$Si_2P_2$ (right), the solid blue line represents the results calculated by PBE, the red dot line represents the results calculated by PBE+SOC.

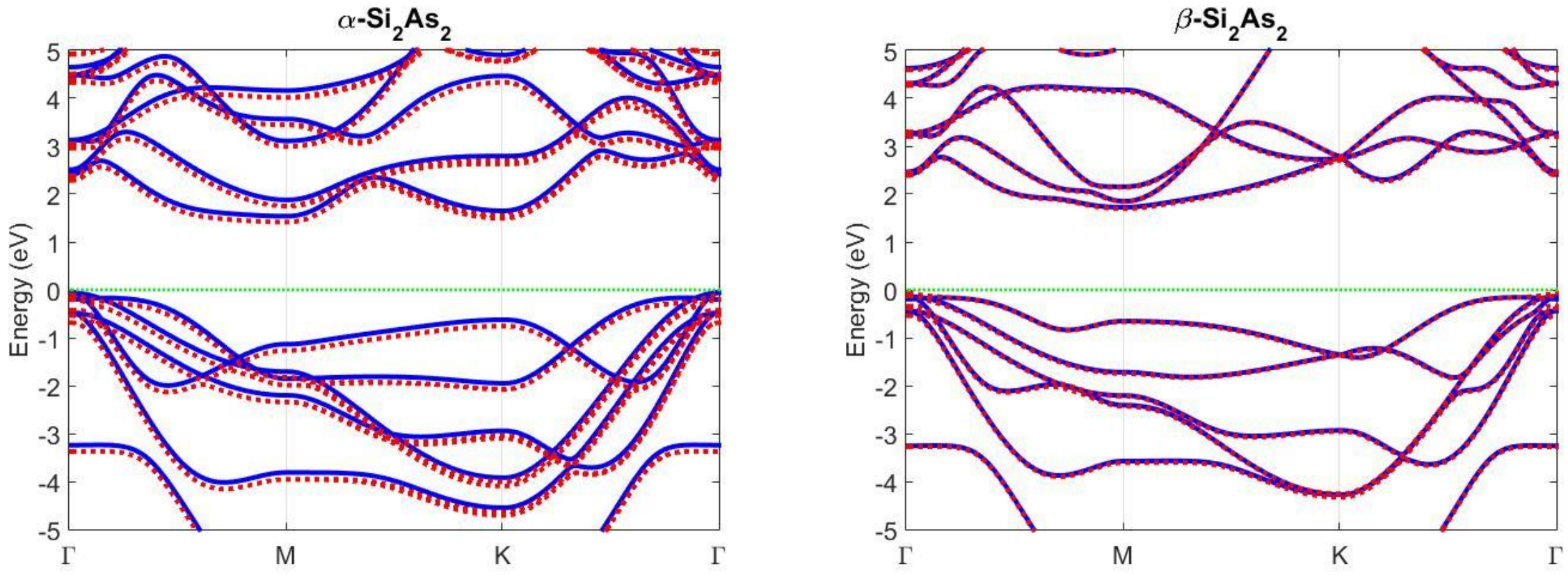

Figure S29. Electronic band of $\alpha$-$Si_2As_2$ (left) and $\beta$-$Si_2As_2$ (right), the solid blue line represents the results calculated by PBE, the red dot line represents the results calculated by PBE+SOC.

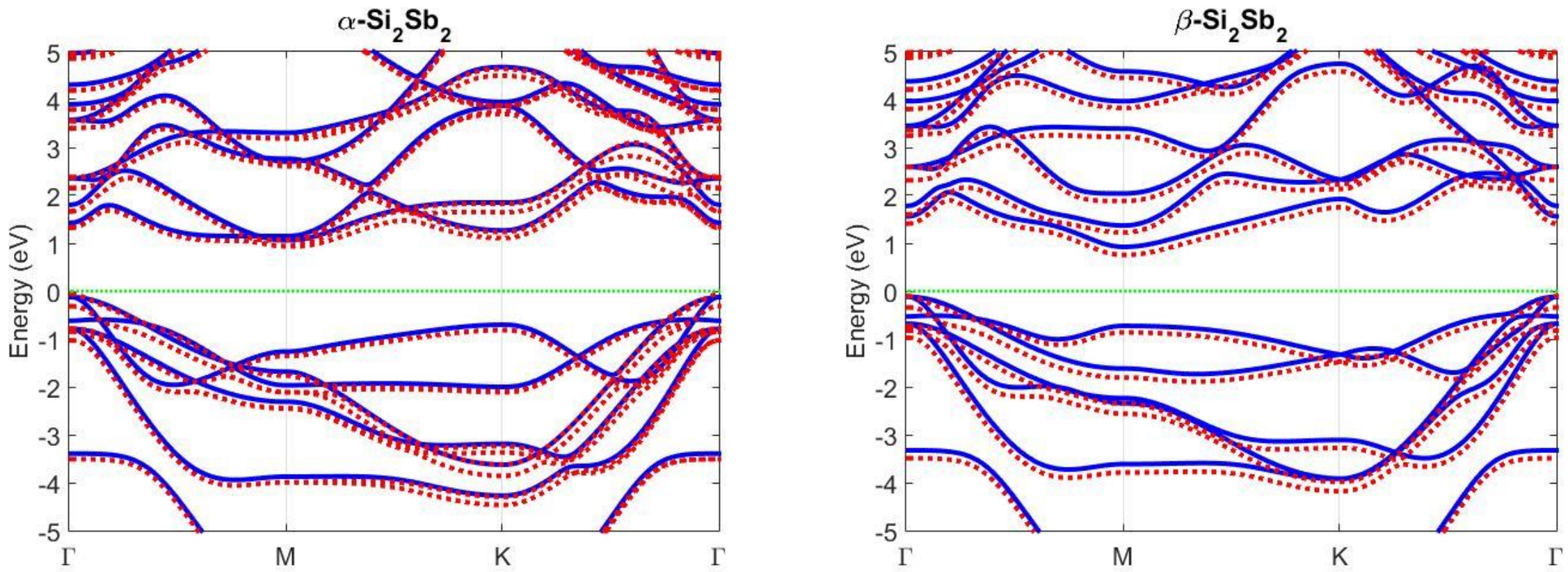

Figure S30. Electronic band of $\alpha$-$Si_2Sb_2$ (left) and $\beta$-$Si_2Sb_2$ (right), the solid blue line represents the results calculated by PBE, the red dot line represents the results calculated by PBE+SOC.

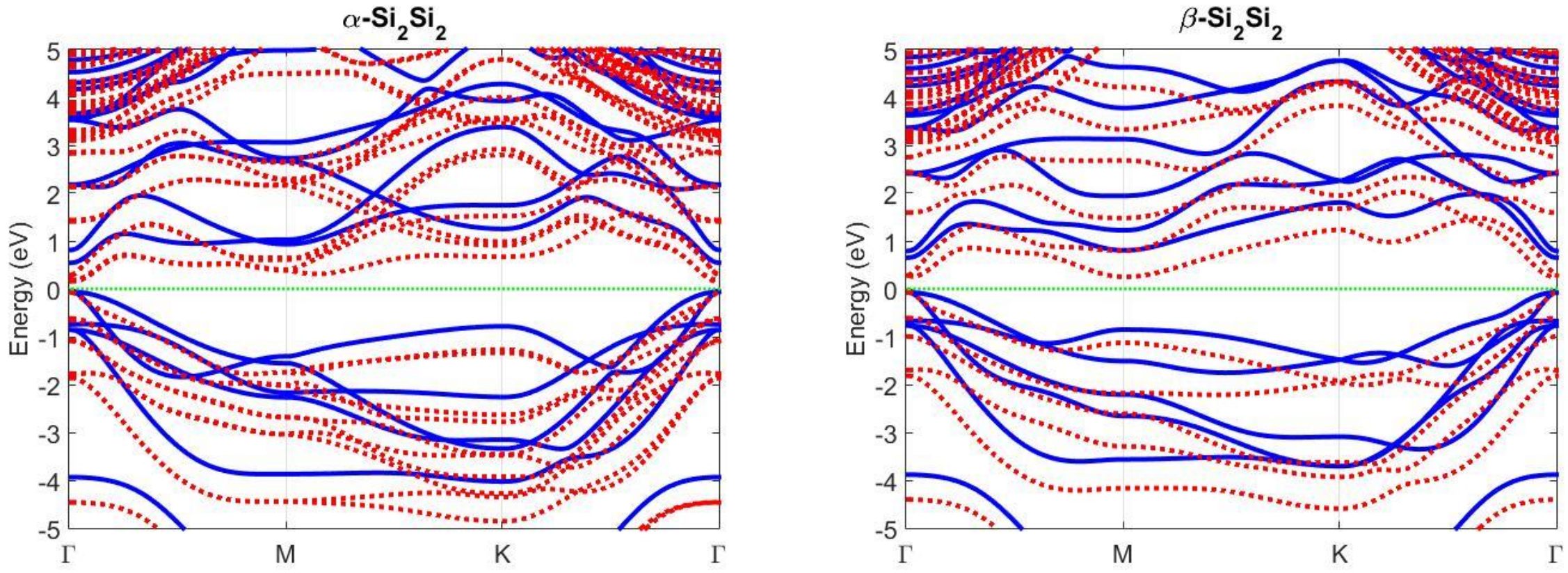

Figure S31. Electronic band of $\alpha$-$Si_2Bi_2$ (left) and $\beta$-$Si_2Bi_2$ (right), the solid blue line represents the results calculated by PBE, the red dot line represents the results calculated by PBE+SOC.

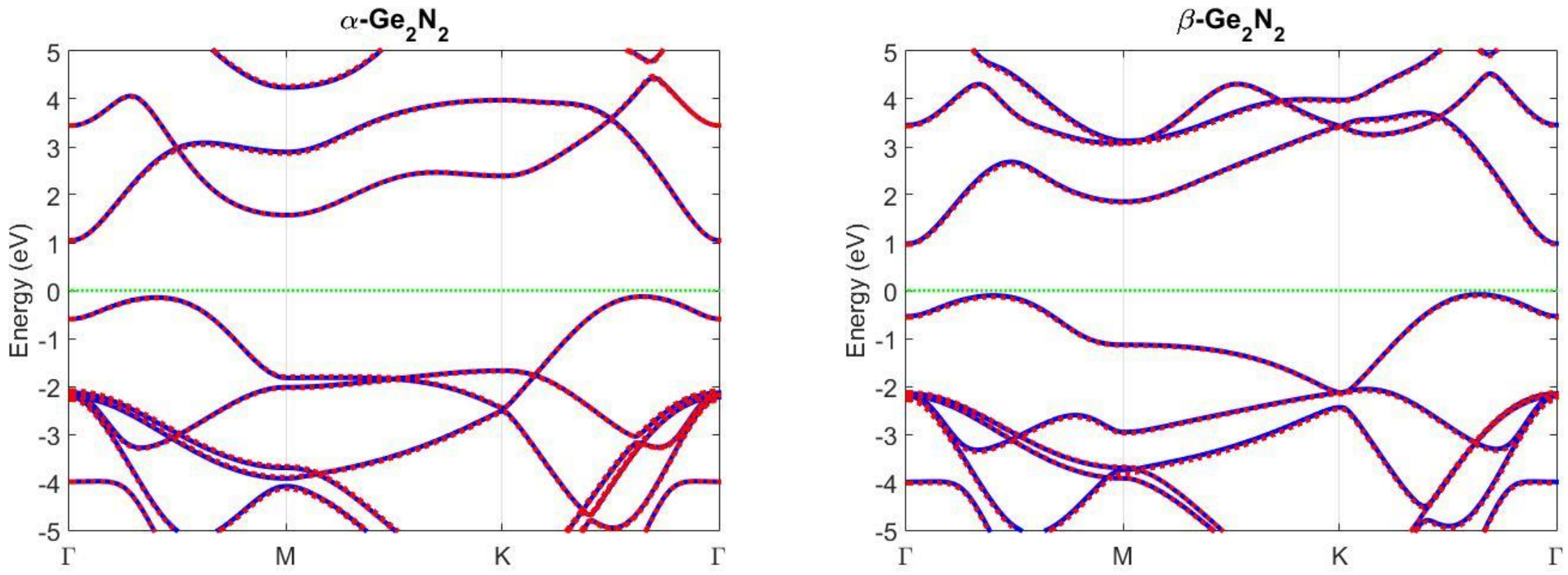

Figure S32. Electronic band of $\alpha$-$Ge_2N_2$ (left) and $\beta$-$Ge_2N_2$ (right), the solid blue line represents the results calculated by PBE, the red dot line represents the results calculated by PBE+SOC.

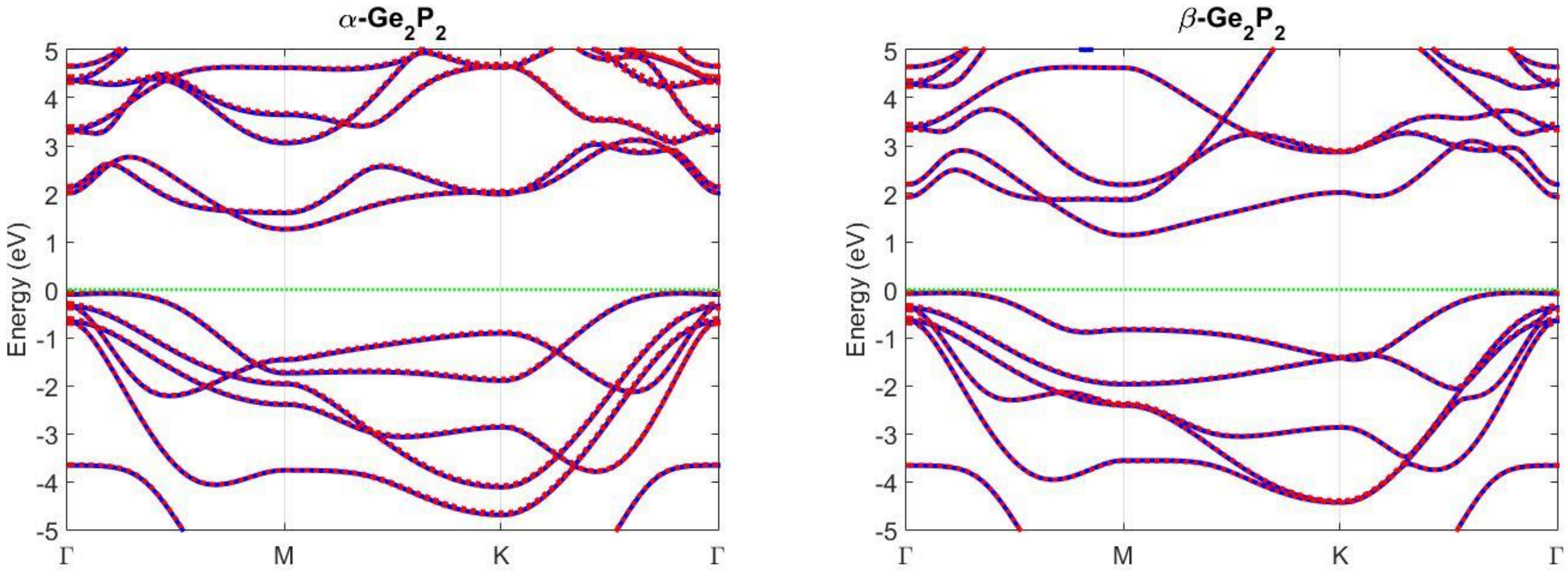

Figure S33. Electronic band of *α*-$Ge_2P_2$ (left) and *β*-$Ge_2P_2$ (right), the solid blue line represents the results calculated by PBE, the red dot line represents the results calculated by PBE+SOC.

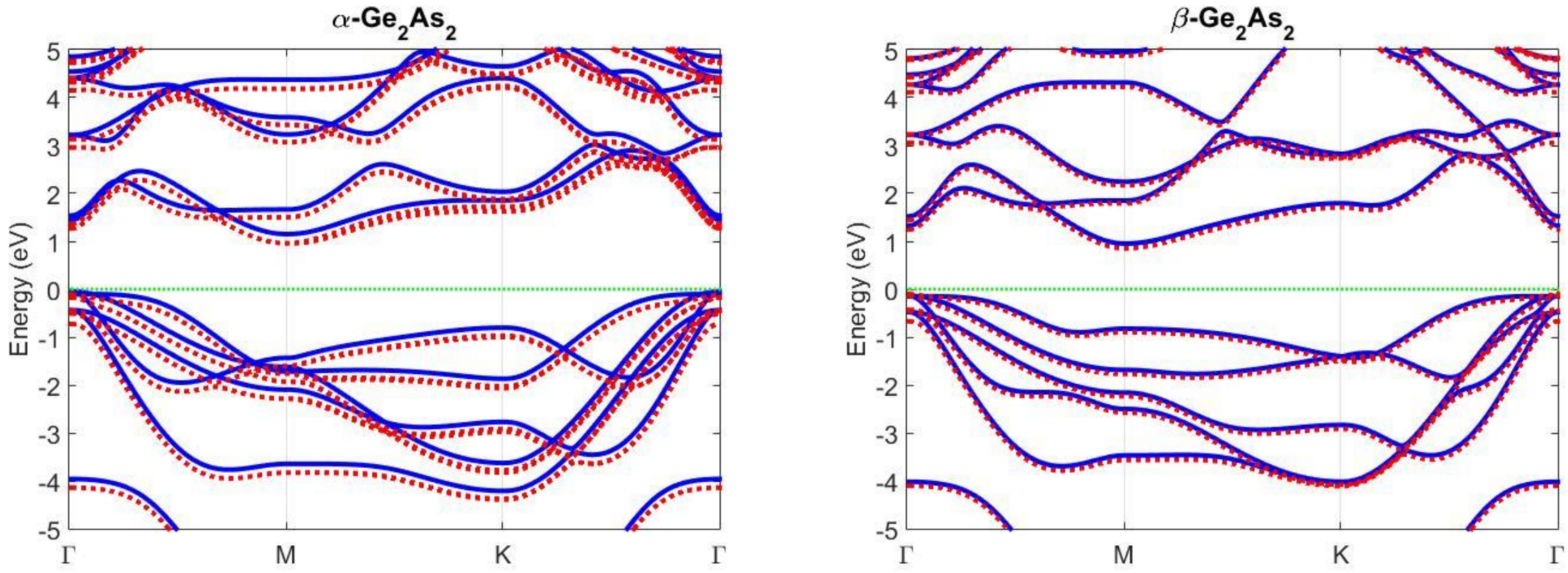

Figure S34. Electronic band of *α*-$Ge_2As_2$ (left) and *β*-$Ge_2As_2$ (right), the solid blue line represents the results calculated by PBE, the red dot line represents the results calculated by PBE+SOC.

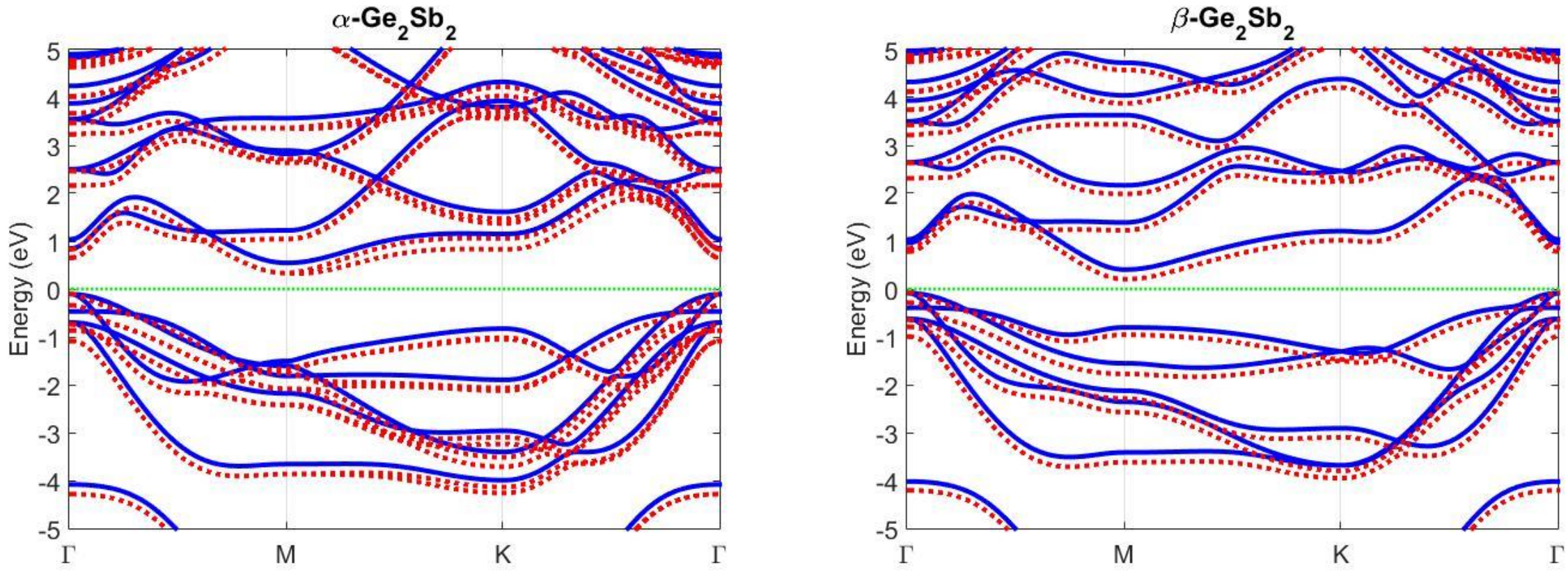

Figure S35. Electronic band of *α*-$Ge_2Sb_2$ (left) and *β*-$Ge_2Sb_2$ (right), the solid blue line represents the results calculated by PBE, the red dot line represents the results calculated by PBE+SOC.

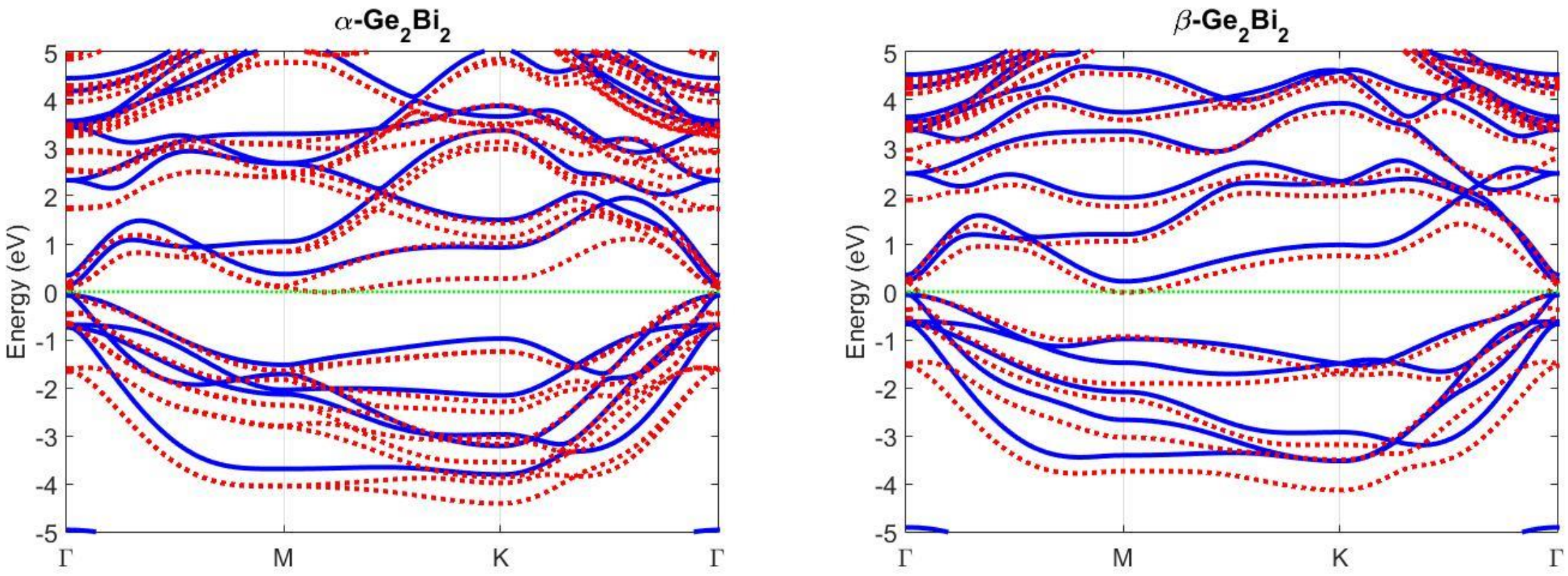

Figure S36. Electronic band of *α*-$Ge_2Bi_2$ (left) and *β*-$Ge_2Bi_2$ (right), the solid blue line represents the results calculated by PBE, the red dot line represents the results calculated by PBE+SOC.

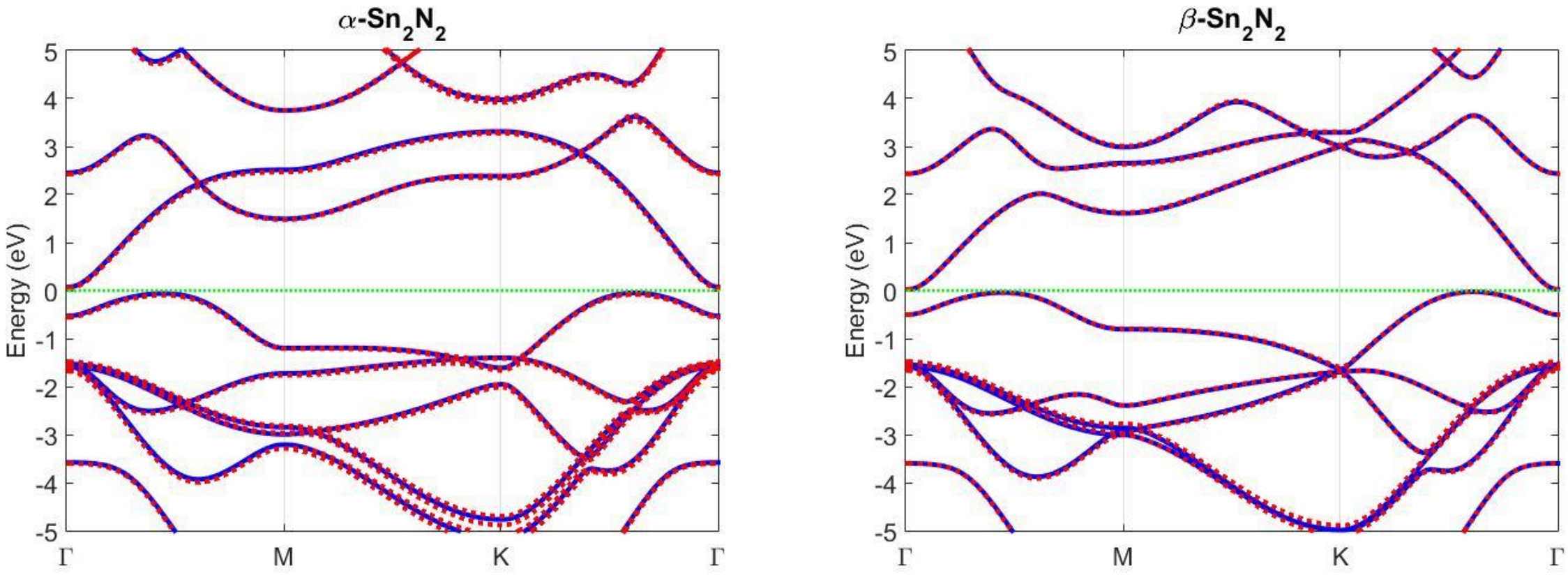

Figure S37. Electronic band of *α*-$Sn_2N_2$ (left) and *β*-$Sn_2N_2$ (right), the solid blue line represents the results calculated by PBE, the red dot line represents the results calculated by PBE+SOC.

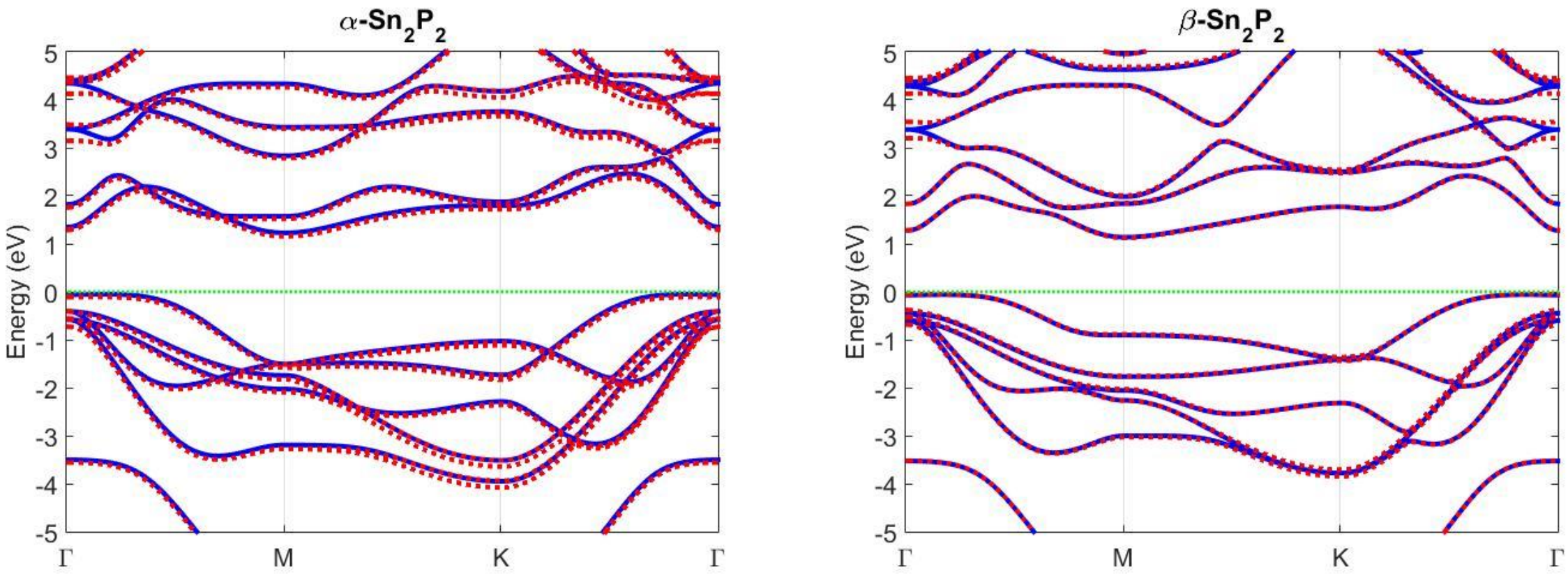

Figure S38. Electronic band of *α*-$Sn_2P_2$ (left) and *β*-$Sn_2P_2$ (right), the solid blue line represents the results calculated by PBE, the red dot line represents the results calculated by PBE+SOC.

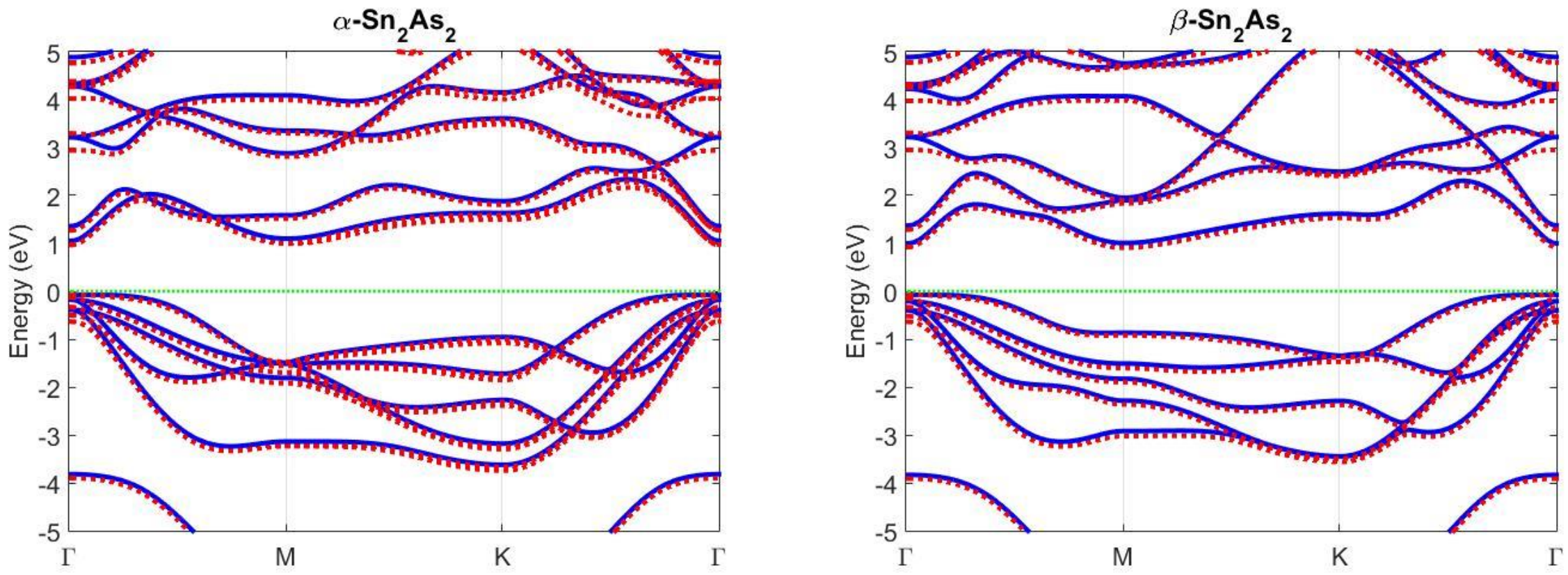

Figure S39. Electronic band of $\alpha$-$Sn_2As_2$ (left) and $\beta$-$Sn_2As_2$ (right), the solid blue line represents the results calculated by PBE, the red dot line represents the results calculated by PBE+SOC.

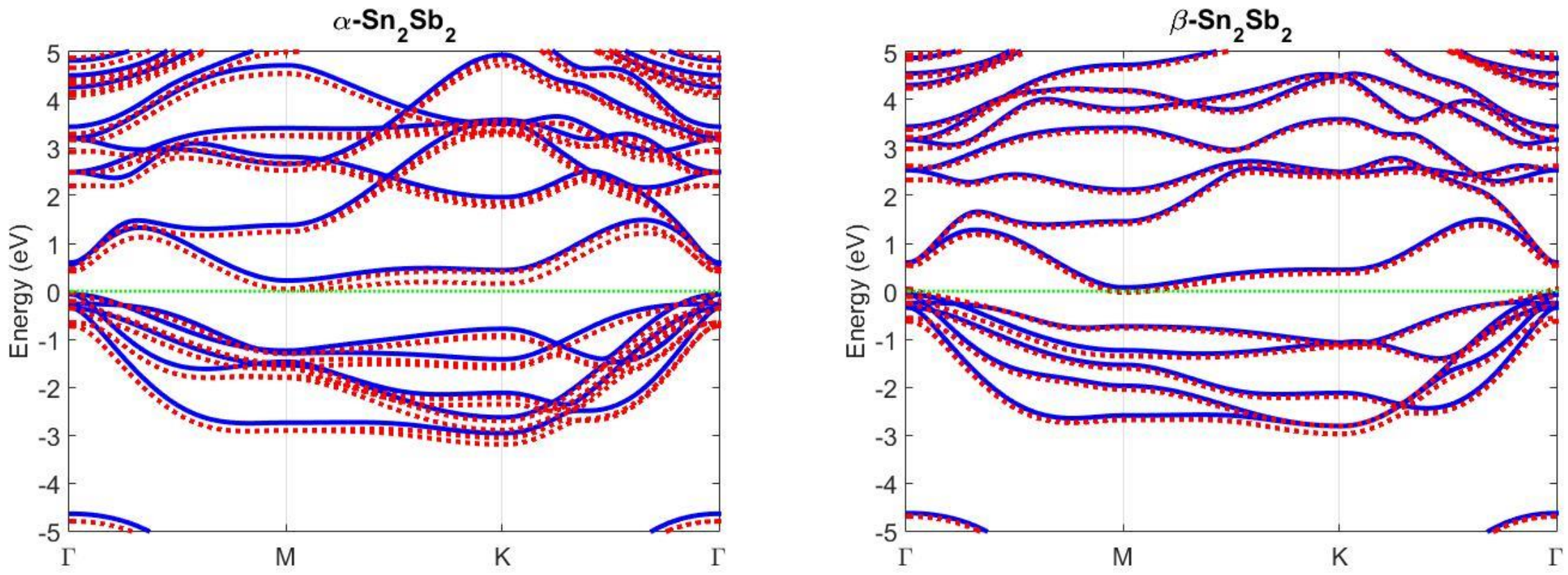

Figure S40. Electronic band of $\alpha$-$Sn_2Sb_2$ (left) and $\beta$-$Sn_2Sb_2$ (right), the solid blue line represents the results calculated by PBE, the red dot line represents the results calculated by PBE+SOC.

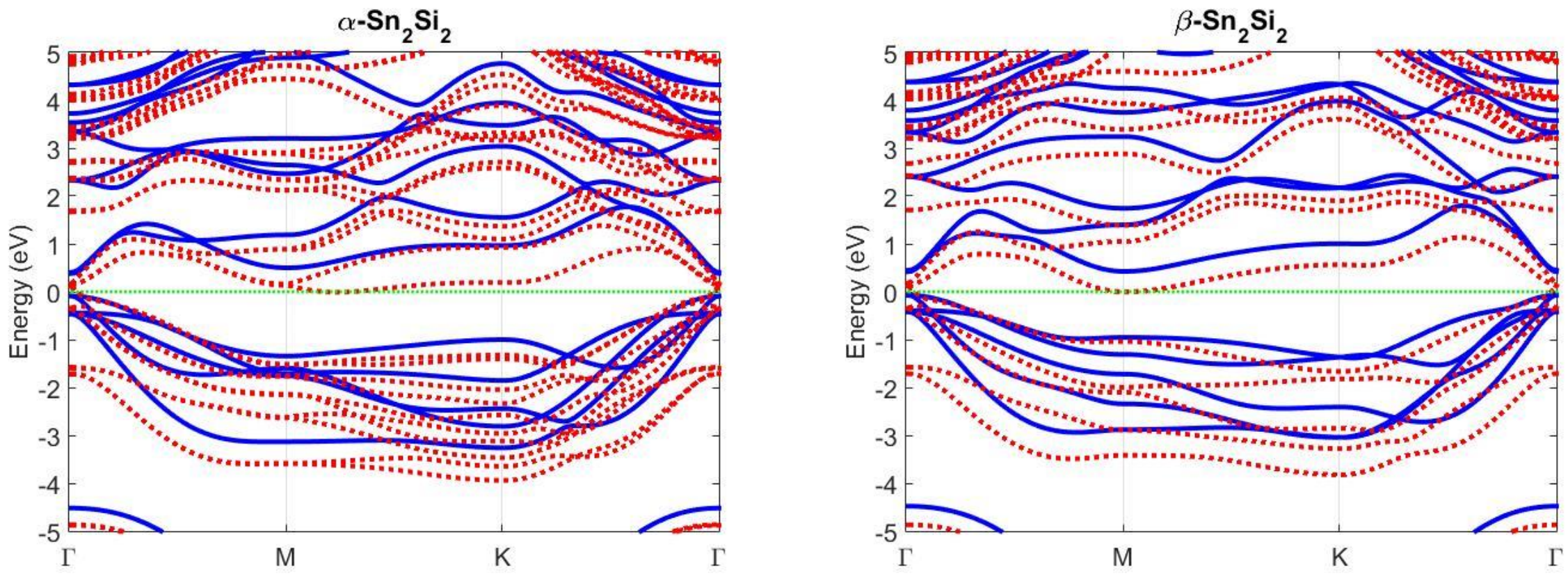

Figure S41. Electronic band of $\alpha$-$Sn_2Bi_2$ (left) and $\beta$-$Sn_2Bi_2$ (right), the solid blue line represents the results calculated by PBE, the red dot line represents the results calculated by PBE+SOC.

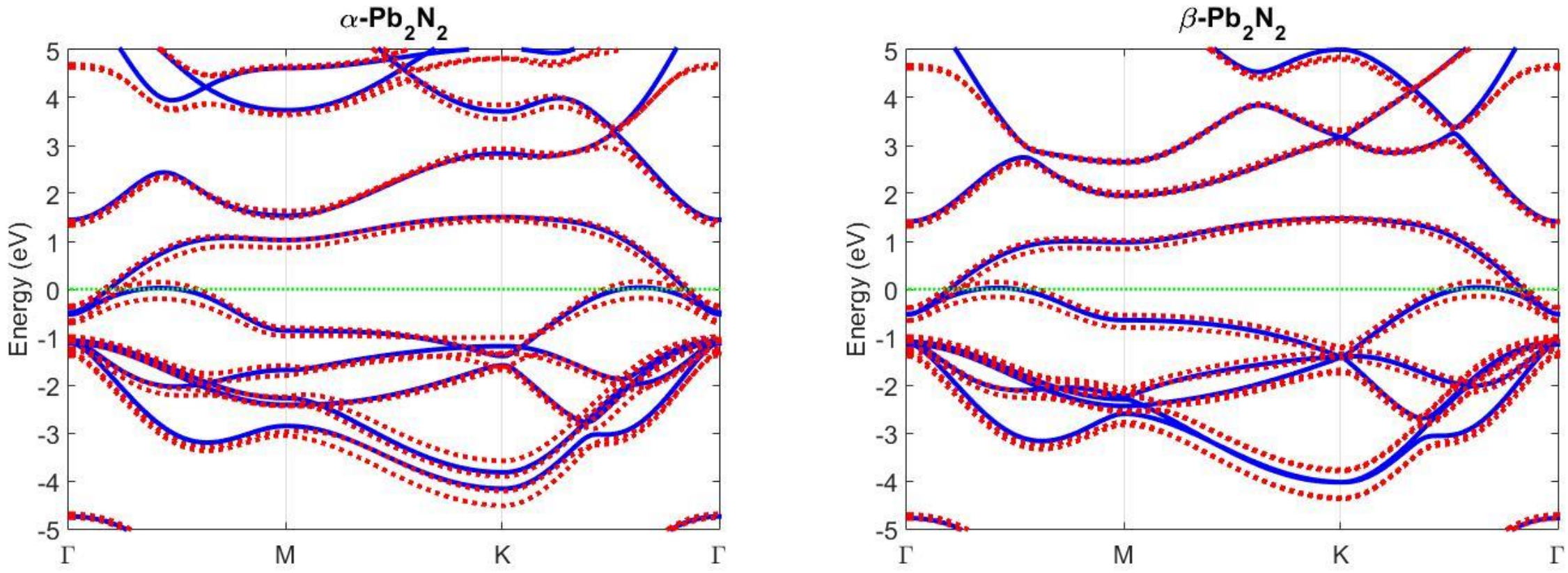

Figure S42. Electronic band of $\alpha$-$Pb_2N_2$ (left) and $\beta$-$Pb_2N_2$ (right), the solid blue line represents the results calculated by PBE, the red dot line represents the results calculated by PBE+SOC.

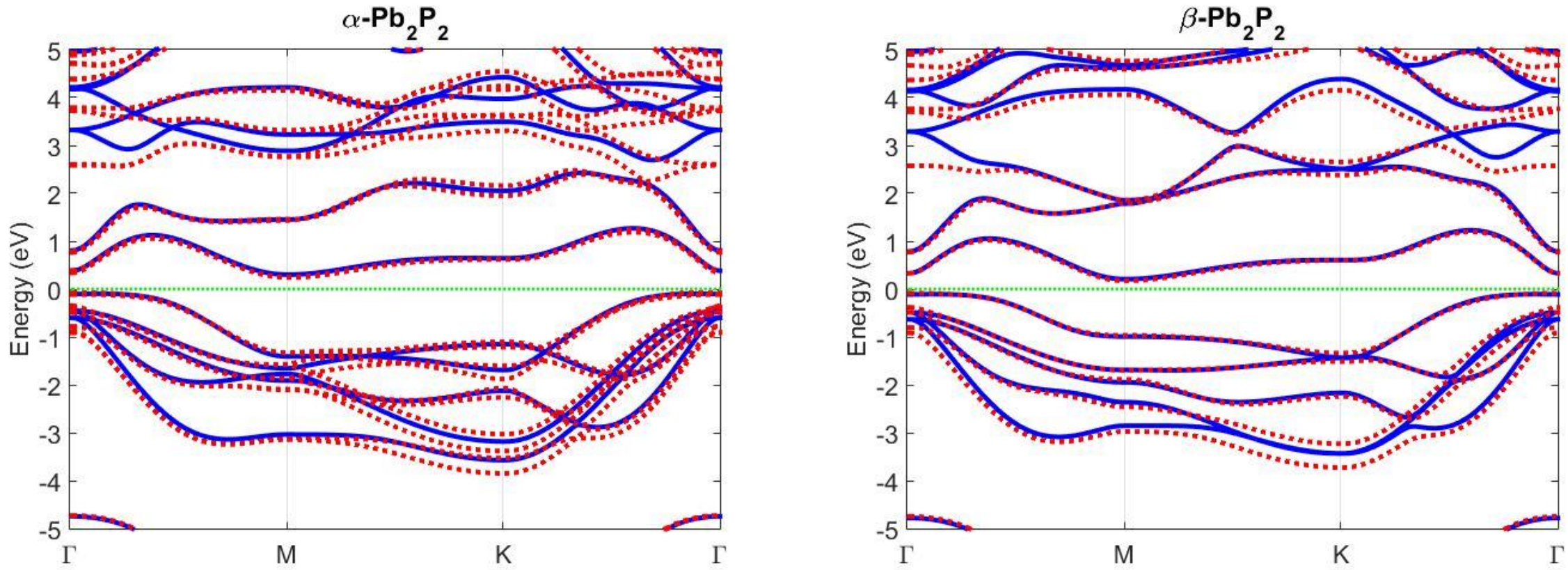

Figure S43. Electronic band of $\alpha$-$Pb_2P_2$ (left) and $\beta$-$Pb_2P_2$ (right), the solid blue line represents the results calculated by PBE, the red dot line represents the results calculated by PBE+SOC.

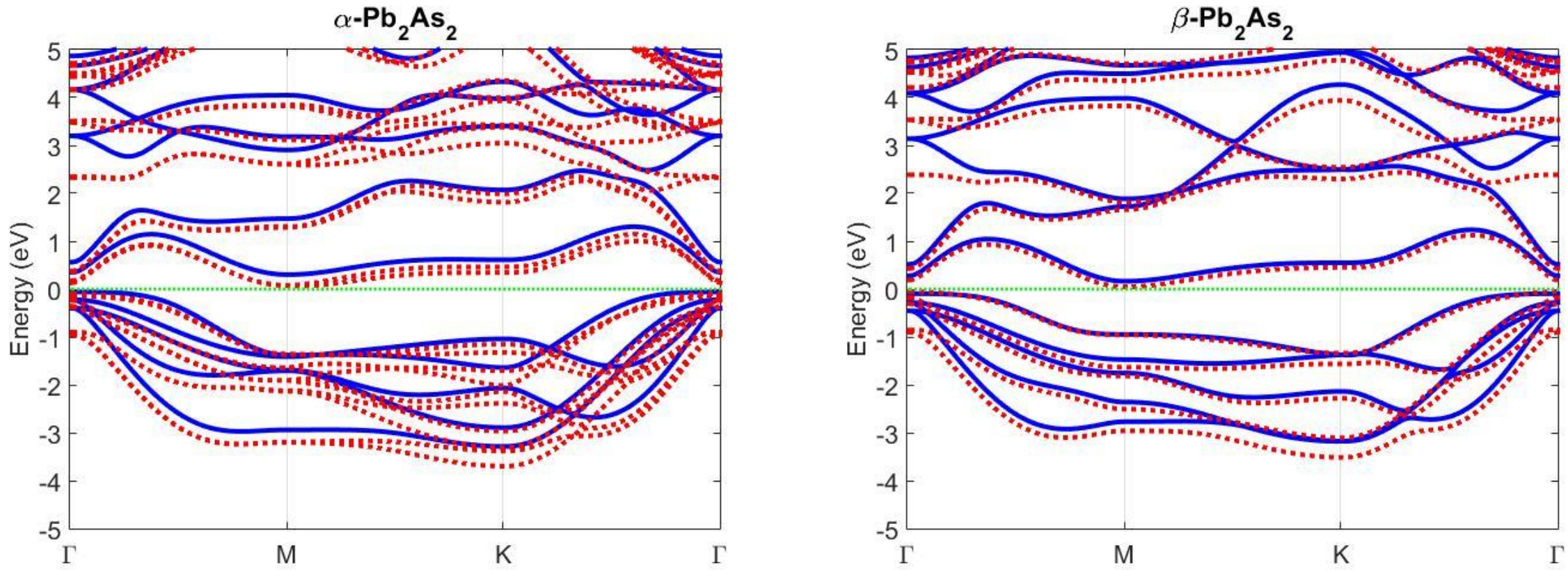

Figure S44. Electronic band of $\alpha$-$Pb_2As_2$ (left) and $\beta$-$Pb_2As_2$ (right), the solid blue line represents the results calculated by PBE, the red dot line represents the results calculated by PBE+SOC.

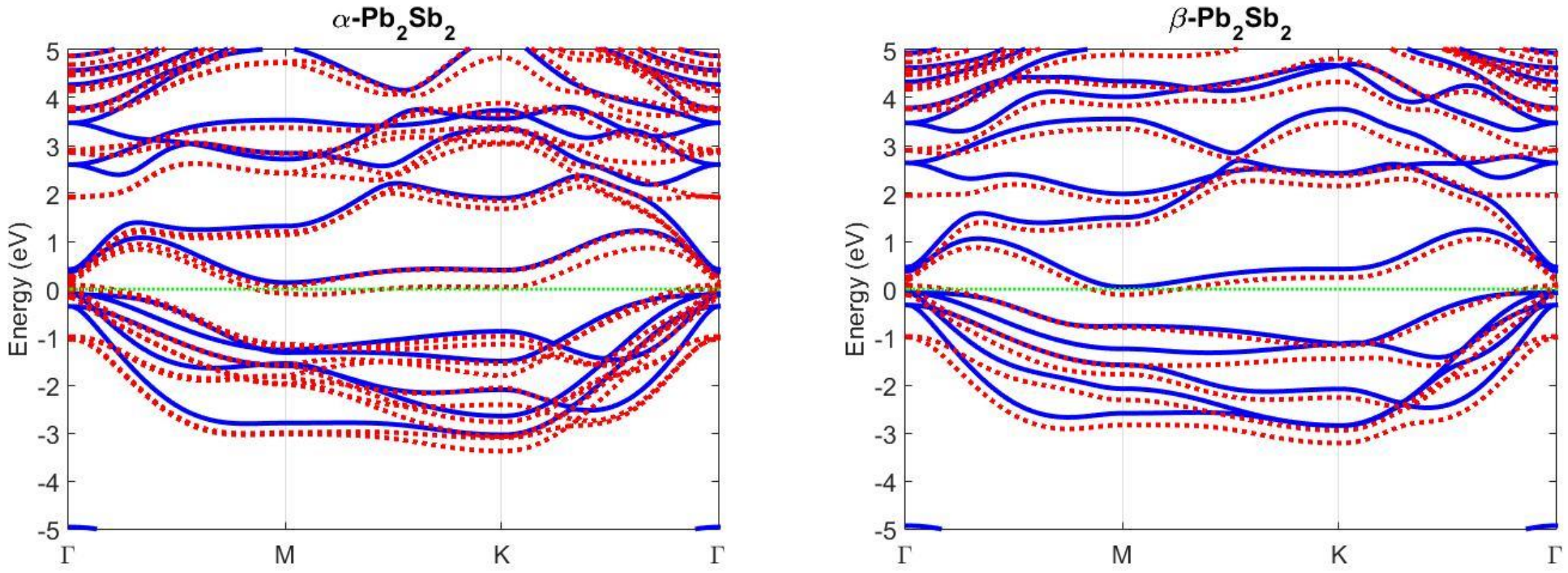

Figure S45. Electronic band of α-$Pb_2Sb_2$ (left) and β-$Pb_2Sb_2$ (right), the solid blue line represents the results calculated by PBE, the red dot line represents the results calculated by PBE+SOC.

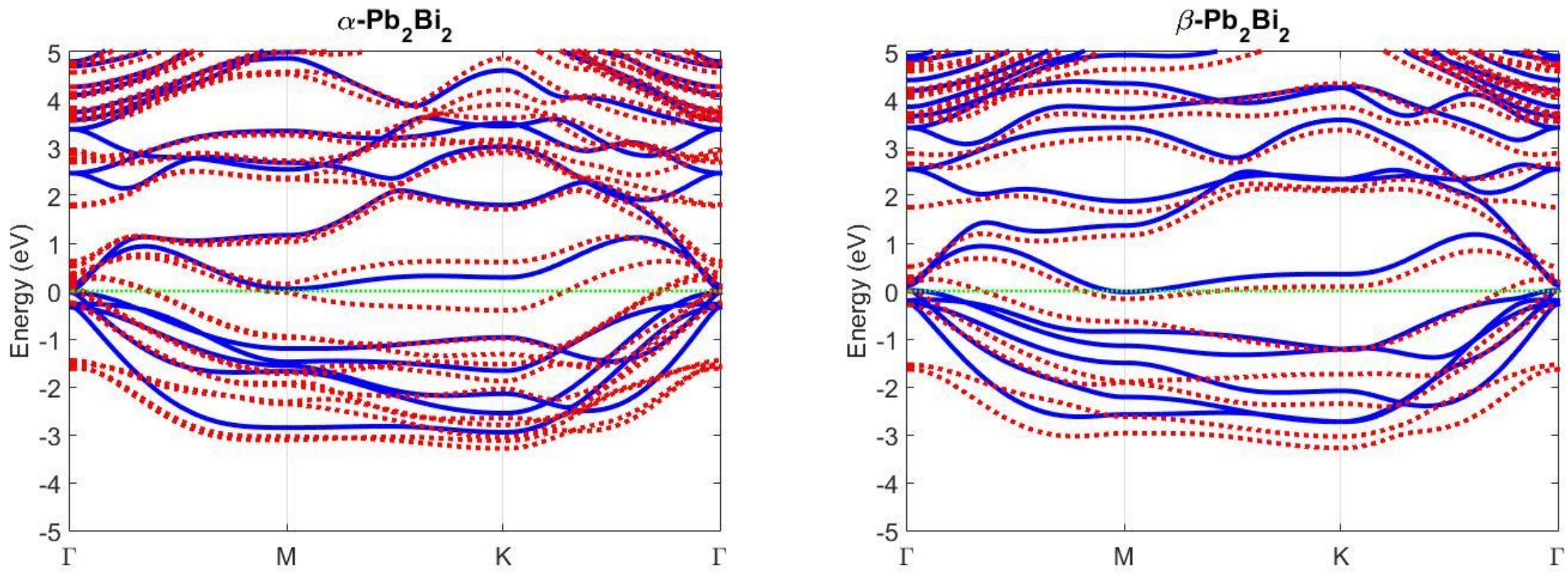

Figure S46. Electronic band of α-$Pb_2Bi_2$ (left) and β-$Pb_2Bi_2$ (right), the solid blue line represents the results calculated by PBE, the red dot line represents the results calculated by PBE+SOC.

## V. Dependence of energy gap on the strain

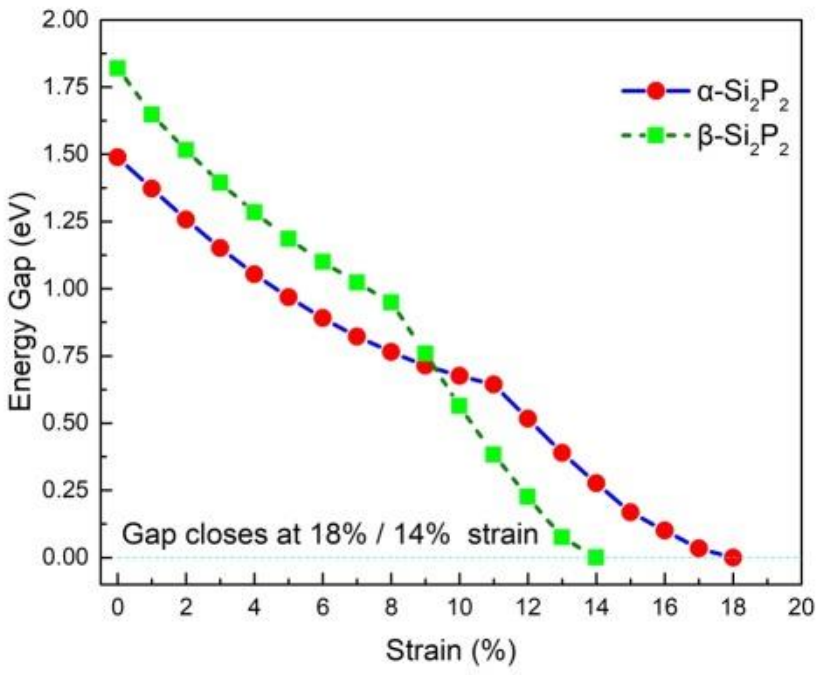

Figure S47. Dependence of energy gap on the strain of *α*-$Si_2P_2$ and *β*-$Si_2P_2$.

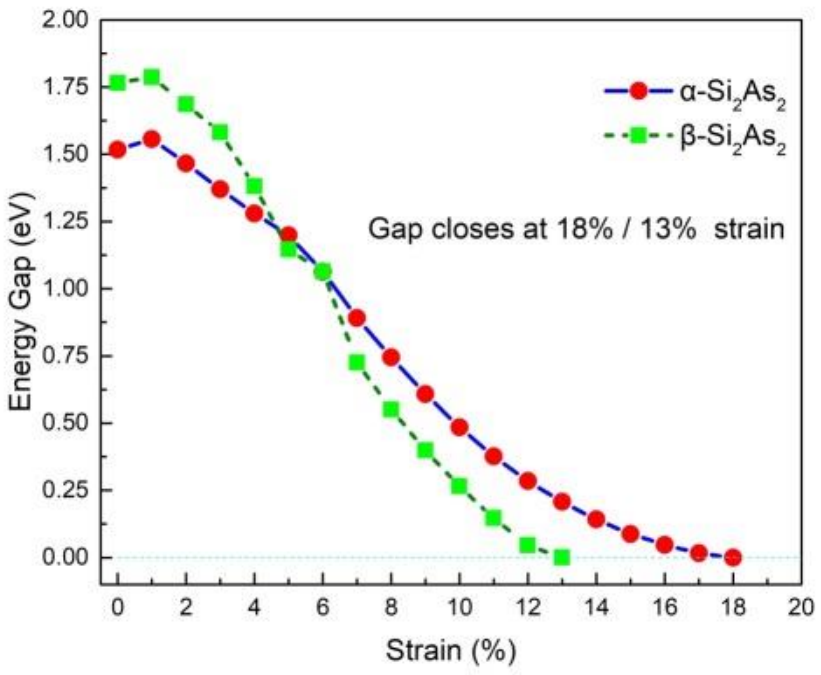

Figure S48. Dependence of energy gap on the strain of *α*-$Si_2As_2$ and *β*-$Si_2As_2$.

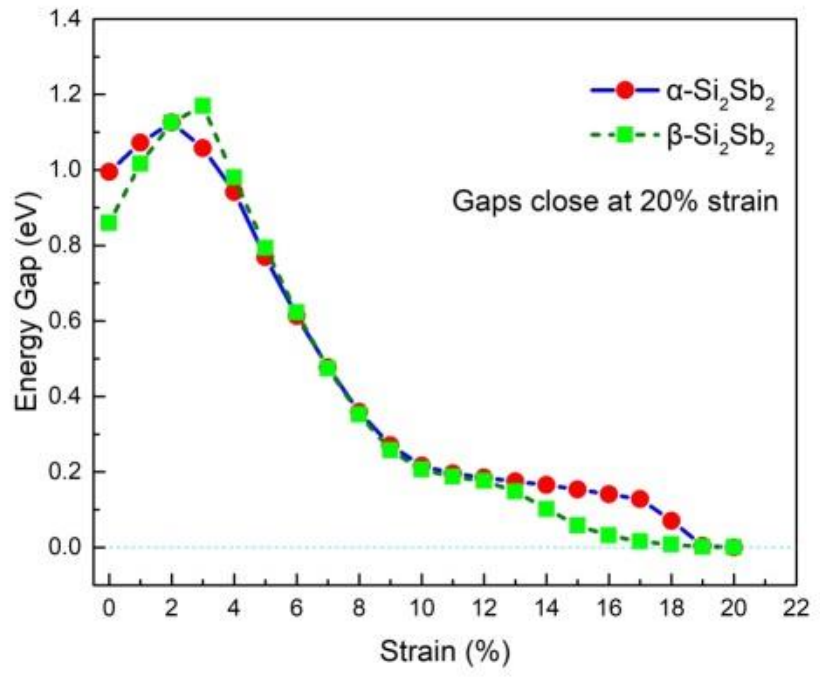

Figure S49. Dependence of energy gap on the strain of *α*-$Si_2Sb_2$ and *β*-$Si_2Sb_2$.

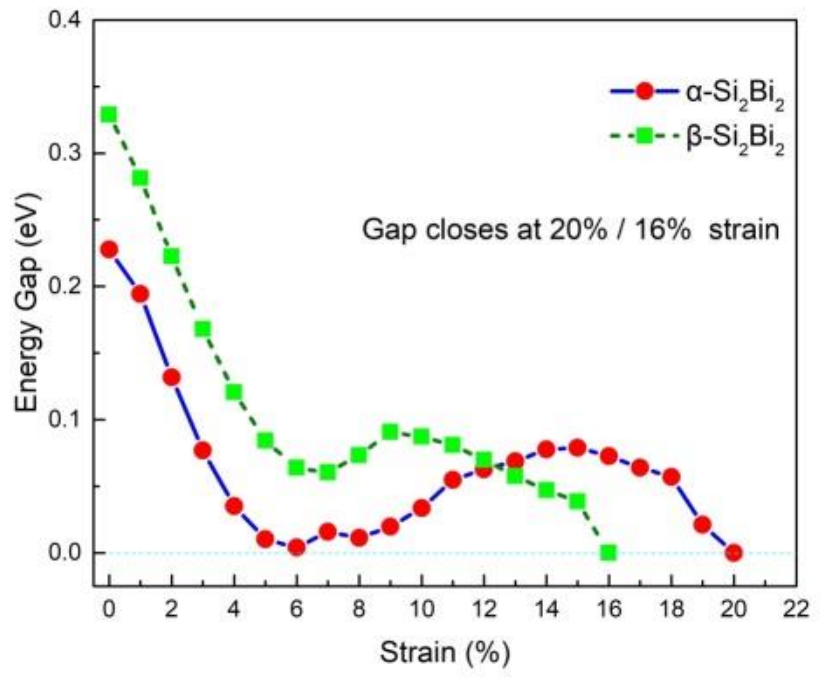

Figure S50. Dependence of energy gap on the strain of *α*-$Si_2Bi_2$ and *β*-$Si_2Bi_2$.

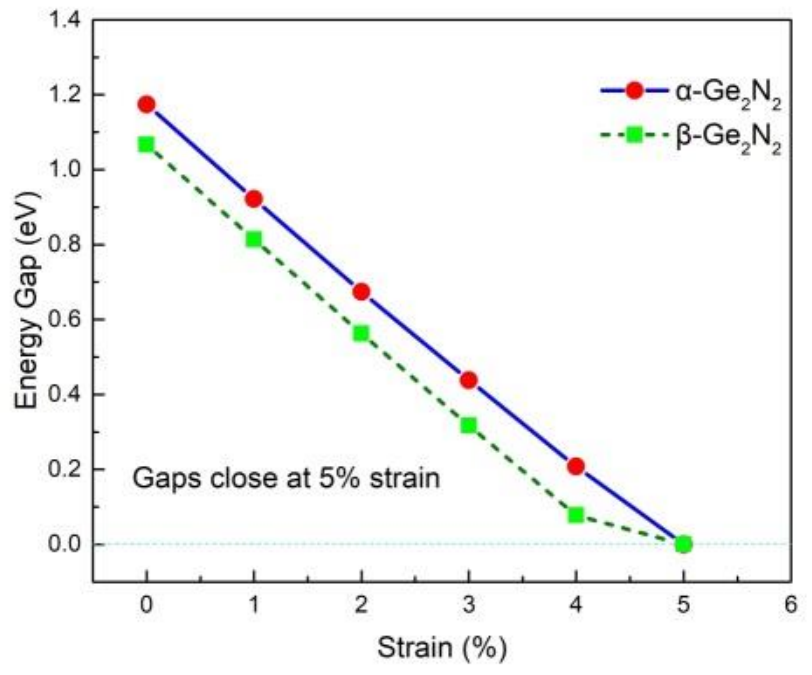

Figure S51. Dependence of energy gap on the strain of *α*-$Ge_2N_2$ and *β*-$Ge_2N$.

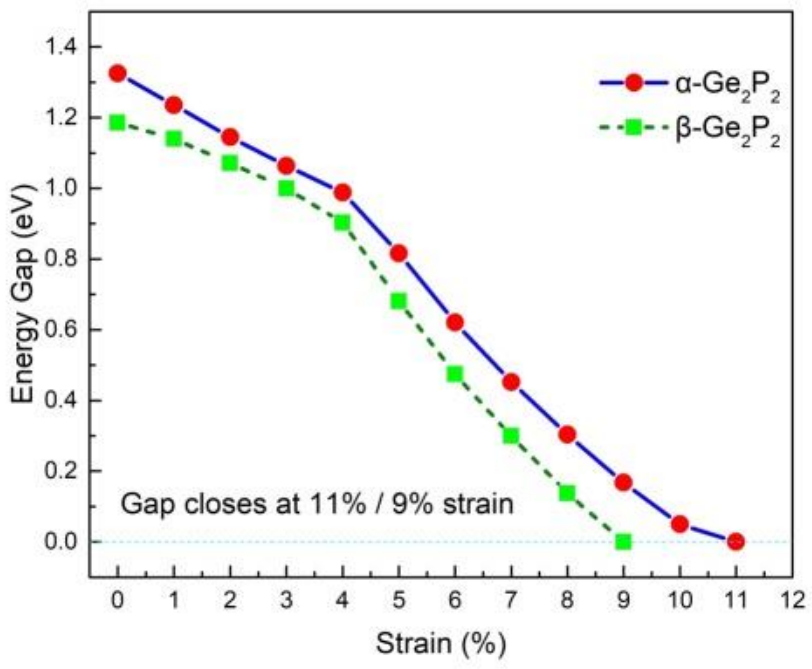

Figure S52. Dependence of energy gap on the strain of *α*-$Ge_2P_2$ and *β*-$Ge_2P_2$.

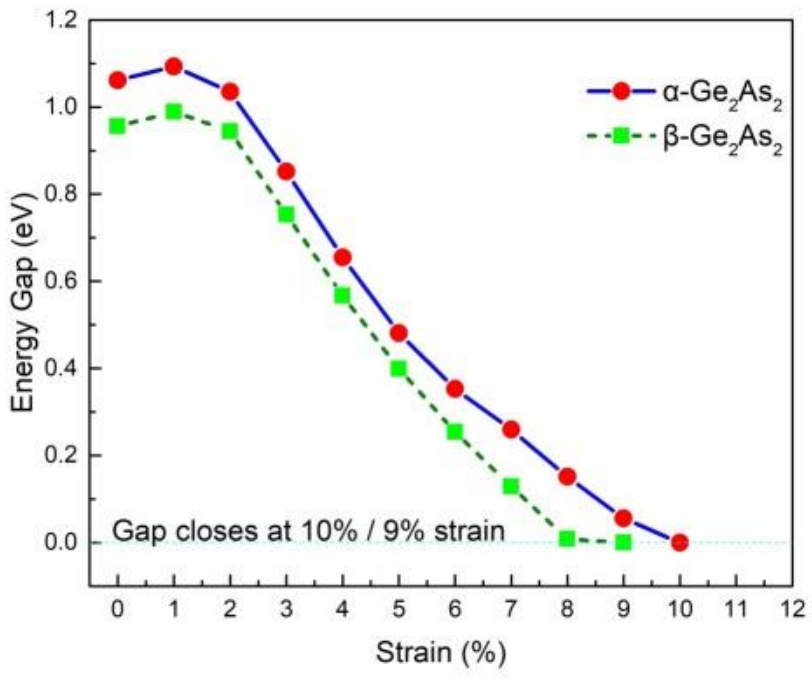

Figure S53. Dependence of energy gap on the strain of *α*-$Ge_2As_2$ and *β*-$Ge_2As_2$.

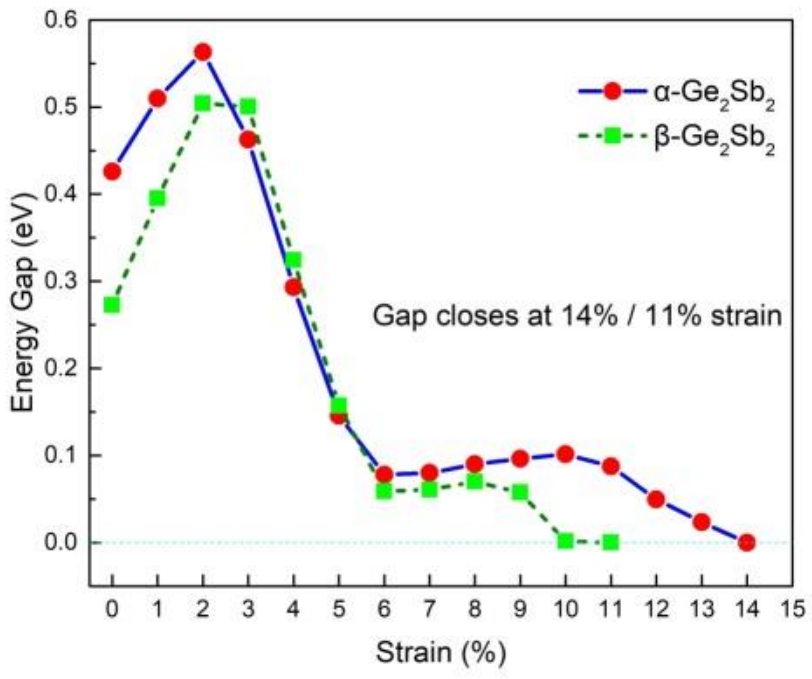

Figure S54. Dependence of energy gap on the strain of *α*-$Ge_2Sb_2$ and *β*-$Ge_2Sb$.

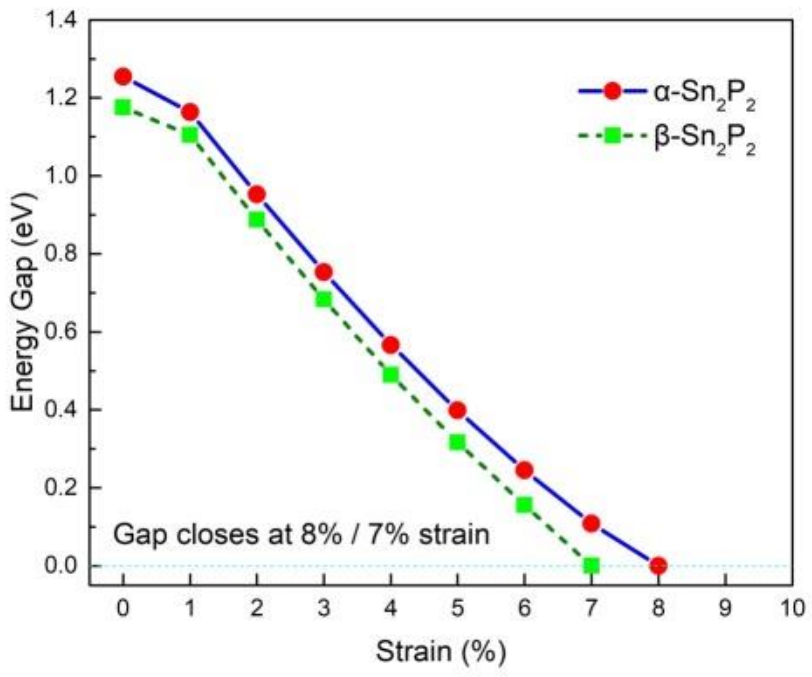

Figure S55. Dependence of energy gap on the strain of $\alpha$-$Sn_2P_2$ and $\beta$-$Sn_2P_2$.

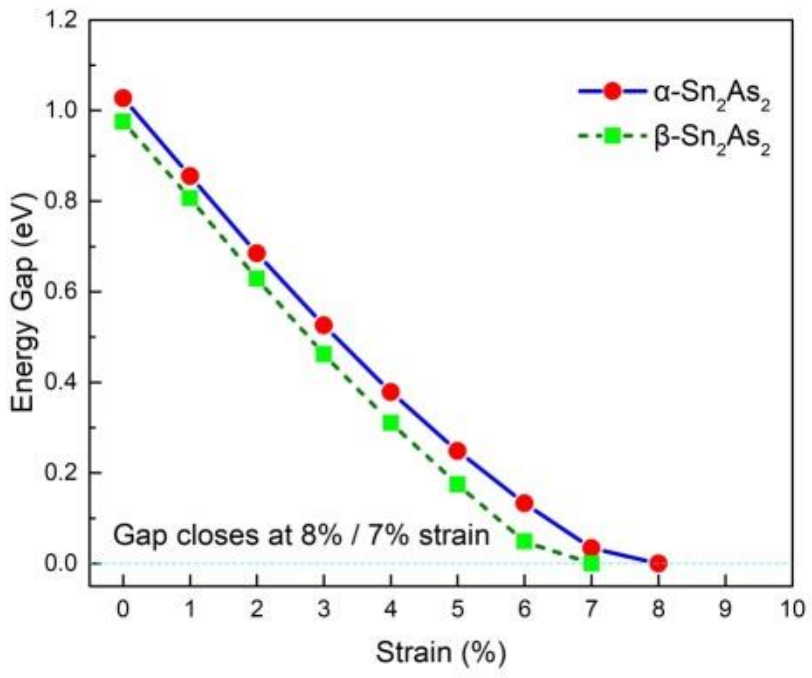

Figure S56. Dependence of energy gap on the strain of α-$Sn_2As_2$ and β-$Sn_2As_2$.

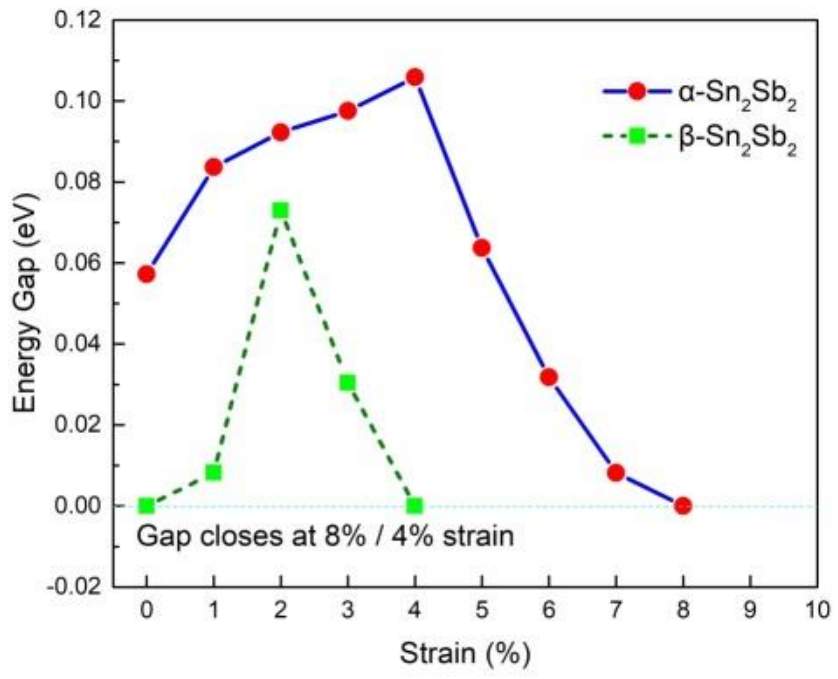

Figure S57. Dependence of energy gap on the strain of $\alpha$-$Sn_2Sb_2$ and $\beta$-$Sn_2Sb_2$.

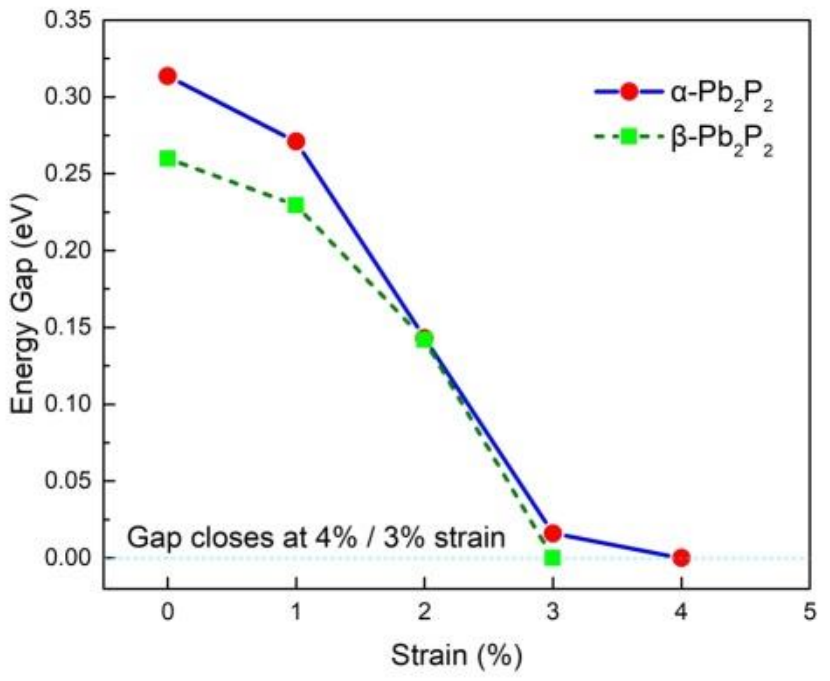

Figure S58. Dependence of energy gap on the strain of $\alpha$-$Pb_2P_2$ and $\beta$-$Pb_2P_2$.

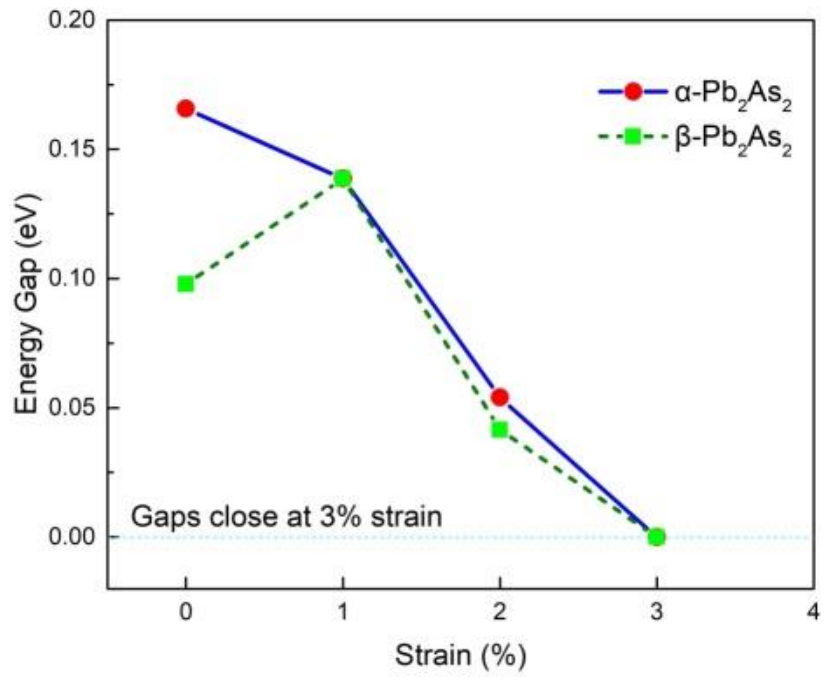

Figure S59. Dependence of energy gap on the strain of $\alpha$-$Pb_2As_2$ and $\beta$-$Pb_2As_2$.

## VI. Electronic band parameters

Table S5. Electronic band parameters of $Si_2Y_2$(Y=N, P, As, Sb, and Bi) in the PBE background.

| Phase | PBE | | | PBE+SOC | | | △ | $\varepsilon_C$(%) |
|---|---|---|---|---|---|---|---|---|
| | Gap1 (eV) | CBM position | VBM position | Gap2 (eV) | CBM position | VBM position | | |
| $\alpha$-$Si_2N_2$ [1] | $1.73^i$ | M | K-Γ | / | / | / | / | $17^a$ |
| $\beta$-$Si_2N_2$ [1] | $1.89^i$ | M | K-Γ | / | / | / | / | $16^a$ |
| $\alpha$-$Si_2P_2$ | $1.489^i$ | M | Γ-K | $1.486^i$ | M | Γ-K | -0.003 | $18^b$ |
| $\beta$-$Si_2P_2$ | $1.820^i$ | M | Γ-K | $1.822^i$ | M | Γ-K | 0.002 | $14^b$ |
| $\alpha$-$Si_2As_2$ | $1.60^i$ | M | Γ | $1.52^i$ | M | Γ | -0.08 | $18^b$ |
| $\beta$-$Si_2As_2$ | $1.86^i$ | M | Γ | $1.77^i$ | M | Γ | -0.09 | $13^b$ |
| $\alpha$-$Si_2Sb_2$ | $1.19^i$ | M | Γ | $0.99^i$ | M | Γ | -0.2 | $20^b$ |
| $\beta$-$Si_2Sb_2$ | $1.04^i$ | M | Γ | $0.86^i$ | M | Γ | -0.18 | $20^b$ |
| $\alpha$-$Si_2Bi_2$ | $0.61^d$ | Γ | Γ | $0.23^d$ | Γ | Γ | -0.38 | $20^b$ |
| $\beta$-$Si_2Bi_2$ | $0.73^d$ | Γ | Γ | $0.33^i$ | M | Γ | 0.4 | $16^b$ |

The superscript *i* represents the direct gap, and *d* represents the indirect gap. Gap1 and Gap2 represent the band gaps calculated by PBE and PBE+SOC, respectively, △=Gap2-Gap1. $\varepsilon_C$ denotes the critical strain from insulativity to metallicity.

[a] Represents the critical stain in the PBE background.

[b] Represents the critical stain in the PBE+SOC background.

Table S6. Electronic band parameters of $Ge_2Y_2$(Y=N, P, As, Sb, and Bi) in the PBE background.

| Phase | PBE | | | PBE+SOC | | | △ | $\varepsilon_C$ (%) |
|---|---|---|---|---|---|---|---|---|
| | Gap1 (eV) | CBM position | VBM position | Gap2 (eV) | CBM position | VBM position | | |
| $\alpha$-$Ge_2N_2$ | $1.172^i$ | Γ | Γ-K | $1.174^i$ | Γ | Γ-K | 0.002 | $5^b$ |
| $\beta$-$Ge_2N_2$ | $1.062^i$ | Γ | Γ-K | $1.067^i$ | Γ | Γ-K | 0.005 | $5^b$ |
| $\alpha$-$Ge_2P_2$ | $1.34^i$ | M | Γ-K | $1.32^i$ | M | Γ-K | -0.02 | $11^b$ |
| $\beta$-$Ge_2P_2$ | $1.201^i$ | M | Γ-K | $1.185^i$ | M | Γ-K | -0.016 | $9^b$ |
| $\alpha$-$Ge_2As_2$ | $1.195^i$ | M | Γ | $1.062^i$ | M | Γ | -0.133 | $10^b$ |
| $\beta$-$Ge_2As_2$ | $1.091^i$ | M | Γ | $0.956^i$ | M | Γ | -0.135 | $9^b$ |
| $\alpha$-$Ge_2Sb_2$ | $0.656^i$ | M | Γ | $0.426^i$ | M | Γ | -0.23 | $14^b$ |
| $\beta$-$Ge_2Sb_2$ | $0.495^i$ | M | Γ | $0.273^i$ | M | Γ | -0.222 | $11^b$ |
| $\alpha$-$Ge_2Bi_2$ | $0.224^d$ | Γ | Γ | 0 | / | / | -0.224 | / |
| $\beta$-$Ge_2Bi_2$ | $0.284^i$ | M | Γ | 0 | / | / | -0.284 | / |

The superscript *i* represents the direct gap, and *d* represents the indirect gap. Gap1 and Gap2 represent the band gaps calculated by PBE and PBE+SOC, respectively, △=Gap2-Gap1. $\varepsilon_C$ denotes the critical strain from insulativity to metallicity.

[b] Represents the critical stain in the PBE+SOC background.

Table S7. Electronic band parameters of $Sn_2Y_2$(Y=N, P, As, Sb, and Bi) in the PBE background.

| Phase | PBE | | | PBE+SOC | | | △ | $\varepsilon_C$ (%) |
|---|---|---|---|---|---|---|---|---|
| | Gap1 (eV) | CBM position | VBM position | Gap2 (eV) | CBM position | VBM position | | |
| $\alpha$-$Sn_2N_2$ | $0.125^i$ | Γ | Γ-K | $0.123^i$ | Γ | Γ-K | -0.002 | $1^b$ |
| $\beta$-$Sn_2N_2$ | $0.050^i$ | Γ | Γ-K | $0.058^i$ | Γ | Γ-K | 0.008 | $1^b$ |
| $\alpha$-$Sn_2P_2$ | $1.280^i$ | M | Γ-K | $1.255^i$ | M | Γ-K | -0.025 | $8^b$ |
| $\beta$-$Sn_2P_2$ | $1.196^i$ | M | Γ-K | $1.175^i$ | M | Γ-K | -0.021 | $7^b$ |
| $\alpha$-$Sn_2As_2$ | $1.128^d$ | Γ | Γ | $1.027^d$ | Γ | Γ | -0.101 | $8^b$ |
| $\beta$-$Sn_2As_2$ | $1.075^d$ | Γ | Γ | $0.975^i$ | Γ | Γ | -0.1 | $7^b$ |
| $\alpha$-$Sn_2Sb_2$ | $0.289^i$ | M | Γ | $0.057^i$ | M-K | Γ | -0.232 | $8^b$ |
| $\beta$-$Sn_2Sb_2$ | $0.151^i$ | M | Γ | 0 | / | / | -0.151 | $4^b$ |
| $\alpha$-$Sn_2Bi_2$ | $0.475^d$ | Γ | Γ | 0 | / | / | -0.475 | / |
| $\beta$-$Sn_2Bi_2$ | $0.493^d$ | Γ | Γ | 0 | / | / | -0.493 | / |

The superscript *i* represents the direct gap, and *d* represents the indirect gap. Gap1 and Gap2 represent the band gaps calculated by PBE and PBE+SOC, respectively, △=Gap2-Gap1. $\varepsilon_C$ denotes the critical strain from insulativity to metallicity.

$^b$ Represents the critical stain in the PBE+SOC background.

Table S8. Electronic band parameters of $Pb_2Y_2$(Y=N, P, As, Sb, and Bi) in the PBE background.

| Phase | PBE | | | PBE+SOC | | | △ | $\varepsilon_C$（%） |
|---|---|---|---|---|---|---|---|---|
| | Gap1 (eV) | CBM position | VBM position | Gap2 (eV) | CBM position | VBM position | | |
| $\alpha$-$Pb_2N_2$ | 0 | / | / | 0 | / | / | 0 | / |
| $\beta$-$Pb_2N_2$ | 0 | / | / | 0 | / | / | 0 | / |
| $\alpha$-$Pb_2P_2$ | $0.400^i$ | M | Γ-M | $0.313^i$ | M | Γ | -0.087 | $4^b$ |
| $\beta$-$Pb_2P_2$ | $0.315^i$ | M | Γ-M | $0.260^i$ | M | Γ | -0.055 | $3^b$ |
| $\alpha$-$Pb_2As_2$ | $0.363^i$ | M | Γ-M | $0.166^i$ | M | Γ | -0.197 | $3^b$ |
| $\beta$-$Pb_2As_2$ | $0.261^i$ | M | Γ | $0.098^i$ | M | Γ | -0.163 | $3^b$ |
| $\alpha$-$Pb_2Sb_2$ | $0.23^i$ | M | Γ | 0 | / | / | -0.23 | / |
| $\beta$-$Pb_2Sb_2$ | $0.120^i$ | M | Γ | 0 | / | / | -0.12 | / |
| $\alpha$-$Pb_2Bi_2$ | $0.064^d$ | Γ | Γ | 0 | / | / | -0.064 | / |
| $\beta$-$Pb_2Bi_2$ | 0 | / | / | 0 | / | / | 0 | / |

The superscript *i* represents the direct gap, and *d* represents the indirect gap. Gap1 and Gap2 represent the band gaps calculated by PBE and PBE+SOC, respectively, △=Gap2-Gap1. $\varepsilon_C$ denotes the critical strain from insulativity to metallicity.

$^b$ Represents the critical stain in the PBE+SOC background.

## VII. Computational details

In the calculation of the force constants for phonon dispersions, the cut-off energy and the size of supercells are shown in Table S9.

Table S9. The cut-off energy and the size of the supercells

| phase | cut-off energy (eV) | size of the supercell | phase | cut-off energy (eV) | size of the supercell | phase | cut-off energy (eV) | size of the supercell |
|---|---|---|---|---|---|---|---|---|
| $\alpha$-$C_2P_2$ | 520 | 6×6×1 | $\alpha$-$Si_2Bi_2$ | 350 | 4×4×1 | $\alpha$-$Sn_2As_2$ | 280 | 4×4×1 |
| $\beta$-$C_2P_2$ | 520 | 6×6×1 | $\beta$-$Si_2Bi_2$ | 350 | 4×4×1 | $\beta$-$Sn_2As_2$ | 280 | 5×5×1 |
| $\alpha$-$C_2As_2$ | 520 | 5×5×1 | $\alpha$-$Ge_2N_2$ | 520 | 4×4×1 | $\alpha$-$Sn_2Sb_2$ | 230 | 4×4×1 |
| $\beta$-$C_2As_2$ | 520 | 5×5×1 | $\beta$-$Ge_2N_2$ | 520 | 4×4×1 | $\beta$-$Sn_2Sb_2$ | 230 | 4×4×1 |
| $\alpha$-$C_2Sb_2$ | 520 | 6×6×1 | $\alpha$-$Ge_2P_2$ | 335 | 4×4×1 | $\alpha$-$Sn_2Bi_2$ | 600 | 4×4×1 |
| $\beta$-$C_2Sb_2$ | 520 | 6×6×1 | $\beta$-$Ge_2P_2$ | 335 | 5×5×1 | $\beta$-$Sn_2Bi_2$ | 600 | 5×5×1 |
| $\alpha$-$C_2Bi_2$ | 600 | 5×5×1 | $\alpha$-$Ge_2As_2$ | 280 | 4×4×1 | $\alpha$-$Pb_2N_2$ | 520 | 4×4×1 |
| $\beta$-$C_2Bi_2$ | 600 | 5×5×1 | $\beta$-$Ge_2As_2$ | 280 | 5×5×1 | $\beta$-$Pb_2N_2$ | 520 | 5×5×1 |
| $\alpha$-$Si_2N_2$ | 450 | 4×4×1 | $\alpha$-$Ge_2Sb_2$ | 230 | 5×5×1 | $\alpha$-$Pb_2P_2$ | 335 | 4×4×1 |
| $\beta$-$Si_2N_2$ | 450 | 4×4×1 | $\beta$-$Ge_2Sb_2$ | 230 | 5×5×1 | $\beta$-$Pb_2P_2$ | 335 | 5×5×1 |
| $\alpha$-$Si_2P_2$ | 335 | 5×5×1 | $\alpha$-$Ge_2Bi_2$ | 230 | 5×5×1 | $\alpha$-$Pb_2As_2$ | 280 | 4×4×1 |
| $\beta$-$Si_2P_2$ | 335 | 5×5×1 | $\beta$-$Ge_2Bi_2$ | 230 | 5×5×1 | $\beta$-$Pb_2As_2$ | 280 | 5×5×1 |
| $\alpha$-$Si_2As_2$ | 320 | 5×5×1 | $\alpha$-$Sn_2N_2$ | 600 | 4×4×1 | $\alpha$-$Pb_2Sb_2$ | 230 | 4×4×1 |
| $\beta$-$Si_2As_2$ | 320 | 5×5×1 | $\beta$-$Sn_2N_2$ | 600 | 4×4×1 | $\beta$-$Pb_2Sb_2$ | 230 | 5×5×1 |
| $\alpha$-$Si_2Sb_2$ | 320 | 5×5×1 | $\alpha$-$Sn_2P_2$ | 335 | 4×4×1 | $\alpha$-$Pb_2Bi_2$ | 138 | 4×4×1 |
| $\beta$-$Si_2Sb_2$ | 320 | 5×5×1 | $\beta$-$Sn_2P_2$ | 335 | 4×4×1 | $\beta$-$Pb_2Bi_2$ | 138 | 4×4×1 |

# VIII. POSCAR files of the structures

POSCAR-*α*-$C_2P_2$

```
     3.00000000000000
       0.8366183838863850     -0.4830218491457943       0.0000000000000000
       0.8366183838863850      0.4830218491457943       0.0000000000000000
       0.0000000000000000      0.0000000000000000      10.0033028495504883
     C      P
       2      2
Direct
   0.0000000000000000   0.0000000000000000   0.0258774575229661
   0.0000000000000000   0.0000000000000000   0.9741225424770340
   0.3333333333333357   0.3333333333333357   0.0549791897473604
   0.3333333333333357   0.3333333333333357   0.9450208102526397
```

POSCAR-*β*-$C_2P_2$

```
     3.00000000000000
       0.8366183838863850     -0.4830218491457943       0.0000000000000000
       0.8366183838863850      0.4830218491457943       0.0000000000000000
       0.0000000000000000      0.0000000000000000      10.0033028495504883
     C      P
       2      2
Direct
   0.0000000000000000   0.0000000000000000   0.0258774575229661
   0.0000000000000000   0.0000000000000000   0.9741225424770340
0.3333333333333357   0.3333333333333357   0.0549791897473604
 -0.3333333333333357   -0.3333333333333357   0.9450208102526397
```

POSCAR-*α*-$C_2As_2$

```
     1.00000000000000
      2.69278639882923355     -1.554680952233637775   0.0000000000000000
      2.69278639882890535      1.554680952234332775   0.0000000000000000
       0.0000000000000000      0.0000000000000000      25.6013886650011280
     C      As
      2      2
Direct
   0.0000000000000000   0.0000000000000000   0.0298020037990700
   0.0000000000000000   0.0000000000000000   0.9701979962009158
   0.3333333333333357   0.3333333333333357   0.0675231569654309
   0.3333333333333357   0.3333333333333357   0.9324768429345743
```

POSCAR-*β*-$C_2As_2$

```
     1.00000000000000
      2.69278639882923355     -1.554680952233637775   0.0000000000000000
      2.69278639882890535      1.554680952234332775   0.0000000000000000
       0.0000000000000000      0.0000000000000000      25.6013886650011280
     C      As
      2      2
Direct
```

0.0000000000000000 0.0000000000000000 0.0298020037990700
0.0000000000000000 0.0000000000000000 0.9701979962009158
0.3333333333333357 0.3333333333333357 0.0675231569654309
-0.3333333333333357 -0.3333333333333357 0.9324768429345743

POSCAR-$\alpha$-$C_2Sb_2$
3.31000000000000
0.8910807025050776 -0.5144656834609223 0.0000000000000000
0.8910807025049883 0.5144656834609004 0.0000000000000000
0.0000000000000000 0.0000000000000000 7.9895035198626543
C Sb
2 2
Direct
0.0000000000000000 0.0000000000000000 0.0288624062787197
0.0000000000000000 0.0000000000000000 0.9711375937212808
0.3333333333333357 0.3333333333333357 0.0700540507434779
0.3333333333333357 0.3333333333333357 0.9299459492565225

POSCAR-$\beta$-$C_2Sb_2$
3.31000000000000
0.8910807025050776 -0.5144656834609223 0.0000000000000000
0.8910807025049883 0.5144656834609004 0.0000000000000000
0.0000000000000000 0.0000000000000000 7.9895035198626543
C Sb
2 2
Direct
0.0000000000000000 0.0000000000000000 0.0288624062787197
0.0000000000000000 0.0000000000000000 0.9711375937212808
0.3333333333333357 0.3333333333333357 0.0700540507434779
-0.3333333333333357 -0.3333333333333357 0.9299459492565225

POSCAR-$\alpha$-$C_2Bi_2$
1.00000000000000
3.1204893349827456 -1.801615358361174475 0.0000000000000000
3.12048933498237745 1.801615358361069225 0.0000000000000000
0.0000000000000000 0.0000000000000000 25.6181855773495286
C Bi
2 2
Direct
0.0000000000000000 0.0000000000000000 0.0290691034198632
0.0000000000000000 0.0000000000000000 0.9709308965801154
0.3333333333333357 0.3333333333333357 0.0740567473954728
0.3333333333333357 0.3333333333333357 0.9259432526044563

POSCAR-$\beta$-$C_2Bi_2$
1.00000000000000
3.1204893349827456 -1.801615358361174475 0.0000000000000000
3.12048933498237745 1.801615358361069225 0.0000000000000000
0.0000000000000000 0.0000000000000000 25.6181855773495286

C Bi
2 2
Direct
0.0000000000000000 0.0000000000000000 0.0290691034198632
0.0000000000000000 0.0000000000000000 0.9709308965801154
0.3333333333333357 0.3333333333333357 0.0740567473954728
-0.3333333333333357 -0.3333333333333357 0.9259432526044563

POSCAR-$\alpha$-$Si_2P_2$
1.00000000000000
3.05639045190647885 -1.7646078501506146 0.0000000000000000
3.05639045190631765 1.764607850150573075 0.0000000000000000
0.0000000000000000 0.0000000000000000 27.1966286625978064
Si P
2 2
Direct
0.0000000000000000 0.0000000000000000 0.0436560124219749
0.0000000000000000 0.0000000000000000 0.9563439875779401
0.3333333333333357 0.3333333333333357 0.0810013235839673
0.3333333333333357 0.3333333333333357 0.9189986764160395

POSCAR-$\beta$-$Si_2P_2$
1.00000000000000
3.05639045190647885 -1.7646078501506146 0.0000000000000000
3.05639045190631765 1.764607850150573075 0.0000000000000000
0.0000000000000000 0.0000000000000000 27.1966286625978064
Si P
2 2
Direct
0.0000000000000000 0.0000000000000000 0.0436560124219749
0.0000000000000000 0.0000000000000000 0.9563439875779401
0.3333333333333357 0.3333333333333357 0.0810013235839673
-0.3333333333333357 -0.3333333333333357 0.9189986764160395

POSCAR-$\alpha$-$Si_2As_2$
1.00000000000000
3.1998494810974334 -1.847433959284871425 0.0000000000000000
3.199849481097536425 1.8474339592831952 0.0000000000000000
0.0000000000000000 0.0000000000000000 26.9957699420950554
Si As
2 2
Direct
0.0000000000000000 0.0000000000000000 0.0437716381548780
0.0000000000000000 0.0000000000000000 0.9562283618451219
0.3333333333333357 0.3333333333333357 0.0847602391142282
0.3333333333333357 0.3333333333333357 0.9152397608857719

POSCAR-$\beta$-$Si_2As_2$
1.00000000000000

```
     3.1998494810974334     -1.847433959284871425    0.0000000000000000
     3.199849481097536425    1.8474339592831952       0.0000000000000000
     0.0000000000000000      0.0000000000000000      26.9957699420950554
   Si    As
     2      2
Direct
  0.0000000000000000  0.0000000000000000  0.0437716381548780
  0.0000000000000000  0.0000000000000000  0.9562283618451219
  0.3333333333333357  0.3333333333333357  0.0847602391142282
-0.3333333333333357 -0.3333333333333357  0.9152397608857719
```

POSCAR-$\alpha$-$Si_2Sb_2$

```
   1.00000000000000
     3.477529132536688375   -2.007752380784829875   0.0000000000000000
     3.47752913253663775     2.007752380784761475   0.0000000000000000
    -0.0000000000000008      0.0000000000000000    26.8867964278114080
   Si    Sb
     2      2
Direct
  0.0000000000000000  0.0000000000000000  0.0438778404244207
  0.0000000000000000  0.0000000000000000  0.9561221595755864
  0.3333333333333357  0.3333333333333357  0.0895659053972646
  0.3333333333333357  0.3333333333333357  0.9104340946027426
```

POSCAR-$\beta$-$Si_2Sb_2$

```
   1.00000000000000
     3.477529132536688375   -2.007752380784829875   0.0000000000000000
     3.47752913253663775     2.007752380784761475   0.0000000000000000
    -0.0000000000000008      0.0000000000000000    26.8867964278114080
   Si    Sb
     2      2
Direct
  0.0000000000000000  0.0000000000000000  0.0438778404244207
  0.0000000000000000  0.0000000000000000  0.9561221595755864
  0.3333333333333357  0.3333333333333357  0.0895659053972646
-0.3333333333333357  -0.3333333333333357  0.9104340946027426
```

POSCAR-$\alpha$-$Si_2Bi_2$

```
   4.10000000000000
      0.8798900703605536     -0.5080047689799765     0.0000000000000000
      0.8798900703605494      0.5080047689799650     0.0000000000000000
     -0.0000000000000001      0.0000000000000000     7.0232546668773681
   Si    Bi
      2      2
Direct
  0.0000000000000000  0.0000000000000000  0.0408136768404271
  0.0000000000000000  0.0000000000000000  0.9591863231595775
  0.3333333333333357  0.3333333333333357  0.0853580790302809
  0.3333333333333357  0.3333333333333357  0.9146419209697303
```

POSCAR-$\beta$-$Si_2Bi_2$

```
     4.10000000000000
        0.8798900703605536    -0.5080047689799765     0.0000000000000000
        0.8798900703605494     0.5080047689799650     0.0000000000000000
       -0.0000000000000001     0.0000000000000000     7.0232546668773681
     Si    Bi
       2     2
Direct
   0.0000000000000000   0.0000000000000000   0.0408136768404271
   0.0000000000000000   0.0000000000000000   0.9591863231595775
   0.3333333333333357   0.3333333333333357   0.0853580790302809
-0.3333333333333357 -0.3333333333333357   0.9146419209697303
```

POSCAR-$\alpha$-$Ge_2N_2$

```
     1.00000000000000
       2.6858814015117245      -1.550694350172467475   0.0000000000000000
       2.685881401511730275     1.5506943501723387     0.0000000000000014
       0.0000000000000036       0.0000000000000018    28.7977799986154288
     Ge    N
      2    2
Direct
   0.0000000000000000   0.0000000000000000   0.0445360843349821
   0.0000000000000000   0.0000000000000000   0.9554639156650027
   0.3333333333333357   0.3333333333333357   0.0676387349246264
   0.3333333333333357   0.3333333333333357   0.9323612650754014
```

POSCAR-$\beta$-$Ge_2N_2$

```
     1.00000000000000
       2.6858814015117245      -1.550694350172467475   0.0000000000000000
       2.685881401511730275     1.5506943501723387     0.0000000000000014
       0.0000000000000036       0.0000000000000018    28.7977799986154288
     Ge    N
      2    2
Direct
   0.0000000000000000   0.0000000000000000   0.0445360843349821
   0.0000000000000000   0.0000000000000000   0.9554639156650027
   0.3333333333333357   0.3333333333333357   0.0676387349246264
  -0.3333333333333357  -0.3333333333333357   0.9323612650754014
```

POSCAR-$\alpha$-$Ge_2P_2$

```
     1.00000000000000
       3.17226503600405695   -1.83150807247685755     0.0000000000000000
       3.172265036003811375   1.83150807247716285     0.0000000000000000
       0.0000000000000000     0.0000000000000000     27.0144259031135050
     Ge    P
     2     2
Direct
   0.0000000000000000   0.0000000000000000   0.0463754999822277
```

```
   0.0000000000000000   0.0000000000000000   0.9536245000177649
   0.3333333333333357   0.3333333333333357   0.0860913358014670
   0.3333333333333357   0.3333333333333357   0.9139086641985259
```

POSCAR-$\beta$-$Ge_2P_2$

```
     1.00000000000000
       3.17226503600405695    -1.83150807247685755       0.0000000000000000
       3.172265036003811375    1.83150807247716285       0.0000000000000000
       0.0000000000000000      0.0000000000000000       27.0144259031135050
     Ge     P
     2      2
Direct
   0.0000000000000000   0.0000000000000000   0.0463754999822277
   0.0000000000000000   0.0000000000000000   0.9536245000177649
   0.3333333333333357   0.3333333333333357   0.0860913358014670
  -0.3333333333333357  -0.3333333333333357   0.9139086641985259
```

POSCAR-$\alpha$-$Ge_2As_2$

```
     1.00000000000000
       3.308544491737140725    -1.91018905292791885      0.0000000000000000
       3.3085444917368041       1.91018905292809205      0.0000000000000000
        0.0000000000000000      0.0000000000000000      24.8348026513106568
     Ge     As
      2     2
Direct
   0.0000000000000000   0.0000000000000000   0.0503191959638081
   0.0000000000000000   0.0000000000000000   0.9496808040361779
   0.3333333333333357   0.3333333333333357   0.0966425407832867
   0.3333333333333357   0.3333333333333357   0.9033574592167131
```

POSCAR-$\beta$-$Ge_2As_2$

```
     1.00000000000000
       3.308544491737140725    -1.91018905292791885      0.0000000000000000
       3.3085444917368041       1.91018905292809205      0.0000000000000000
        0.0000000000000000      0.0000000000000000      24.8348026513106568
     Ge     As
      2     2
Direct
   0.0000000000000000   0.0000000000000000   0.0503191959638081
   0.0000000000000000   0.0000000000000000   0.9496808040361779
   0.3333333333333357   0.3333333333333357   0.0966425407832867
-0.3333333333333357  -0.3333333333333357   0.9033574592167131
```

POSCAR-$\alpha$-$Ge_2Sb_2$

```
     1.00000000000000
       3.571305899704054975    -2.061894427885734525    -0.0000000000000028
       3.57130589870062875      2.0618944265292214       0.0000000000000042
        0.0000000000000000      0.0000000000000010      27.9919743476202534
     Ge     Sb
```

2 2
Direct
0.0000000000000000 0.0000000000000000 0.0446703061143147
0.0000000000000000 0.0000000000000000 0.9554069217168412
0.3333333333333357 0.3333333333333357 0.0895737796231723
0.3333333333333357 0.3333333333333357 0.9105033723662791

POSCAR-$\beta$-$Ge_2Sb_2$
1.00000000000000
3.571305899704054975 -2.061894427885734525 -0.0000000000000028
3.57130589870062875 2.0618944265292214 0.0000000000000042
0.0000000000000000 0.0000000000000010 27.9919743476202534
Ge Sb
2 2
Direct
0.0000000000000000 0.0000000000000000 0.0446703061143147
0.0000000000000000 0.0000000000000000 0.9554069217168412
0.3333333333333357 0.3333333333333357 0.0895737796231723
-0.3333333333333357 -0.3333333333333357 0.9105033723662791

POSCAR-$\alpha$-$Ge_2Bi_2$
1.00000000000000
3.691898116261825175 -2.131518371244824025 0.0000000000000000
3.691898116261731025 2.131518371244842675 0.0000000000000000
0.0000000000000000 0.0000000000000000 27.2970182528660921
Ge Bi
2 2
Direct
0.0000000000000000 0.0000000000000000 0.0456521176766244
0.0000000000000000 0.0000000000000000 0.9543478823233678
0.3333333333333357 0.3333333333333357 0.0933182337000747
0.3333333333333357 0.3333333333333357 0.9066817662999325

POSCAR-$\beta$-$Ge_2Bi_2$
1.00000000000000
3.691898116261825175 -2.131518371244824025 0.0000000000000000
3.691898116261731025 2.131518371244842675 0.0000000000000000
0.0000000000000000 0.0000000000000000 27.2970182528660921
Ge Bi
2 2
Direct
0.0000000000000000 0.0000000000000000 0.0456521176766244
0.0000000000000000 0.0000000000000000 0.9543478823233678
0.3333333333333357 0.3333333333333357 0.0933182337000747
-0.3333333333333357 -0.3333333333333357 0.9066817662999325

POSCAR-$\alpha$-$Sn_2N_2$
1.00000000000000
2.961570375750981025 -1.709863453663774025 0.0000000000000000

2.961570375750992125 1.709863453663735375 0.0000000000000000
0.0000000000000000 0.0000000000000000 25.4964114750307900
Sn N
2 2
Direct
0.0000000000000000 0.0000000000000000 0.0583626132257341
0.0000000000000000 0.0000000000000000 0.9416373867743063
0.3333333333333357 0.3333333333333357 0.0870608867702451
0.3333333333333357 0.3333333333333357 0.9129391132297298

POSCAR-$\beta$-$Sn_2N_2$
1.00000000000000
2.961570375750981025 -1.709863453663774025 0.0000000000000000
2.961570375750992125 1.709863453663735375 0.0000000000000000
0.0000000000000000 0.0000000000000000 25.4964114750307900
Sn N
2 2
Direct
0.0000000000000000 0.0000000000000000 0.0583626132257341
0.0000000000000000 0.0000000000000000 0.9416373867743063
0.3333333333333357 0.3333333333333357 0.0870608867702451
-0.3333333333333357 -0.3333333333333357 0.9129391132297298

POSCAR-$\alpha$-$Sn_2P_2$
1.00000000000000
3.4219651314743511 -1.9756724897968343 0.0000000000000000
3.4219651314683488 1.9756724897896818 0.0000000000000000
0.0000000000000000 0.0000000000000000 27.9061349396144713
Sn P
2 2
Direct
0.0000000000000000 0.0000000000000000 0.0517690944281229
0.0000000000000000 0.0000000000000000 0.9482309055718420
0.3333333333333357 0.3333333333333357 0.0935128711615948
0.3333333333333357 0.3333333333333357 0.9064871288381781

POSCAR-$\beta$-$Sn_2P_2$
1.00000000000000
3.4219651314743511 -1.9756724897968343 0.0000000000000000
3.4219651314683488 1.9756724897896818 0.0000000000000000
0.0000000000000000 0.0000000000000000 27.9061349396144713
Sn P
2 2
Direct
0.0000000000000000 0.0000000000000000 0.0517690944281229
0.0000000000000000 0.0000000000000000 0.9482309055718420
0.3333333333333357 0.3333333333333357 0.0935128711615948
-0.3333333333333357 -0.3333333333333357 0.9064871288381781

POSCAR-$\alpha$-$Sn_2As_2$

```
     1.00000000000000
       3.543520771516774025   -2.045852671314388975   0.0000000000000000
       3.543520771516788225    2.0458526713143712     0.0000000000000000
       0.0000000000000000      0.0000000000000000    29.3994156300147687
     Sn    As
      2     2
Direct
   0.0000000000000000   0.0000000000000000   0.0489564718281513
   0.0000000000000000   0.0000000000000000   0.9510435281718402
   0.3333333333333357   0.3333333333333357   0.0912109207509854
   0.3333333333333357   0.3333333333333357   0.9087890792490299
```

POSCAR-$\beta$-$Sn_2As_2$

```
     1.00000000000000
       3.543520771516774025   -2.045852671314388975   0.0000000000000000
       3.543520771516788225    2.0458526713143712     0.0000000000000000
       0.0000000000000000      0.0000000000000000    29.3994156300147687
     Sn    As
      2     2
Direct
   0.0000000000000000   0.0000000000000000   0.0489564718281513
   0.0000000000000000   0.0000000000000000   0.9510435281718402
   0.3333333333333357   0.3333333333333357   0.0912109207509854
-0.3333333333333357   -0.3333333333333357   0.9087890792490299
```

POSCAR-$\alpha$-$Sn_2Sb_2$

```
     1.00000000000000
       3.7949779723496051     -2.1910315539079348     0.0000000000000000
       3.7949779723491468      2.1910315539066585     0.0000000000000000
       0.0000000000000002      0.0000000000000000    28.9659127945051722
     Sn    Sb
      2     2
Direct
   0.0000000000000000   0.0000000000000000   0.0507544752408575
   0.0000000000000000   0.0000000000000000   0.9517129312883560
   0.3333333333333357   0.3333333333333357   0.0976120300719327
   0.3333333333333357   0.3333333333333357   0.9048553864335136
```

POSCAR-$\beta$-$Sn_2Sb_2$

```
     1.00000000000000
       3.7949779723496051     -2.1910315539079348     0.0000000000000000
       3.7949779723491468      2.1910315539066585     0.0000000000000000
       0.0000000000000002      0.0000000000000000    28.9659127945051722
     Sn    Sb
      2     2
Direct
   0.0000000000000000   0.0000000000000000   0.0507544752408575
   0.0000000000000000   0.0000000000000000   0.9517129312883560
```

0.3333333333333357 0.3333333333333357 0.0976120300719327
-0.3333333333333357 -0.3333333333333357 0.9048553864335136

POSCAR-$\alpha$-$Sn_2Bi_2$
1.00000000000000
3.90439686116838035 -2.25420457883116355 0.0000000000000000
3.9043968611661386 2.254204578826708 0.0000000000000000
0.0000000000000003 0.0000000000000000 27.3651496665047489
Sn Bi
2 2
Direct
0.0000000000000000 0.0000000000000000 0.0534734730418469
0.0000000000000000 0.0000000000000000 0.9489939334873496
0.3333333333333357 0.3333333333333357 0.1046120631251983
0.3333333333333357 0.3333333333333357 0.8978553533802553

POSCAR-$\beta$-$Sn_2Bi_2$
1.00000000000000
3.90439686116838035 -2.25420457883116355 0.0000000000000000
3.9043968611661386 2.254204578826708 0.0000000000000000
0.0000000000000003 0.0000000000000000 27.3651496665047489
Sn Bi
2 2
Direct
0.0000000000000000 0.0000000000000000 0.0534734730418469
0.0000000000000000 0.0000000000000000 0.9489939334873496
0.3333333333333357 0.3333333333333357 0.1046120631251983
-0.3333333333333357 -0.3333333333333357 0.8978553533802553

POSCAR-$\alpha$-$Pb_2N_2$
1.00000000000000
3.152967822486948175 -1.820366821053765575 0.0000000000000000
3.152967822486964175 1.82036682105379265 0.0000000000000000
0.0000000000000000 0.0000000000000000 26.6253221809360312
Pb N
2 2
Direct
0.0000000000000000 0.0000000000000000 0.0596082803201259
0.0000000000000000 0.0000000000000000 0.9403917196798659
0.3333333333333357 0.3333333333333357 0.0892077311039359
0.3333333333333357 0.3333333333333357 0.9107922688960629

POSCAR-$\beta$-$Pb_2N_2$
1.00000000000000
3.152967822486948175 -1.820366821053765575 0.0000000000000000
3.152967822486964175 1.82036682105379265 0.0000000000000000
0.0000000000000000 0.0000000000000000 26.6253221809360312
Pb N
2 2

Direct
0.0000000000000000 0.0000000000000000 0.0596082803201259
0.0000000000000000 0.0000000000000000 0.9403917196798659
0.3333333333333357 0.3333333333333357 0.0892077311039359
-0.3333333333333357 -0.3333333333333357 0.9107922688960629

POSCAR-$\alpha$-$Pb_2P_2$
1.00000000000000
3.5664225783782908 -2.05907503567105765 0.0000000000000000
3.566422578378206425 2.059075035670562925 0.0000000000000000
0.0000000000000000 0.0000000000000000 29.0505717665422445
Pb P
2 2
Direct
0.0000000000000000 0.0000000000000000 0.0526436393207049
0.0000000000000000 0.0000000000000000 0.9473563606793014
0.3333333333333357 0.3333333333333357 0.0944879345040668
0.3333333333333357 0.3333333333333357 0.9055120654959324

POSCAR-$\beta$-$Pb_2P_2$
1.00000000000000
3.5664225783782908 -2.05907503567105765 0.0000000000000000
3.566422578378206425 2.059075035670562925 0.0000000000000000
0.0000000000000000 0.0000000000000000 29.0505717665422445
Pb P
2 2
Direct
0.0000000000000000 0.0000000000000000 0.0526436393207049
0.0000000000000000 0.0000000000000000 0.9473563606793014
0.3333333333333357 0.3333333333333357 0.0944879345040668
-0.3333333333333357 -0.3333333333333357 0.9055120654959324

POSCAR-$\alpha$-$Pb_2As_2$
1.00000000000000
3.677677983482954725 -2.1233083737575389 0.0000000000000000
3.677677983482653175 2.1233083737565872 0.0000000000000000
0.0000000000000000 0.0000000000000000 27.3195087628506776
Pb As
2 2
Direct
0.0000000000000000 0.0000000000000000 0.0557389838268677
0.0000000000000000 0.0000000000000000 0.9442610161731174
0.3333333333333357 0.3333333333333357 0.1027734402906814
0.3333333333333357 0.3333333333333357 0.8972265597093336

POSCAR-$\beta$-$Pb_2As_2$
1.00000000000000
3.677677983482954725 -2.1233083737575389 0.0000000000000000
3.677677983482653175 2.1233083737565872 0.0000000000000000

```
      0.0000000000000000      0.0000000000000000     27.3195087628506776
   Pb     As
     2      2
Direct
  0.0000000000000000   0.0000000000000000   0.0557389838268677
  0.0000000000000000   0.0000000000000000   0.9442610161731174
  0.3333333333333357   0.3333333333333357   0.1027734402906814
-0.3333333333333357  -0.3333333333333357  0.8972265597093336
```

POSCAR-$\alpha$-$Pb_2Sb_2$

```
   1.00000000000000
      3.9193672531710119 75   -2.2628477385762546 5    0.0000000000000000
```

```
   1.00000000000000
      3.919367253171011975   -2.262847738576254 65     0.0000000000000000
      3.919367253243375875    2.262847738704010675    0.0000000000000000
      0.0000000000000000      0.0000000000000000     27.7386661521943836
   Pb     Sb
     2      2
Direct
  0.0000000000000000   0.0000000000000000   0.0547209410739452
  0.0000000000000000   0.0000000000000000   0.9452790589260548
  0.3333333333333357   0.3333333333333357   0.1048009881253003
  0.3333333333333357   0.3333333333333357   0.8951990118746931
```

POSCAR-$\beta$-$Pb_2Sb_2$

```
   1.00000000000000
      3.919367253171011975   -2.262847738576254 65     0.0000000000000000
      3.919367253243375875    2.262847738704010675    0.0000000000000000
      0.0000000000000000      0.0000000000000000     27.7386661521943836
   Pb     Sb
     2      2
Direct
  0.0000000000000000   0.0000000000000000   0.0547209410739452
  0.0000000000000000   0.0000000000000000   0.9452790589260548
  0.3333333333333357   0.3333333333333357   0.1048009881253003
 -0.3333333333333357  -0.3333333333333357   0.8951990118746931
```

POSCAR-$\alpha$-$Pb_2Bi_2$

```
   1.00000000000000
     4.009153338919828625   -2.314685759449021325   0.0000000000000000
     4.009153338919734475    2.3146857594469621     0.0000000000000000
      0.0000000000000000      0.0000000000000000     31.1987269062439410
   Pb     Bi
     2      2
Direct
  0.0000000000000000   0.0000000000000000   0.0483810534543669
  0.0000000000000000   0.0000000000000000   0.9516189465456331
  0.3333333333333357   0.3333333333333357   0.0943381093385243
  0.3333333333333357   0.3333333333333357   0.9056618906615963
```

POSCAR-$\beta$-$Pb_2Bi_2$

```
   1.00000000000000
     4.009153338919828625    -2.314685759449021325    0.0000000000000000
     4.009153338919734475     2.3146857594469621      0.0000000000000000
     0.0000000000000000      0.0000000000000000     31.1987269062439410
   Pb    Bi
     2      2
Direct
  0.0000000000000000   0.0000000000000000   0.0483810534543669
  0.0000000000000000   0.0000000000000000   0.9516189465456331
  0.3333333333333357   0.3333333333333357   0.0943381093385243
 -0.3333333333333357  -0.3333333333333357   0.9056618906615963
```